**И. З. ШКУРЧЕНКО**

# СТРОЕНИЕ СОЛНЦА И ПЛАНЕТ СОЛНЕЧНОЙ СИСТЕМЫ С ТОЧКИ ЗРЕНИЯ МЕХАНИКИ БЕЗЫНЕРТНОЙ МАССЫ I

В данной монографии (1973 - 1974) автор использует теорию механики безынертной массы для исследования строения небесных тел солнечной системы. Строение планет и Солнца является единственной причиной их осевого вращения или отсутствия вращения, присутствия или отсутствия спутников и атмосфер. Оно есть один из основных климатических факторов для каждой планеты и Солнца, который определяет тип климата и его возможные изменения. Понимание этих процессов позволит знать перспективы эволюции Солнца и планет, включая Землю. Монография была разделена на две части редактором в 2007. Поскольку автор разработал некоторые теоретические положения "Механики жидкости и газа, или механики безынертной массы" (1971), то Часть I содержит эти изменения. Часть II содержит исследование, которое дает нам новое знание и новое значение накопленных знаний и информации о Солнце и планетах солнечной системы. Монография адресована специалистам в области теоретической и практической гидродинамики и смежных наук. Она будет полезна для астрономов, синоптиков и геологов.

**I. Z. SHKURCHENKO**

# THE STRUCTURE OF THE SUN AND THE PLANETS OF THE SOLAR SYSTEM FROM THE VIEWPOINT OF MECHANICS OF THE INERTLESS MASS I

In this monograph (written in 1973−1974) the author uses the theory of mechanics of the inertless mass to investigate the structure of heavenly bodies of the solar system. The structure of the Sun and planets is the sole reason of the character of their axial rotation, presence or absence of satellites and atmospheres. This structure is one of the main climatic factors for each planet and Sun. It determines the climate and its possible changes. Understanding these processes is very important for determining perspectives of the evolution of the Sun and the planets, including the Earth. This monograph was divided into two parts by editor in 2007. Since author has developed some theoretical positions of "Mechanics of liquids and gas, or mechanics of the inertless mass" (1971), the first Part contains these changes. The Part II contains the investigation that gives us new results and new meaning of the stored information about the Sun and the planets of the solar system. This monograph is addressed to specialists in the field of theoretical and practical hydrodynamics and adjacent sciences. It will be useful for astronomers, meteorologists and geologists.

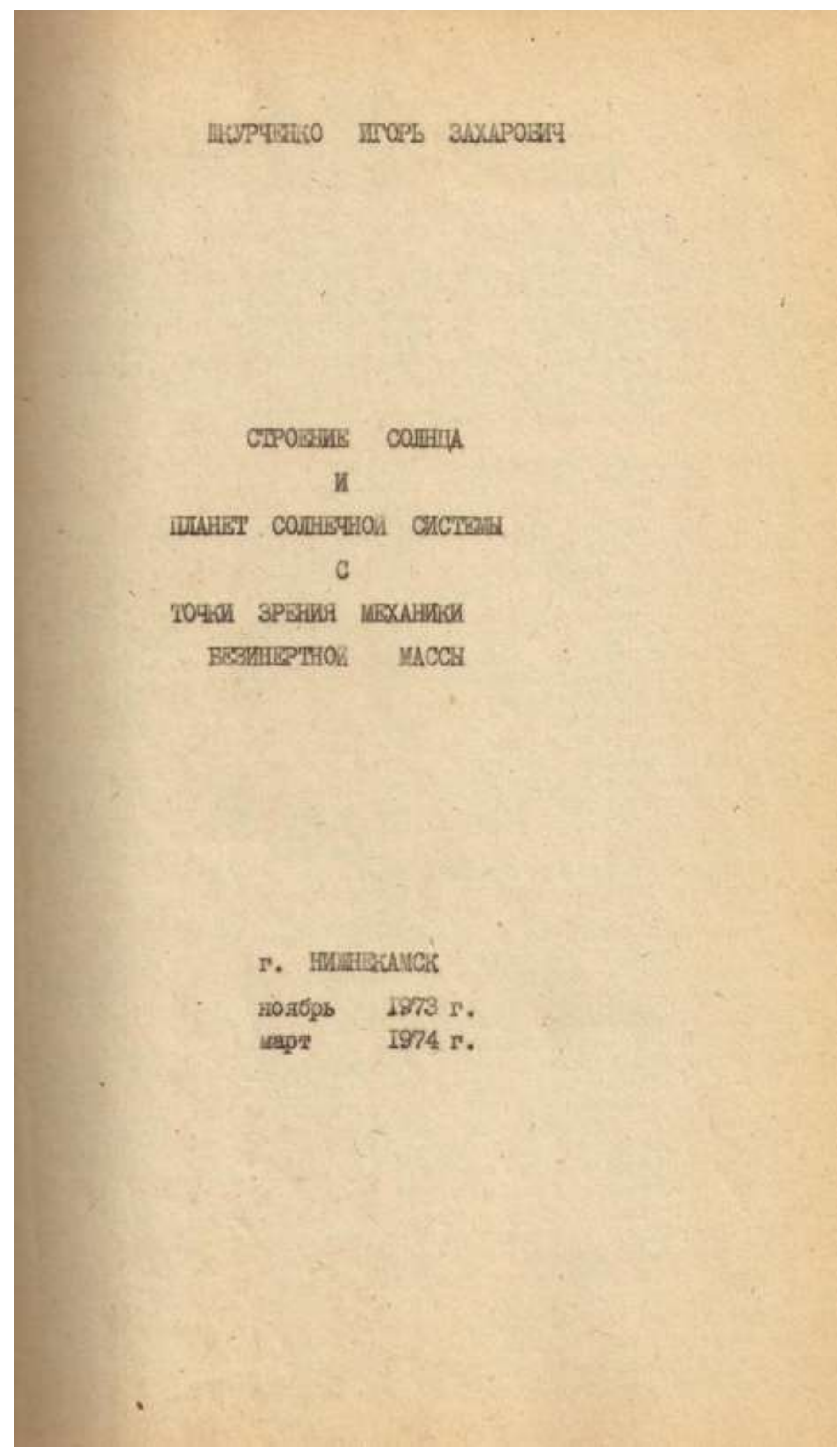

ШКУРЧЕНКО ИГОРЬ ЗАХАРОВИЧ

СТРОЕНИЕ СОЛНЦА
И
ПЛАНЕТ СОЛНЕЧНОЙ СИСТЕМЫ
С
ТОЧКИ ЗРЕНИЯ МЕХАНИКИ
БЕЗИНЕРТНОЙ МАССЫ

г. НИЖНЕКАМСК
ноябрь 1973 г.
март 1974 г.

*титульный лист авторской рукописи*

## ЧАСТЬ I[1]

[1] Файл «Строение Солнца и планет солнечной системы с точки зрения механики безынертной массы» был разделён редактором на две части, т.е. два файла, по техническим причинам.

**СОДЕРЖАНИЕ ЧАСТИ I:**

## ОТ АВТОРА

...Если они поймут, что в реальной основе мечты лежит человеческий труд, порою даже очень тяжёлый, и что всякое интересное рождается в реальных жизненных условиях, то их жизнь придёт к норме человеческого существования. Тогда они получат истинное наслаждение от своего жизненного пребывания на Земле. Человеческая забота об их будущем заставляет меня писать данную работу даже в таких условиях, в которых я сейчас нахожусь.

В своём путешествии мы посетим планеты солнечной системы и само Солнце. Посмотрим на их строение с позиций механики безынертной массы. Чтобы сделать путешествие реальным, мы должны будем хорошо усвоить положения и законы этой механики. Поэтому начнём с того, что сначала усвоим для себя в необходимой степени механику безынертной массы. Это будут первые и самые большие трудности в нашем путешествии, которые надо преодолеть, чтобы увидеть необычную красоту гармоничного построения планет солнечной системы.

***

***<u>От редактора</u>*:** Рукопись *Строение Солнца и планет солнечной системы с точки зрения механики безынертной массы(1974г)*, вторая после работы *Механика жидкости и газа, или механика безынертной массы (1971 г.),* написана автором для своих детей (об авторе см. *Предисловие редактора* к работе [1]). Дело в том, что к тому времени автор лишился возможности профессионально проводить исследования, поскольку был уволен из КБ. Нашёл работу, но по другой специальности, связанную с многомесячными командировками. Поэтому дети росли без него. В этих обстоятельствах автору было не до формы, т.к. даже в таких условиях, нужно было зафиксировать свои мысли на бумаге и в то же время просветить детей, пока это ещё позволял сделать их возраст.

Т.к. рукопись носит неформальный характер, редактору пришлось изъять из текста то, что прямо не относится к теме, в том числе от главы *От автора* редактор оставил только два абзаца. Сейчас читать полный текст неуместно. Прошу помнить, что везде, где редактором оставлено местоимение *вы*, оно означает обращение автора непосредственно к своим детям. Автор старается объяснить механику безынертной массы таким образом, чтобы она стала понятна подросткам, подобно тому, как механику Ньютона преподают в старших классах. Механика безынертной массы должна была стать таким же необходимым школьным предметом, как механика твёрдого тела, и по той же причине. Следовательно, взрослым данные автором объяснения должны быть тем более понятны.

В оригинале работы «Строение Солнца и планет солнечной системы» автором был дан расчёт крыльчатки центробежного насоса. Редактор переместил этот расчёт в *Приложение* к основному труду - *«Механика жидкости и газа, или механика безынертной массы»*. По этому расчёту читатель, уже знакомый с этой работой, может судить, что текст этой второй рукописи совсем не детский, «на вырост», как автор ни старался.

Пояснения автора своей теории помогут правильно понять механику безынертной массы. Первая часть текста - не просто краткое изложение основных положений механики безынертной массы. Поэтому редактор считает, что читателю, у которого появилось серьёзное намерение дальше разбираться с новой теорией, абсолютно необходимо прочитать эту часть текста. Кроме того, будут более понятны соображения автора, изложенные во второй части, содержащей описание строения небесных тел.

Что касается слова *путешествие*, то это способ привлечения внимания этих детей к серьёзным вещам. Собственно, расхождения тут нет. Путешествия или, говоря иначе, экспедиции людей всегда носили именно *исследовательский характер*, кроме последнего времени, когда путешествия выродились в бессмысленный туризм, в развлечение, так что уже появились космические туристы. В общем, путешествие значит – исследование. Исследование планет, включая Землю, планеты и Солнце в данном случае. Таким образом, автор приглашает к исследованию. Редактор - тоже. О результатах этого путешествия - судить читателю.

Отмечу, что если бы рукопись не была сокращена редактором, то её стиль мог бы послужить хорошим образцом для подражания учителям и преподавателям. Как и зачем нужно преподавать сложное просто, на вырост, разбудить в детях дух первооткрывателей, а не задушить его рутиной, не примитивно в смысле забавно или «оригинально», ибо действительность не нуждается в искусственных украшениях, чтобы вызывать неподдельное восхищение. Всё подобное так же нелепо и вредно делать, как на цветущем лугу прятать полевые цветы под бумажными, чтобы «произвести впечатление» и «привить» детям некую «любовь к природе». Поэтому редактор оставил достаточно, чтобы отчасти показать эту ценность рукописи, которая станет актуальной, когда преподаватели вновь начнут испытывать естественную тревогу не за неких детей, а за будущее, которое в их руках, и за прошлое, т. к. недоросли обязательно обесценивают всю прожитую не только вами, но и всеми предыдущими поколениями, жизнь, чего и врагу не пожелаешь. Всё сокращённое мной имеет научную ценность, но иного плана, поэтому напомню, что было сокращено по той причине, что не относится прямо к теме механики, и будет ждать своего часа.

P.S. В предыдущей редакции редактор допустил ошибки, в том числе смысловые. В этой редакции они были устранены по мере их обнаружения. Читатель должен использовать v2 работ И.З. Шкурченко, см. https://arxiv.org/search/physics?searchtype=author&query=Shkurchenko%2C+I

В настоящее время редактор готовит для публикации исправленную версию «Механики жидкости и газа, или механики безынертной массы», которая скоро появится в архиве.

## Глава 1. ОСНОВНЫЕ ЗАКОНЫ И ПОЛОЖЕНИЯ МЕХАНИКИ ЖИДКОСТИ И ГАЗА, ИЛИ МЕХАНИКИ БЕЗЫНЕРТНОЙ МАССЫ

Механика жидкости и газа, или механика безынертной массы, является наукой, которая изучает одно из основных свойств массы, проявляющихся в реальных жизненных условиях. Теперь попробуем представить себе это определение в допустимой для нашего понимания форме.

Вы уже имеете некоторое представление о массе. Поэтому вы знаете, что содержание всех окружающих вас тел составляет масса. Под механикой мы обобщённо понимаем те физические действия, которые мы должны уметь совершать с этими телами: заставлять их двигаться различными способами, останавливаться в нужный момент и оставлять их в этом состоянии на протяжении нужного нам времени. Поэтому механика именно с этой точки зрения смотрит на окружающую вас материальную природу. Материальный мир очень разнообразен, и формы его движения многочисленны. Механика твёрдого тела была получена и написана Ньютоном. Изучив её законы, вы сможете использовать различные механические состояния твёрдых тел для своих нужд. Современная механика своим высоким уровнем развития обязана механике Ньютона. На основе этой механики им же была создана механика небесных тел, благодаря которой стали возможны космические полёты. Это стало возможным потому, что законы механики твёрдого тела выразили общность механических состояний для любых твёрдых тел. Поэтому каждый изучивший эти законы может получить для себя необходимое движение или состояние покоя, интересующего его твёрдого тела. В механике жидкостей и газов мы тоже будем стремиться к подобному.

Если вы к жидкости и газу попытаетесь применить законы механики твёрдого тела, то у вас ничего из этого не выйдет, например, когда вы задумаете лопатой черпать воду или воздух, а землю вы свободно перебрасываете лопатой с места на место. Поэтому, когда вам надо заполнить камеры вашего велосипеда воздухом, то вы прибегаете к помощи насоса. Твёрдое тело, в какую бы тонкую муку не было бы измельчено, вы всегда сможете взять руками. Правда, жидкость можно черпать, например, стаканами, но если вы будете находиться под водой, то вам тоже придётся прибегнуть к помощи насоса. Эти примеры говорят о том, что законы механики твёрдого тела не приемлемы для жидкостей и газов. Таким образом, мы сузили себе задачу, отбросив твёрдые тела, которые тоже обладают массой.

В механике твёрдого тела под понятием твёрдость, которое относится к разряду свойств тел, определяется принадлежность тел к общим механическим движениям. Например, по вашим понятиям, подушка – мягкая, но она в своих механических движениях тоже подчиняется законам механики твёрдого тела. Следуя данной аналогии, мы можем жидкие и газообразные тела выделить тоже одним свойством, например, свойством текучести. Таким образом, мы выделили предмет для изучения механических движений жидкостей и газов. Ибо текучесть является общим свойством и для жидкостей, и для газов. Основное различие между жидкостью и газом заключается в том, что жидкость практически не сжимаема, а газы сжимаемы. Но это различие не является таким коренным, что разделяло бы общность в движении жидкостей и газов. Поэтому мы его можем учесть позже в определённых расчётах.

Текучесть действительно является общим свойством жидкостей и газов. Ибо нам, будучи погружёнными в воду или воздух, в том и другом случае придётся пользоваться для различных их перемещений одними и теми же приспособлениями и механизмами, например насосами. Тем самым мы подчеркнули, что законы механики жидкости и газа, или механики безынертной массы, будут действительны только в том случае, когда мы находимся полностью погружёнными в жидкости или газы. В действительной ситуации мы не всегда можем находиться в подобных условиях. Да в этом нет никакой необходимости. Подобное разъяснение нужно лишь для того, чтобы вы поняли, каким образом люди отвлекаются от различных реальных условий движения жидкостей и газов, чтобы выявить конечную общность в механических движениях жидкостей и газов.

В реальных условиях вы видите, например, воду в виде рек, морей, океанов и текущую из водопроводного крана. Жидкости бывают различными: ими может быть и ртуть, и вода, и расплавленный металл, и сжатый воздух. Движение воздуха вы встречаете в виде ветра, урагана и смерча. Звук, который вы слышите, тоже передаётся при помощи особого движения жидкости или газа. Всё это природное разнообразие жидкостей и газов, как по химическому составу, так и по движению можно привести к более общим понятиям. Для этого нам не придётся погружаться в жидкости и газы, а мы просто должны будем стать на позиции механики безынертной массы и сделать обзор будущего предмета своих исследований. Тогда с этих позиций мы увидим следующее:

что всё пространство, как бы оно ни было велико, заполнено однородной жидкостью, а жидкость представляет собой массу, свойством которой является текучесть.

Вот это пространство, заполненное жидкостью, назовем средой. Используя это название, мы в дальнейшем можем обойтись без дополнительных разъяснений.

Всякая среда может содержать в себе различную жидкость. Это различие мы определим плотностью, которую принято обозначать буквой $\rho$. Количественной величиной плотности является количество массы в единице объёма. Эта величина нам понадобится для количественной характеристики среды.

Далее, чтобы ещё больше упростить понятие среды, мы полагаем, что жидкость, заполняющая пространство, не обладает вязкостью и не обладает сжимаемостью, т.е. она несжимаема. Этим упрощением мы выравниваем все жидкости по текучести, потому что мы знаем, например, что мёд более вязкий, чем растительное масло, а масло более вязкое, чем вода, и приравниваем таким образом газы к жидкостям, лишая их сжимаемости.

Теперь мы получили сравнительно простую модель жидкостей и газов, которая по своим свойствам близка к некоторым реальным жидкостям, например, к такой жидкости, как вода. С подобными упрощениями жидкость, наполняющую пространство, принято считать идеальной. Следовательно, среда будет представлять собой пространство, полностью заполненное идеальной жидкостью.

Коль мы для своих исследований приняли среду, или пространство заполненное идеальной жидкостью, то мы должны хорошо представлять себе механическое движение жидкости в выбранной нами среде. О движении твёрдого тела вы уже имеете представление: оно имеет скорость, ускорение и пройденный путь за определённое время, который принято называть траекторией. Мы легко можем представить механическое движение твёрдых тел, так как можем наблюдать его. Несмотря на такую лёгкость, человечеству потребовался долгий путь развития, чтобы полностью выявить механические движения твёрдых тел. Окончательную и обобщённую форму этих движений получил Ньютон. С механическим движением жидкостей и газов дело обстоит несколько хуже, хотя человек научился строить корабли, самолёты и ракеты. На данном этапе человеческого развития люди имеют весьма приблизительное понятие о движении жидкостей и газов, ибо за ним очень трудно наблюдать. Поэтому, используя понятие среды и известные наблюдения за движением жидкостей и газов, мы постараемся выразить, пока в общей форме, понятие об их движении. Тогда движение идеальной жидкости мы можем представить себе в общем виде в следующей форме:

1. что при движении перемещается вся масса жидкости, заключённая в исследуемом пространстве;

2. что границы этого пространства определяют нам форму потока движущейся массы;

3. что количественной величиной любого движущегося потока является расход массы в единицу времени.

Здесь мы представляем себе движение жидкости как перемещение её массы в пространстве, полностью заполненном этой жидкостью.

Такое мы можем себе представить, хотя это не так просто сделать. Ибо масса жидкости не представляет собой общий монолит твёрдого тела, а состоит из каких-то мельчайших структурных единиц, которые перемещаются в этом едином потоке. Сами по себе эти структурные единицы занимают незначительный объём и обладают пространственной подвижностью, которая определяется свойством текучести жидкостей. В настоящее время эту структурную единицу жидкости определяют как атом или молекулу.

Пространственно же любое движение жидкости мы можем представить себе по границам, в которых размещается движущийся объём жидкости, который мы называем потоком. Ибо в этом потоке мы лишь подразумеваем себе движущиеся структурные единицы, а осмысленно понимаем его, как движение сплошной однородной массы без всяких структурных единиц в границах исследуемого потока. Поэтому пространственно движение жидкости мы представляем себе как поток, границы которого мы видим. В наших исследованиях мы должны для этого совместить границы среды с границами движущегося потока. В этом случае мы не только сможем представить себе поток пространственно, но и измерить его количественно. Количественной величиной потока является расход массы в единицу времени. Количество массы определяется плотностью, то есть количеством массы в единице объёма. Поэтому расход массы в единицу времени будет выглядеть как некоторый объём, заполненный массой с соответствующей плотностью. Отсюда следует, что движение жидкости в общем случае выражается как вытеснение текучей массы из объёма, а для твёрдого тела оно выражается как перемещение определённого объёма в пространстве. В этом заключается качественное различие между движением твёрдого и жидкого тела. Это значит, что механическое движение жидкости и газа есть новый тип механического движения массы, отличный от механического движения твёрдого тела.

Даже при таком специфичном определении среды и её движения вы, пожалуй, скажите, что вам всё это известно и зачем утверждать это как новое. Да, действительно, вам в какой-то степени всё было привычно, например, открой водопроводный кран, и потечет вода, если в речке бурное течение, то не следует лезть в воду. Но вам другое непонятно. Вы слышите, что строятся плотины, новые самолёты или ракеты, новые турбины и т.д. Все эти современные достижения даются человечеству за счёт больших физических и материальных затрат. Ибо им приходится ставить сотни и тысячи экспериментов, чтобы улучшить хоть на немного технические достижения прошлых лет. Лишь сотни и тысячи экспериментов дают возможность найти небольшое новое для технического использования движения жидкостей и газов, поскольку люди в настоящее время имеют неправильное представление о движении жидкостей и газов. Прежде всего, это неправильное в понимании людей связано с тем, что они объясняют движение жидкостей и газов с точки зрения механики твёрдого тела. Современные ученые всеми силами стараются приспособить законы механики твёрдого тела к движению жидкостей и газов. Позже вы в этом сможете убедиться, когда сопоставите механику твёрдого тела с механикой безынертной массы. Это значит, что современная механика жидкости и газа даёт неправильное истолкование движению жидкостей и газов. Поэтому она является большим тормозом в развитии современной техники и тем самым заставляет всё человечество делать большие материальные затраты на достижение каждого нового технического уровня. Это значит, что новые достижения в технике идут за счёт опыта, накопленного при создании конструкций прошлого, и современных экспериментальных данных. Так что для вас движение жидкостей и газов является потребительски привычным, а для современных ученых и инженеров оно является абсолютно новым. Поэтому мне приходится так старательно всё вам объяснять.

Будем считать, что вы познакомились и поняли основное, или общее, условие механики жидкости и газа, или механики безынертной массы, то есть вы усвоили, что объектом исследования механики является среда, или пространство заполненное идеальной жидкостью, что движение жидкостей выражается потоком текучей массы, который движется в границах обозначенной среды. Теперь мы можем перейти к законам и основным положениям механики жидкости и газа.

Вы спросите, почему именно к законам? Постараюсь ответить на этот вопрос. Мы получили свой объект исследования в виде среды, содержание которой составляет текучая масса, то есть текучесть мы выделили как основное свойство массы среды. Коль эта масса текучая, то нас будет интересовать вопрос, каким образом можно заставить эту массу сохранять нужное нам механическое состояние покоя или движения. Для этого нам необходимо будет сориентировать свои действия относительно среды. Ведь мы понимаем всё лишь в том случае, когда сможем хотя бы мысленно представить свои действия относительно интересующего нас объекта.

Механические действия мы можем совершить над любым интересующим нас объектом не такие, какие нам заблагорассудятся, а такие, которые требует сам объект в зависимости от своих свойств. Свойства же объектов и механические действия, которые можно и должно совершать над ними, имеют природное происхождение. Поэтому свойства и соответствующие им механические действия определяются самой природой этих объектов и существуют независимо от нашего сознания, то есть объективно. Поэтому эти наши механические действия мы не можем сами придумывать, а мы их можем только находить, познавая свойства.

В окружающих нас жизненных условиях мы наблюдаем текучую массу в самых разнообразных механических состояниях, которые часто кажутся нам несовместимыми друг с другом. Например, что может быть общего между звуковой волной, течением реки, взрывом пороха и спокойным воздухом? В то же время общее в механическом движении текучей массы всегда присутствует, несмотря на всё многообразие встречающихся нам форм движения жидкостей и газов, которое определено самой природой вещей. Вот это общее мы должны найти.

Для поисков общего в механическом состоянии текучей массы человек использует не только простые наблюдения за её естественным механическим состоянием, но и ставит специальные эксперименты, использует для этого известные приборы и при необходимости создаёт новые.

Поиски эти могут продолжаться в течение десятков и даже сотен лет, и лишь в определённое время определённому человеку, наконец, удаётся найти это общее, используя опыт своей исследовательской работы и тот опыт, который накопило всё человечество в своей исследовательской работе за сравнительно продолжительное время своего существования. Вот такие большие усилия тратит всё человечество для поиска общего в определённых свойствах исследуемых объектов. Когда это общее найдено, его излагают в виде определённой словесной формулировки. В одних случаях может быть достаточно одной такой формулировки, в других случаях их может быть несколько, чтобы выразить полностью всё необходимое общее в исследуемых объектах. Затем каждую такую формулировку называют законом. Это делается для того, чтобы утвердить найденное общее как общее для всех, чтобы современники и будущие поколения людей не тратили своих сил и средств на его поиск. Поэтому само определение *закон* обозначает обязательное для исполнения, и формулировки с добавлением к их названию термина *закон*, выражающие общность в исследуемых объектах, становятся обязательными для исполнения. В этом случае каждый, кто сомневается в этих законах, может проверить их экспериментом, чтобы убедиться для себя в их правильности. Ибо всякая природная общность выявляется на основе наблюдений, если даже таковые ведутся с помощью приборов. Сформулированное общее всегда содержит сущность нового, ранее неизвестного человечеству, и в дальнейшем становится основой для количественной разработки в виде математических зависимостей этого общего. Ибо оно выявляет для человека новые действия, которые ранее были ему неизвестны, а новые действия требуют новой определённой математической обработки, вернее, новые действия дают возможность для количественной обработки. Поэтому законы различных наук различны и их количественные зависимости тоже различны. Для примера вы можете взять такие науки, как химия, физика, биология, астрономия, и убедиться в этом. Это значит, что законы являются фундаментом любой науки, на основе которых в последующем строится всё научное здание под названием наука с соответствующим наименованием. Всё это вместе взятое называется теорией той или иной науки.

Как мы выяснили выше, действительно новое в любой науке выражается в её законах, которые показывают общность исследуемых объектов. Всё остальное в таких науках по отношению к её законам уже не будет новым. Это всё остальное в виде количественных зависимостей и других положений будет либо вытекать из найденной общности, либо будет являться определённым пояснением к этому общему. В настоящее время существует много наук, где не найдена общность в изучаемых явлениях природы. В таких науках содержится в основном опыт исследований, накопленный человечеством, в виде эмпирических зависимостей, поясняющих положений, и всё это располагается в определённой последовательности. Подобное положение в науке выражает лишь определённую ступень в её развитии. Поэтому оно требует от человечества дальнейших поисков необходимых общностей.

К настоящему времени современная механика жидкости и газа тоже относится к разряду подобных наук. Ибо её теория строится на общности с теорией механики твёрдого тела. Поэтому среду представляли в виде небольших объёмов, которые способны сохранять механическое состояние покоя и движения в соответствии с законами механики твёрдого тела. По этой причине к настоящему времени теория механики жидкости и газа была не верна, и пользоваться ею для практических расчётов просто невозможно. Лишь некоторые количественные зависимости, типа уравнения Бернулли, относятся к разряду действительной механики жидкости и газа. Только эти зависимости приносили некоторую пользу при практическом использовании этой теории и давали возможность в определённой степени развивать технику. Но основное развитие техника получила с помощью эксперимента. При тех кажущихся больших темпах развития техники, они всё равно довольно низкие, так как механика жидкости и газа, конечно, общепризнанная и современная, является большим тормозом в развитии науки и техники. Все эти положения старых и новых механик вы сможете сопоставить немножко позднее, когда ещё подрастете и подучитесь. Сейчас вы просто должны понять, что новым является только общность свойств изучаемых объектов, выраженных в виде законов. Вот эти законы являются открытием для человечества, несмотря на то, что какая-либо наука уже имеет некоторый материал в виде зависимостей и разъясняющих положений. В данной работе мы будем излагать законы и зависимости, которые уже были изложены мной в работе под названием *Механика жидкости и газа, или механика безынертной массы* в 1971 году.

***

Теперь нам остаётся вспомнить, что мы имеем дело со средой, или пространством, заполненным идеальной жидкостью. Нам следует записать для неё общность в виде законов, которые бы содержали общие способы механического обращения со средой.

Начнём с первого закона. Первый закон называется *законом сохранения состояния*. Он формулируется следующим образом:

***жидкости и газы сохраняют энергию покоя и установившегося движения только в силовом поле и изменяют её лишь при изменении этого поля.***

Сделаем пояснение к этому закону.

Мы в нашем законе утверждаем, что жидкости и газы образуют среду только в силовом поле. На этом основании вы согласитесь, что с помощью силового поля можно организовать движение жидкости и газа и что только силовым полем оно организуется, то есть никаким другим способом оно не может быть организовано.

Мы рассмотрели связь жидкости с силовым полем в движущемся потоке. Ибо здесь можно наиболее наглядно показать эту связь. В состоянии покоя жидкости и газы среды могут находиться тоже только под воздействием силового поля. Ведь силы и там тоже присутствуют. Их выражают как давление. Что означает силу, приложенную к единице площади. Коль силы в покоящейся жидкости присутствуют, то они могут быть организованы только силовым полем. Вам должно быть это ясным из вышеизложенного. Поэтому мы можем утверждать, что способ механического воздействия для жидкостей и газов находящихся в состоянии покоя и движения одинаков и выражается в воздействии, как в том, так и в другом случае, силовым полем.

Классификацию силовых полей мы здесь разбирать не будем по той простой причине, что мы просто не знаем все виды силовых полей, которые предстоит ещё изучить. Для наших целей нам необходимо будет знать только то, что все силовые поля бывают с направленным и ненаправленным действием, или, выражаясь математическим языком, силовые поля бывают векторными и скалярными. Примером векторного силового поля может служить гравитационное поле Земли, а примером действия скалярного силового поля могут служить сосуды, в которых жидкости и газы находятся под давлением.

Теперь уточним их силовое воздействие. Для чего снова вернемся к магниту и к его силовому полю. Возьмём , например, какой-то определённый магнит и несколько одинаковых железных шариков. Если мы теперь к этому магниту будем подносить на определённое расстояние каждый из этих шариков, держа их в руке, то каждый раз мы будем ощущать силу одной и той же величины. При таком силовом воздействии не участвуют никакие промежуточные предметы между шариком и магнитом. В этом заключается основная особенность силового взаимодействия в силовом поле, которое отличается от силового взаимодействия твёрдых тел. В твёрдых телах это взаимодействие происходит непосредственно в точке соприкосновения этих тел. В силовом поле взаимодействие, или связь, одностороннее. Другой отличительной особенностью силового поля будет то обстоятельство, что величина действующих сил на притягиваемый ими предмет будет зависеть только от напряженности силового поля. В твёрдых же телах сила взаимодействия зависит, прежде всего, от массы этих тел и скорости их движения. Это вам известно. Здесь была описана не абсолютно точная картина силового взаимодействия в силовом поле, а лишь смоделированная, которая нам помогла наглядно представить это взаимодействие.

Сама же сила, какая бы она ни была, рассматривается как величина, которая действует в определённый момент времени. Существование сил во времени определяет энергия.

В технической системе единиц она измеряется в килограммометрах, а в международной – в джоулях. Энергия имеет одинаковую размерность с работой. Под работой мы понимаем работу, которая уже выполнена кем-то или чем-то. Под энергией мы понимаем тоже работу, но не простую, а располагаемую, то есть такую работу, которую мы можем получить от имеющегося источника энергии. Работа, совершаемая в настоящий момент, называется мощностью. Количественно она определяется количеством работы в единицу времени. Все эти три характеристики одинаковой размерности имеют разное назначение в науке и технике. Поэтому определяются различными зависимостями.

Основным назначением работы и её зависимостей служит то обстоятельство, что работу, как уже совершенное, измеряют и подсчитывают для определения того, насколько полно используется тот или иной источник механической энергии. Поэтому с помощью её характеристик определяют просто количество совершенной работы и коэффициент её использования, который в технике называют коэффициентом полезного действия.

Мощность характеризует фронт использования того или иного источника энергии. Она отвечает на вопрос, как быстро используется энергия источника энергии. Энергия же, как будущий работник, проявляет себя через силу, которую мы можем ощутить и замерить. Непосредственным проявлением энергии является сила, которая служит как бы работником энергии, то есть является её непосредственным исполнителем.

Как выше мы определили, действенной силой в жидкостях и газах являются силы силового поля, т.е. силы, которыми силовое поле притягивает или выталкивает массу жидкости среды. Поэтому масса жидкости является как бы промежуточным телом, с помощью которого силовое поле проявляет свою энергию. По аналогии с тем примером, когда мы держим в руке железный предмет и ощущаем силу, с которой притягивается этот предмет магнитом.

Поскольку масса жидкостей и газов является лишь промежуточным телом в силовом поле, то это тело будет характеризовать энергию силового поля, в котором оно находится, а не свою собственную. По этой причине мы в первом законе определили, что жидкости и газы сохраняют энергию силового поля. Энергию эту они сохраняют и в состоянии покоя, и в состоянии установившегося движения. Как видим, это определение является общим для покоящихся и движущихся жидкостей, то есть оно выражает определённую общность, которая, в свою очередь, отвечает на вопрос, каким способом или чем мы должны действовать на жидкость, чтобы изменить её то или иное механическое состояние в нужном для нас направлении. Количественные зависимости для выражения энергии мы получим ниже. Для нас сейчас важно только одно: нам надо выяснить общий способ воздействия на жидкость. Мы его выяснили. Теперь мы знаем, что на жидкость можно действовать только путем изменения силового поля. Конкретно мы понимаем это тогда, когда становимся способными выразить свои действия. Ибо всё мы понимаем через конкретность выражения собственных действий относительно рассматриваемого.

Подобный закон можно получить только непосредственным наблюдением за состоянием жидкостей и газов. Поэтому он является не придуманным, а объективным законом природы, который мы нашли путем непосредственного наблюдения.

Теперь перейдем ко второму закону механики безынертной массы, который называется *уравнением сил расходного вида движения.* Запишем его формулировку:

***динамические силы давления $P_{\text{дин}}$ равны произведению расхода массы в единицу времени M на линейную скорость W и делённому на площадь сечения потока F, или***

$$P_{\text{дин}} = \frac{M}{F} W. \qquad (1)$$

где $P_{\text{дин}}$ – является силой, действующей на единицу площади сечения потока, $W$ – линейная скорость, которая измеряется *количеством длины в единицу времени* (выд. – *ред.*); $M$ – расход массы в единицу времени, которая движется в исследуемом потоке жидкости[2], $F$ – площадь сечения потока жидкости.

Этот закон выражает общую зависимость для сил, действующих в потоке жидкости. Коль мы её возвели в ранг закона, то она должна выражать общность, присущую любому механическому состоянию жидкости и газа.

Теперь постараемся выяснить, является ли эта сила общей для любых механических состояний жидкости и газа. Для чего возьмём среду с покоящейся и движущейся жидкостью. Коль мы в нашем законе нашли силу для движущейся жидкости, то вы, наверное, согласитесь с тем, что она является общей для любой движущейся жидкости. Состояние покоя жидкости мы должны понимать как состояние застывшего движения, то есть аналогично состоянию покоя твёрдых тел.

---

[2] По техническим причинам редактор обозначает *расход массы* не буквой М с точкой наверху, а буквой М без точки. Буква m будет являться обозначением *массы.*
Как видно из текста, движется масса, в смысле, вещество среды, с определённой скоростью, хотя в других текстах автор употребляет, например, выражение «пройдёт расход массы». Как может расход массы проходить? Расход массы - это количество массы, которое прошло через площадь сечения в единицу времени. Поэтому, например, можно сказать, что через данную площадь сечения пройдёт 5 кг массы в единицу времени. И вот эти 5 кг образуют объём, разумеется, условный объём, а не реальный, необходимый тоже для количественной характеристики движения, работы. Или можно сказать так, что на данной площади сечения произойдёт расход 5 кг массы с такой-то скоростью. Редактор считает, что наиболее подходящая терминология установится со временем.

Тогда зависимость для силы станет общей для жидкости, находящейся в состоянии покоя или движения.

Например, общность силы для твёрдых тел выражается вторым законом Ньютона, который гласит, что сила равна массе твёрдого тела, умноженной на ускорение. В наших земных условиях состояние покоя твёрдого тела мы воспринимаем через силу как вес данного тела. В этом случае мы выражаем вес этого тела тоже как массу, умноженную на ускорение. В этом случае ускорение у нас равно 9,81 *м/сек* $^2$. Как мы знаем, ускорение является характеристикой движущегося тела, а мы применяем её к покоящемуся телу. Следовательно, мы принимаем состояние покоя твёрдого тела как состояние застывшего движения. Аналогичным образом ведет себя жидкость, находящаяся в состоянии покоя в поле земного тяготения. Разница здесь будет заключаться в том, что в зависимости для сил твёрдого тела присутствует ускорение, а в зависимости для сил жидкостей его нет, там имеется скорость. Поэтому твёрдое тело, брошенное с определённой высоты, в поле земного тяготения будет падать с постоянным ускорением $g$ = 9,81 *м/сек* $^2$, а жидкость, тоже брошенная с высоты, в поле земного тяготения будет падать с постоянной скоростью $w = \sqrt{9{,}81} = 3{,}132$ *м/с.*[3]

Как видите, здесь мы наблюдаем коренное различие между жидкостью и твёрдым телом. Это различие вы можете легко проверить в своих условиях. Падение твёрдого тела вам нечего проверять, так как оно неоднократно было проверено и в современных условиях применяется для практических расчётов. Падение жидкости вы можете проверить следующим образом: для этого вам придётся взять, например, стеклянную трубку, длиной в несколько метров. Заполнить её водой. Трубка должна стоять вертикально по отношению к земной поверхности. Затем вы должны будете отметить на этой трубке два уровня на расстоянии нескольких метров друг от друга. Затем вы открываете конец трубки и замеряете время, за которое вода пройдет через отмеченные вами уровни. После чего вы легко найдете скорость движения и с помощью несложных вычислений сверите её с вышеуказанной цифрой, т.е. $w$ = 3,132 *м/с.* Можно это различие проверить другими способами, но мы здесь дали наиболее простой из них, который требует очень простых приспособлений.[4]

Вот мы и выяснили общность динамических сил давления $P_{дин}$ и убедились, что она должна относиться к разряду законов. Зависимость (1) динамических сил давления относится к количественным величинам, которые могут быть получены только экспериментальным путем, а не абстрактными вычислениями. Поэтому каждый сомневающийся может убедить себя только с помощью эксперимента. В общем, эксперимент является главным судьей этого закона.

Добавим, что зависимость для динамических сил давления (1) известна в газовой динамике сравнительно давно под названием реактивной тяги. Поэтому она применялась для вычисления только реактивной тяги. По этой причине она имела частный характер, так как для других механических состояний жидкостей и газов её не применяли. Наша заслуга здесь состоит в том, что мы её возвели в ранг закона, определив её как общность для любых механических состояний жидкости и газов.

В механике твёрдого тела так же существует понятие количества движения:

$$K = m \cdot W, \tag{2}$$

где $K$ – количество движения, $m$ – масса, $W$ – скорость.

Если принять в уравнении (2) массу постоянной, а скорость переменной, то мы можем взять первую производную от количества движения по времени. Подобные приёмы применяются в высшей математике. Поскольку вы её не знаете, то мы выразим это с точки зрения вашего понимания. Тогда смысл будет таков: приняв скорость переменной, а массу постоянной в уравнении (2), мы тем самым определили, что скорость $W$ и количество движения $K$ мы можем разделить на время, а массу $m$ нет. Тогда получим:

---

[3] См. гл. 2.3 «Состояние покоя среды в векторном силовом поле» работы [1]. Автор говорит о движении вещества среды под действием сил земного притяжения, т.е. «падение жидкости в жидкости, газа – в газе», что им будет разъяснено далее, когда будет рассматриваться строение и динамика атмосфер планет и Земли в т.ч., и недр небесных тел. Понятно, что, например, капли дождя падают с ускорением.

[4] Насколько редактор помнит слова автора, нижний конец трубки в этом опыте должен находиться под водой, причём резервуар должен быть настолько большим, чтобы подъём уровня был практически нулевым.

$$\frac{dK}{dt} = m\frac{dW}{dt}. \quad (3)$$

В уравнении (3) мы разделили количество движения $K$ и скорость движения $W$ на время $t$, как оно обозначается в высшей математике. Тогда количество движения делённое на время будет ни что иное, как сила $R$, а скорость делённая на время будет ускорением $a$. Если мы эту расшифровку подставим в уравнение (3), то получим:

$$R = m\cdot a.$$

Это уравнение означает, что сила $R$ равна массе, умноженной на ускорение $a$, то есть мы получили второй закон Ньютона. Если мы теперь снова разделим количество движения $K$ и массу $m$ на время, для чего нам придётся принять массу $m$ переменной, а скорость постоянной, то получим:

$$\frac{dK}{dt} = \frac{dm}{dt}W. \quad (4)$$

Как мы знаем, количество движения, делённое на время, есть сила $R$, а масса $m$, делённая на время $t$, есть расход массы в единицу времени $M$. Подставим нашу расшифровку в уравнение (4), получим:

$$R = M\cdot W. \quad (5)$$

Мы получили наше уравнение (1), которое гласит, что сила равна расходу массы в единицу времени умноженному на скорость. Если мы разделим уравнение (5) на площадь потока $F$, то получим:

$$\frac{R}{F} = \frac{M}{F}W.$$

Сила $R$, делённая на площадь потока $F$, есть ни что иное, как динамическое давление $P_{дин}$. Подставим его в уравнение, получим:

$$P_{дин} = \frac{M}{F}W.$$

Таким образом, мы снова получили уравнение динамических сил давления (1), но через уравнение количества движения. Что указывает на одинаковое происхождение сил твёрдого тела и динамических сил давления жидкостей и газов. Следовательно, мы получили вторичное подтверждение в правильности того, что динамические силы давления являются общими для любых механических состояний жидкостей и газов.

[*Отвлечёмся немного, ребята, и сделаем паузу отдыха. Ведь это не фантастическая или приключенческая повесть, которая приятно щекочет ваши ощущения, тогда вы чувствуете себя необыкновенными людьми, способными совершать чудеса. Но эти ощущения длятся недолго, лишь до тех пор, пока вы читаете эти повести. Поэтому потребность в ощущении необычного заставляет вас поглощать десятки и сотни томов подобных повестей. Тем самым вы отвлекаетесь от реальной жизни и перестаёте ощущать её красоту. Ведь в реальной жизни всё происходит не по щучьему велению, а под влиянием ваших действий и сил. Поэтому все ваши физические затраты в свою очередь требуют от вас соответствующих знаний и умения. Это необходимо для того, чтобы ваша деятельность не затрачивалась на переливание из пустого в порожнее и чтобы она не выглядела для окружающих смешно. Когда вы свою жизнь превращаете в сплошную читку фантастики и приключений или ей подобной литературы и лишь изредка обращаетесь к реальной действительности, то вы начинаете ощущать её для себя как что-то неприятное. Ибо вам кажется всё не таким да не этаким. Так как вам часто влетает за небрежно выполненную работу или за другую подобную деятельность, вы, в свою очередь, перестаёте понимать правильность своих действий, и единственной вашей реакцией на всё окружающее являются ваши капризы и оскорбленная поза. В конечном итоге всё это вам мешает нормально жить и получать удовлетворение от реальной жизни. Данная работа затрагивает, прежде всего, ваше мышление. По этой причине вам несколько трудно будет читать её и вы постараетесь поскорее от неё отделаться. Возможно, совсем не возьмете в руки, так как вам за это никто двоек не будет ставить, и вам некого будет бояться. Так уж поставлено обучение, после которого утрачивается жажда к* *знаниям. Если вы всё же преодолеете своё нежелание и постараетесь самостоятельно проработать данную работу, то вы получите немалое удовлетворение, осознав весь смысл её до конца. Прежде всего, вы увидите, что в самом тексте работы заложено стремление к такой форме изложения материала, которая была бы наиболее удобна и доступна для осмысленного понимания содержания данной работы при возможно минимальном её объёме. Далее, осмысленно осознав содержание работы, вы проникните в определённую сущность жидкостей и газов, которая даст вам ключ к пониманию происходящих вокруг вас различных явлений. Что, в свою очередь, даст вам возможность испытать человеческую гордость за себя как истинного повелителя природы, способного создавать машины и аппараты на основе уже известных вам законов движения жидкостей и газов. Главное заключается в том, что вы получите нужное и правильное направление в своей деятельности, касающейся определённых областей вашей практики. Поэтому данная работа не отвлекает вас от реальной жизни и тем самым повышает вашу жизнедеятельность, после которой вы всегда будете получать удовлетворенность. Наша пауза отдыха, пожалуй, вам покажется нравоучением. Но что поделаешь, если нет других способов убедить вас в хорошем.]*[5]

Теперь перейдем к продолжению нашей работы. На очереди у нас третий закон. В механике безынертной массы он называется *формальным принципом связи вида движения с формой уравнений неразрывности и движения*. Название уравнений неразрывности и движения взято из старой механики жидкости и газа. В нашей работе они будут называться *уравнениями движения и уравнениями сил*. Большинство из этих уравнений будет иметь совсем иную форму, чем она дана в старой механике жидкостей и газов. В последующем вы увидите, что эта связь вида движения жидкостей и газов проявляется, в основном, с формой уравнений движения. В связи с новыми дополнениями дадим третьему закону несколько измененное название. Назовем его так: *формальный принцип связи вида движения с формой уравнений движения.* Тем самым мы привели в соответствие название закона с его формулировкой. Запишем её:

***отсутствие или наличие в уравнении движения параметров пространства и времени определяет его связь с тем или иным видом движения.***

Смысл этого закона заключается в том, что он даёт возможность по количеству сочетаний параметров пространства и времени определить полное число видов движения жидкостей и газов. Таких сочетаний из параметров можно сделать четыре:

1. отсутствуют параметры пространства и времени;
2. отсутствует параметр времени и присутствует параметр пространства;
3. отсутствует параметр пространства и присутствует параметр времени;
4. присутствуют параметры пространства и времени.

Это значит, что для жидкостей и газов существуют всего четыре вида движения. В старой механике жидкостей и газов их определяли бесчисленным множеством.

Вы теперь понимаете, что третий закон может быть получен путем непосредственного наблюдения за движением жидкостей и газов. Наблюдая за различными видами движения жидкостей и газов, мы установили, что жидкости и газы имеют четыре вида движения. Установить четыре вида движения для жидкостей и газов – дело непростое. Вы, наверное, понимаете, что исследования в области механики жидкости и газа проводятся давно. Первые сознательные записи исследований в этой области науки насчитывают многие сотни лет. К первым исследователям можно отнести Архимеда. И в настоящее время многие ученые различных стран мира занимаются исследованиями в области механики жидкости и газа. До сих пор никто из них не пришел к такому простому выводу, что жидкости и газы имеют всего-навсего четыре вида движения. Вы можете судить сами, какое важное дело мы сделали, избавив людей от безуспешных поисков.

Как мы знаем, что всякое движение происходит в пространстве и во времени. Движение жидкостей и газов тоже происходит в пространстве среды и во времени. Вот эту связь между движением жидкости и пространством и временем мы выразили в третьем законе. Тем самым определили ещё одну общность для жидкостей и газов.

Все четыре вида движения жидкостей и газов были известны сравнительно давно, хотя всё равно продолжают считать, что видов их движения – бесчисленное множество. Поэтому название для каждого вида движения, за исключением расходного вида движения, взяты из старой механики жидкости и газа. Сделаем перечень наименований для каждого вида движения:

1. *установившийся вид движения*
   (отсутствуют параметры пространства и времени);
2. *плоский установившийся вид движения*

[5] В квадратных скобках редактор оставил часть авторского текста.

(отсутствует параметр времени и присутствует параметр пространства);

3. *расходный вид движения*
(отсутствует параметр пространства и присутствует параметр времени);
4. *акустический вид движения*
(присутствуют параметры пространства и времени).

Установившийся вид движения вы можете увидеть, наблюдая движение жидкости на прямом участке трубы или морские и океанические течения. Плоский установившийся вид движения вы можете наблюдать в виде водоворотов на реке, вихрей в летние жаркие дни, и, наконец, вы можете наблюдать его в собственной ванне, когда вода сливается в открытую воронку. При расходном виде движения надувается ваш футбольный мяч или камера велосипеда, происходит движение поршня в цилиндре мотора внутреннего сгорания. В этом случае жидкости и газы перемещаются всем объёмом. Наконец, четвертый вид движения, который мы назвали акустическим, тоже хорошо вам известен. Этот вид движения позволяет вам слышать звуки на расстоянии. Его же вы наблюдаете в виде огромных волн при штормовом состоянии морей и можете видеть его в виде расходящихся кругов волн от брошенного вами камня на неподвижной глади тихого озера, и т.д.

Вот мы и познакомились с четырьмя видами движения жидкостей и газов. Но в вашей душе, возможно, ещё остался червячок сомнения, а действительно ли, что жидкости и газы обладают только четырьмя видами движения? Это говорит о том, что наших доказательств для третьего закона ещё недостаточно. Окончательно вы убедитесь в его правильности лишь после того, когда мы запишем все необходимые количественные зависимости для всех четырёх видов механического движения жидкостей и газов. Тогда вы сможете количественно проверить любые встречающиеся вам в повседневной практике виды движения жидкостей и газов. После чего вы окончательно убедитесь в правильности третьего закона, и в вашей душе исчезнет червячок сомнения.

Как вы знаете, для любого механического движения должны быть записаны необходимые зависимости. Для движущихся жидкостей таких уравнений должно быть два. Одно из них должно описывать непосредственно само движение жидкости или газа, другое – силы, действующие на жидкость при конкретном её движении. Дальше мы постараемся записать эти уравнения для каждого вида движения.

Прежде, чем приступить к записи этих уравнений, мы сначала должны будем оговорить способы их условной записи. В комплекс приёмов записи и изображения входят система координат и приёмы изображения движения и равновесия механических состояний жидкостей и газов. Также мы должны оговорить условия движения и состояния покоя жидкостей и газов.

Для механики безынертной массы принимаем следующие условия движения и состояния покоя:

1. пространство считаем неподвижным и полностью заполненным жидкостью или газом, которое выше мы определили как среду;

2. время считаем непрерывным и постоянным в своём движении;

3. различные дополнительные факторы берутся из конкретных условий движения и равновесия.

Комплекс приёмов записи и изображения движения и равновесия будет заключаться в следующем:

1. за систему единиц принимаются общепринятые системы единиц для записи пространства, времени, силы, массы, скорости и т.д.

2. для изображения условий движения и равновесия принимается плоскость (поверхность), для которой конкретно записываются условия движения и равновесия. В соответствии со вторым законом механики безынертной массы эта плоскость (поверхность) располагается обязательно перпендикулярно направлению движения;

3. условия движения и равновесия рассматриваются в одних плоскостях относительно других, т.е. мы в каждом конкретном случае для исследования движения потока размещаем в нем плоскости и поверхности в необходимых для исследования местах. Затем делаем запись пространственного расположения этих плоскостей относительно избранной системы координат.

Теперь мы можем переходить к непосредственному получению уравнений движения и сил для четырех видов движения жидкостей и газов. Раздел, исследующий движение, во всех механиках называется динамикой. Вот мы и займемся динамикой. Законы, полученные нами выше, позволяют это сделать.

## Глава.2 ДИНАМИКА. УРАВНЕНИЯ ДВИЖЕНИЯ И СИЛ

### *а) Установившийся вид движения*

Общее наглядное представление об установившемся виде движения вы уже имеете. К этому виду движения мы отнесли движение жидкости на прямых участках труб, морские и океанические течения и другие им подобные формы движения жидкости, которые мы можем наблюдать в нашей повседневной практике. Характерной особенностью потока установившегося движения жидкости является лишь одно обстоятельство: на протяжении всего потока мы имеем либо одинаковые площади сечения, либо мы имеем различные площади сечения в различных местах потока.

Для своих исследований мы возьмём установившийся поток жидкости, который на протяжении своей длины имеет различные площади сечения. Ибо подобный поток является более общим по сравнению с потоком постоянного сечения. Это значит, что и зависимости этого потока будут более общими по отношению к зависимостям потока с постоянным сечением. Далее нам необходимо будет действовать в соответствии с вышепринятыми положениями механики безынертной массы. Тогда мы будем вынуждены поступить следующим образом: мы отбросим материальные границы принятого для исследования потока и заменим их границами среды. Изобразим такой поток на рис. 1. Помимо тех упрощений, которые нам уже известны для среды, замена материальных границ границами среды для движущейся жидкости даёт дополнительные упрощения. В этом случае мы не будем учитывать взаимодействие движущейся жидкости с материальной стенкой потока. В реальных условиях такое взаимодействие всегда существует. В конечном итоге, это взаимодействие называют потерей энергии. Подобное упрощение приводит лишь к лучшему выявлению картины движения жидкости, а не к её искажению, т.е. подобное упрощение способствует нашему стремлению её выявить. По этой причине такая замена нам просто необходима. В дальнейшем, когда мы будем рассматривать другие виды движения жидкости, мы тоже будем заменять материальные границы потоков границами среды. Перейдём к непосредственному получению уравнений движения установившегося потока.

Сам установившийся поток мы теперь можем наблюдать, сколько нам захочется, ибо он ограничен границами среды, а они неподвижны в пространстве. Сечение потока может быть круглым или прямоугольным, как показано на рис. 1, но оно обязательно должно быть симметричным. Поскольку симметричность потока тоже является обязательным условием установившегося вида движения. Отклонение от симметрии будет приводить к искажению картины установившегося движения. В данном потоке ось симметрии совпадает с линией тока. Форму установившегося потока жидкости мы берём из реальных условий, то есть из условий реального наблюдения. Вышеизложенными условиями мы лишь выделяем истинную форму движения установившегося вида движения.

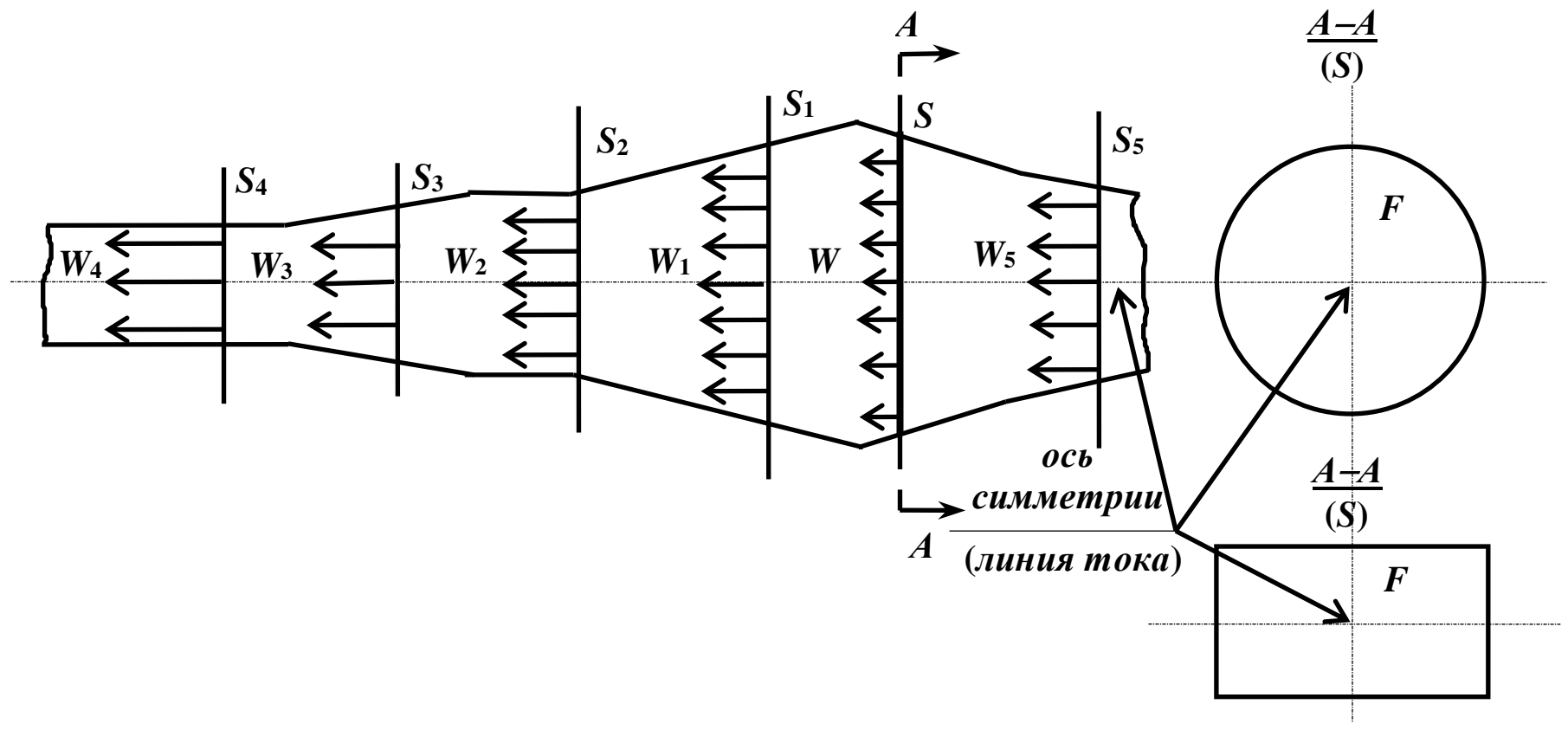

рис. 1

Количественная сторона установившегося вида движения определяется третьим законом и подчёркивается непосредственно названием этого вида движения. Ибо основной характеристикой любого потока жидкости является расход массы в единицу времени *M*. Отсутствие параметров пространства и времени для установившегося вида движения означает для потока жидкости постоянство количественного расхода массы в единицу времени. Это условие говорит о том, что в любом сечении потока, например в сечениях $S$, $S_1$, $S_2$, $S_3$ и т.д. (см. рис.1), в любое время расход массы в единицу времени будет постоянным и равным определённой величине. Это положение является основным условием для количественного получения и определения уравнения движения установившегося вида движения.

Далее мы определим уравнение движения установившегося потока, например, в плоскости $S$ (рис. 1). Эта плоскость располагается перпендикулярно оси симметрии потока, или его линии тока, так как эти понятия одинаковы. Согласно второму закону механики безынертной массы, жидкость будет двигаться перпендикулярно нашей плоскости $S$. Плоскость сечения $S$ будет больше площади сечения нашего потока. Тогда площадь, смачиваемая потоком, будет располагаться в этой плоскости $S$. Площадь, смачиваемую потоком, назовем площадью сечения потока и обозначим её буквой $F$. Вот на этой площади сечения потока $F$ мы будем рассматривать движение потока. Движение жидкости в потоке характеризуется линейной скоростью $W$, плотностью $\rho$ и площадью сечения потока $F$. Тогда, используя характеристики потока жидкости и постоянство расхода количества массы в единицу времени для установившегося потока, мы можем записать уравнение движения. Для чего нам необходимо будет выразить расход количества массы в единицу времени $M$, который является постоянным для данного потока, через характеристики потока. Тогда мы получим:

$$M = FW\rho = \text{const}. \qquad (2)^{6}$$

Уравнение (2) является уравнением движения установившегося вида движения жидкости. Если мы теперь запишем это же уравнение (2) для плоскости сечения $S_1$, то оно тоже будет справедливым для него в том случае, если мы подставим в уравнение (2) характеристики, которые присущи потоку в этом сечении. А расход массы в этом сечении потока будет такой же, как в сечении потока в плоскости $S$. Коль расход массы в единицу времени в сечениях потока для плоскостей $S$ и $S_1$ одинаков, тогда мы можем записать его таким образом:

$$M = FW\rho = F_1W_1\rho = \text{const}. \qquad (3)$$

На рис. 1 мы видим, что площадь сечения потока $F_1$ в плоскости $S_1$ меньше площади сечения потока $F$ в плоскости $S$. Согласно этой разнице должны измениться другие характеристики потока, чтобы произведение их осталось неизменной величиной, поскольку расход массы в единицу времени есть величина постоянная. Плотность $\rho$ тоже есть величина постоянная, так как мы приняли плотность среды постоянной. Тогда у нас должна измениться скорость движения потока $W_1$. Она станет больше, чем скорость $W$ в сечении потока $S$. Отсюда следует, что при увеличении площади сечения потока скорость движения жидкости падает и, наоборот, при уменьшении площади сечения потока его скорости возрастают. Вот мы и получили общее уравнение движения (2) для установившегося вида движения, что значит, оно пригодно для любых случаев установившегося вида движения жидкостей и газов.

Как видим, в этом уравнении в явном виде не присутствуют параметры пространства и времени. Привычную для нас, например, ось координат $X$ мы можем совместить с осью симметрии потока. Затем, приняв одну из плоскостей сечения потока, например плоскость $S$, за начало отсчёта, мы можем замерить и записать расстояние от этой плоскости до любой интересующей нас плоскости сечения $S_1$, $S_2$, $S_3$ и т.д., но эти величины не входят в уравнение движения. Поэтому мы говорим, что параметр пространства отсутствует в уравнении движения (2) установившегося вида движения.

Отметим, что уравнение движения установившегося вида движения (2) известно в существующей механике жидкости и газа как уравнение неразрывности. Вы, наверное, понимаете, коль мы приняли среду как пространство, полностью заполненное идеальной жидкостью, то это определение автоматически включает в себя понятие неразрывности и сплошности среды. В механике давалось подобное определение среды, но механическое движение воспринималось и понималось как перемещение точки или тела в пространстве. Видите, какая нелепость, которая сохранилась и по настоящее время. По этой причине они уравнение движения отнесли просто к характеристикам среды, а не к её движению, и свели его до косвенной характеристики движения жидкости и газов. Мы получили для установившегося вида движения уравнение движения. Для полной количественной характеристики этого вида движения необходимо получить ещё уравнение сил.

[6] Уравнение идет под номером два, хотя оно первое в разделе динамики. Чтобы далее в тексте не менять номера уравнений и ссылки на соответствующие уравнения, редактор оставляет данный номер. Вероятно, уравнением (1) следует считать математическую формулировку второго закона.

Установившийся вид движения жидкости вам уже известен. Это прямолинейный поток жидкости, любая площадь сечения которого имеет ось симметрии (см. рис. 1). Поток может иметь различные площади своего сечения, то есть он может и сужаться, и расширяться, но во всех случаях сохраняет параллельное движение жидкости относительно оси симметрии, которую мы называем линией тока. Ибо она указывает положение направленного движения установившегося потока жидкости. По этой причине плоскости сечения потока $S$, $S_1$, $S_2$, $S_3$ и т.д., расположенные перпендикулярно оси симметрии потока, одновременно будут расположены перпендикулярно направлению скорости движения потока. Направление действия динамических сил давления $P_{дин}$ потока совпадает с направлением скорости потока. По этой причине плоскости сечения потока $S$ (см. рис. 1) будут расположены тоже перпендикулярно направлению действия динамических сил давления $P_{дин}$ потока.

В установившемся потоке жидкости также присутствуют статические силы давления $P_{ст}$. Действие этих сил в любой точке одновременно распространяется во всех направлениях от этой точки, то есть действие этих сил относится к разряду скалярных величин, а действие динамических сил давления относится к разряду векторных величин, то есть величин, имеющих направление. Поэтому плоскость сечения потока будет испытывать полную силу действия статического давления. Для исследования статических сил можно располагать плоскость как угодно в потоке, так как действие их на плоскость будет во всех случаях одинаковым. В нашем случае мы принимаем для наших исследований плоскости $S$ специально расположенные таким образом лишь для того, чтобы одновременно ощутить действие и динамических, и статических сил давления.

В механике обычным методом для выявления действия сил служит метод сечения. Он заключается в том, что в исследуемом объекте плоскостью или поверхностью какая-то часть, конечно, мысленно отсекается, а действие сил отброшенной части конструкции заменяется соответствующими силами. Что даёт возможность выявить количественно действие исследуемых сил. Этот метод мы и применим для выявления уравнения сил установившегося вида движения жидкости.

В механике применяется принцип независимости действия сил, то есть каждая сила и её действие рассматриваются самостоятельно, а затем результаты складываются. В нашем случае мы можем, согласно этому принципу, рассматривать действие динамических и статических сил давления раздельно. Для выявления уравнения сил принимаем тот же поток жидкости, который нами показан на рис. 1, но для наших исследований мы возьмём не весь поток, а лишь его часть, в которой расположена плоскость сечения $S$, и покажем эту часть на рис. 2.*а*.

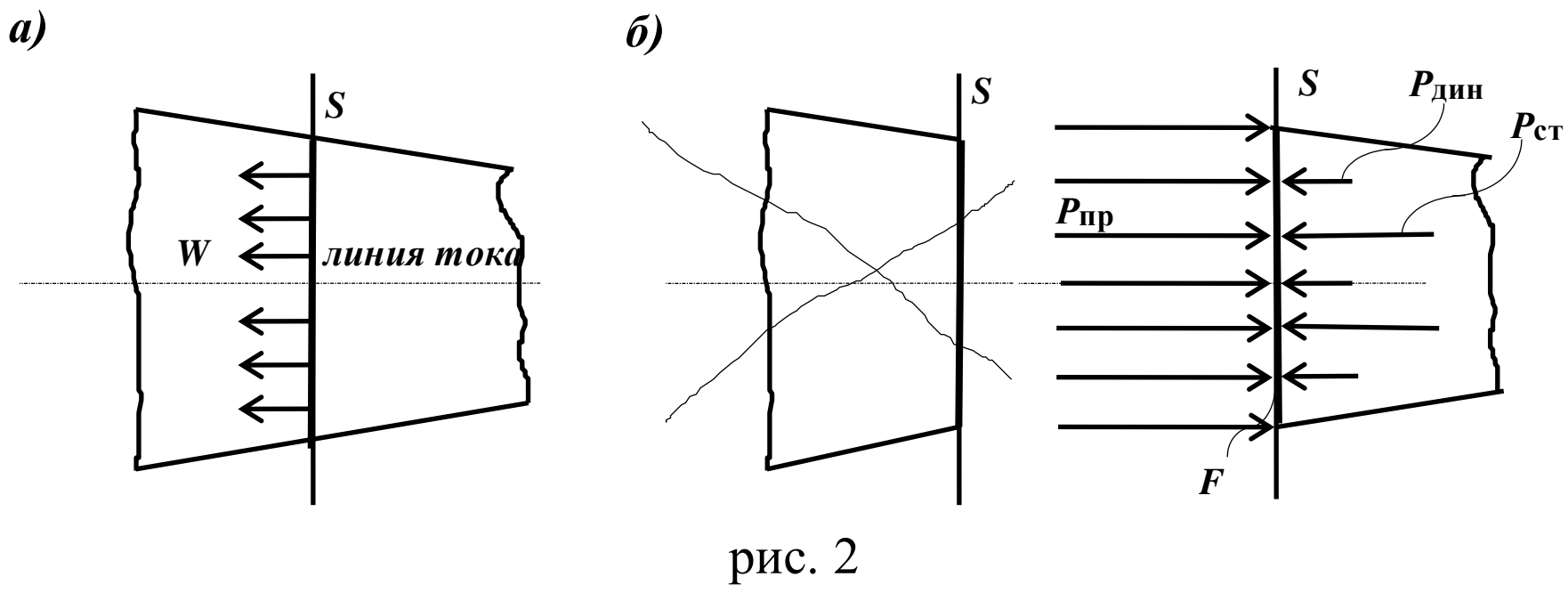

рис. 2

Затем эту часть рассекаем по плоскости $S$. Левую часть потока отбрасываем, а действие левой, отброшенной, части заменяем принятыми силами давления $P_{пр}$, которые будут приложены с левой стороны плоскости $S$ (см. рис. 2.*б*). Далее из условия равновесия мы можем записать уравнение для сил. Оно будет выглядеть таким образом:

$$FP_{пр} = FP_{дин} + FP_{ст}\,. \tag{4}$$

Это уравнение выражает тот смысл, что динамические $P_{дин}$ и статические $P_{ст}$ силы давления первой части потока уравновешиваются принятыми $P_{пр}$ силами давления левой, отброшенной, части потока. В этом уравнении статические силы давления $P_{ст}$ мы можем найти из условия энергетического состояния жидкости. Уравнение энергий мы получим ниже. Динамические силы давления мы можем получить из второго закона механики безынертной массы. Они будут равны:

$$P_{дин} = \frac{M}{F} W.$$

Площадь сечения потока $F$ в плоскости $S$ мы можем измерить. Зная количественную величину динамических и статических сил давления, мы можем определить и принятые силы давления по уравнению (4). Уравнение (4) будет нашим искомым уравнением сил установившегося вида движения.

Уравнение (4) мы можем представить несколько в ином виде, если заменим динамические силы давления через их значение, получим:

$$FP_{пр} = MW + FP_{ст}. \qquad (5)$$

В уравнении (5) мы можем заменить расход массы в единицу времени $M$ характеристиками потока жидкости по уравнению (2), тогда получим:

$$FP_{пр} = F\rho W^{2} + FP_{ст}. \qquad (6)$$

Мы можем получить ещё одну разновидность уравнения сил. Для этого отнесём силы давления к единице площади сечения потока, то есть разделим левую и правую часть уравнения (6) на площадь сечения потока $F$. Получим:

$$P_{пр} = \rho W^{2} + P_{ст}. \qquad (7)$$

Уравнения сил (5), (6) и (7) являются разновидностью уравнения сил установившегося вида движения жидкости (4). Так что все эти зависимости в одинаковой степени пригодны для применения, но наиболее ходовое значение в практике будет иметь уравнение (7). Оно просто более удобно к применению.

Отметим, что статические силы давления в сечении $S$ потока находятся в равновесном состоянии, так как они действуют и с правой, и с левой стороны площади сечения потока $F$ с одинаковой силой. Динамические силы давления $P_{дин}$ уравновешиваются только расходом движущейся жидкости. Тогда мы можем заменить принятые силы давления $P_{пр}$ на две силы: на принятые динамические силы давления $P_{пр.дин}$ и на статические силы давления $P_{ст}$. В этом случае получим:

$$P_{пр} = P_{пр.дин} + P_{ст}. \qquad (8)$$

Заменим в уравнении (4) принятые силы давления по уравнению (8). Мы получим:

$$FP_{пр.дин} + FP_{ст} = FP_{дин} + FP_{ст}. \qquad (9)$$

В уравнении (9) произведение площади сечения потока $F$ на статические силы давления $P_{ст}$ для правой и левой части есть величина постоянная. Поэтому мы можем её сократить. Тогда уравнение (9) примет вид:

$$FP_{пр.дин} = FP_{дин}. \qquad (10)$$

Уравнение (10) тоже будет уравнением сил, записанным относительно динамических сил давления.

В уравнении (10) мы можем заменить динамические силы давления по второму закону механики безынертной массы. Тогда получим:

$$FP_{пр.дин} = MW. \qquad (11)$$

В уравнении (11) расход массы в единицу времени $M$ можем заменить по уравнению движения (2). Тогда получим:

$$FP_{\text{пр.дин}} = F\rho W^{2}. \tag{12}$$

Затем приведём динамические силы давления к единице площади сечения потока. Для чего разделим правую и левую части уравнения (12) на площадь сечения потока $F$, получим:

$$P_{\text{пр.дин}} = \rho W^{2}. \tag{13}$$

Уравнение (13) тоже будет уравнением сил установившегося вида движения, записанным относительно динамических сил давления.

Будем считать, что мы получили необходимые уравнения движения и сил для установившегося вида движения жидкостей и газов, которые пригодны для практических расчётов.

### *б) Плоский установившийся вид движения*

Теперь перейдем к получению уравнений движения и сил для плоского установившегося вида движения жидкостей и газов. Для установившегося вида движения мы давали уравнения с выводом. В последующем мы этого не будем делать. Чтобы сократить текст, мы просто дадим смысловое объяснение и запишем необходимые зависимости. Таким же образом мы будем поступать с оставшимися видами движения.

Наглядно вы уже знакомы с плоским установившимся движением жидкости и газа. Нам остаётся лишь придать ему осмысленное содержание. Изобразим этот вид движения на рис. 3 в виде цилиндрического потока жидкости. Этот поток составит нам среду, или пространство полностью заполненное жидкостью и ограниченное границами среды, которые не вносят никаких дополнительных помех и изменений в движение жидкости в этом потоке.

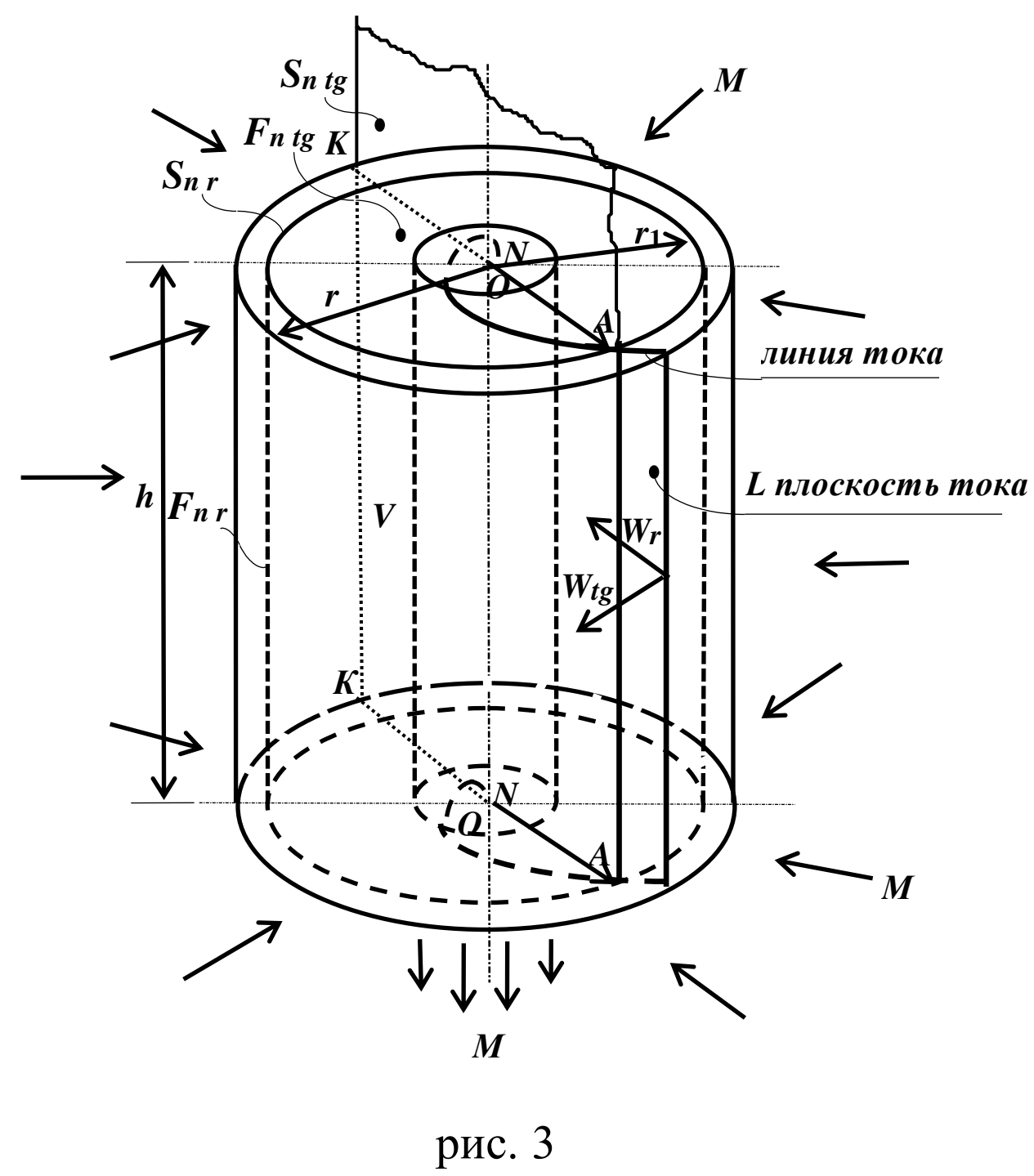

рис. 3

На рис. 3 мы показали поток в таком виде, в каком вы его привыкли видеть. Жидкость с постоянным расходом массы $M$ подходит к потоку из окружающей среды. Затем эта жидкость проходит через цилиндрический объём потока и вытекает из него через отверстие в потоке с тем же постоянным расходом массы $M$, который притекает к потоку через его внешнюю границу. Ведь цилиндр потока, как вы знаете, находится по всей своей высоте в окружающей его жидкости. Мы просто приняли, что жидкость движется со стороны внешней границы цилиндра потока к внутренней его границе, но данный поток может быть организован и обратным движением жидкости, то есть движением жидкости со стороны внутренних границ потока к его внешним границам. Подобная специфика не влияет на общий смысл уравнений движения и сил плоского установившегося вида движения жидкости.

Поскольку расход массы жидкости в единицу времени для потока есть величина постоянная во времени, то этот вид движения тоже имеет название установившегося. Но сам поток имеет цилиндрическую форму, потому что движение жидкости происходит одновременно в двух направлениях.

Одним таким направлением будет движение по направлению радиуса цилиндра. Назовем это движение радиальным. Вторым одновременным движением жидкости будет движение, направленное перпендикулярно к радиальному движению или к радиусу цилиндра. Назовем это движение тангенциальным. На рис. 3 скорость радиального движения мы обозначим как $W_r$, а скорость тангенциального движения как $W_{tg}$. Хотя эти скорости взаимно перпендикулярны, но они лежат в одной плоскости, которая расположена перпендикулярно к оси цилиндра $O$ - $O$. Поэтому мы данное установившееся движение дополнительно назвали плоским. Общее же название подобного потока звучит как плоский установившийся вид движения.

В точке $A$ (рис. 3), которая расположена на расстоянии радиуса $r_1$ от оси потока $O$ - $O$, тангенциальная плоскость $S_{tg}$ и радиальная плоскость $S_r$ будут взаимно пересекаться. Площадь сечения потока в точке $A$ очень мала, она составляет незначительную площадь сечения потока. Незначительная часть жидкости потока будет поступать в точку $A$ через радиальную плоскость $S_r$, в которой площадь сечения точки $A$ будет занимать бесконечно малую площадь $\Delta F_r$. Эта же незначительная часть жидкости потока будет уходить из точки $A$ через тангенциальную плоскость сечения $S_{tg}$ потока, в которой площадь сечения точки $A$ будет занимать бесконечно малую площадь сечения $\Delta F_{tg}$. Расход массы жидкости в единицу времени, поступающий через радиальную площадь сечения $\Delta F_r$ точки $A$ и убывающий через тангенциальную площадь сечения $\Delta F_{tg}$ той же точки, будет постоянным и равным бесконечно малой величине. По условию постоянства расхода массы потока во времени $\Delta M_r = \Delta M_{tg}$. Тогда для точки $A$ (рис. 3) мы будем вынуждены записать два уравнения движения: для тангенциального и радиального движения жидкости. Запишем эти уравнения:

$$\Delta M_r = \Delta F_r W_r \rho,$$

$$\Delta M_{tg} = \Delta F_{tg} W_{tg} \rho. \qquad (14)$$

Затем через точку $A$ проведём цилиндрическую поверхность $S_{n\ r}$ с радиусом $r_1$. По этой поверхности расположим такое количество радиальных сечений точки $A$, чтобы они образовали цилиндрическую площадь сечения в радиальном направлении. Тогда мы получим полную поверхность площади сечения потока в радиальном направлении $F_{n\ r}$. Полная площадь сечения в радиальном направлении будет равна высоте потока $h$ умноженной на длину окружности. Запишем её в количественном выражении:

$$F_{n\ r} = h \cdot 2\pi r_1. \qquad (15)$$

Тогда уравнение движения для площади сечения потока в радиальном направлении примет вид:

$$M_r = F_{n\ r} W_r \rho \ = 2\pi r_1 h \cdot W_{1r} \rho. \qquad (16)$$

Мы здесь взяли площадь сечения в радиальном направлении с определённым радиусом $r_1$. Но мы можем с, таким же успехом, взять эту площадь сечения с другим радиусом. Во всех случаях уравнение (16) будет справедливо. Мы здесь отметим лишь ту количественную разницу, что радиальная скорость потока $W_r$ будет расти по мере уменьшения радиуса $r$. Это происходит потому, что с уменьшением радиуса радиальная площадь сечения потока будет уменьшаться. Поскольку плотность $\rho$ жидкости потока и расход массы в единицу времени $M$ есть величины постоянные для данного потока, то уменьшение площади потока будет компенсироваться увеличением скорости потока $W_r$, и – наоборот. Следовательно, на наружней поверхности потока радиальная скорость будет иметь минимальное значение, а на внутренней – максимальное.

Чтобы сделать уравнение (16) более общим, запишем его в таком виде:

$$M_r = 2\pi rh \cdot W_r \rho. \tag{17}$$

Уравнение (17) будет общим уравнением радиального движения потока плоского установившегося вида движения жидкости и газа.

Теперь нам остаётся получить общее уравнение движения для тангенциального движения жидкости потока.

В тангенциальной плоскости потока лежит радиус и ось потока. Проведём тангенциальную плоскость $S_{n\ tg}$ через ось потока, расположив её по радиусу, как это показано на рис. 3. Затем расположим тангенциальные площади сечения потока (точки $A$) таким образом, чтобы они образовали полную площадь сечения потока $F_{n\ tg}$ в тангенциальном направлении. Эта площадь будет равна высоте потока $h$, умноженной на ширину потока $KN$, то есть

$$F_{n\ tg} = h \cdot (KN). \tag{18}$$

Тогда тангенциальное уравнение движения потока жидкости будет иметь вид:

$$M_{tg} = F_{n\ tg} W_{tg} \rho = h \cdot (KN) W_{tg} \rho. \tag{19}$$

Уравнение (19) будет общим уравнением тангенциального движения в плоском установившемся потоке. Ибо тангенциальная скорость движения $W_{tg}$ будет одинакова для всего сечения потока, так как эта площадь сечения потока не зависит от изменения величины радиуса потока.

Следовательно, для плоского установившегося потока жидкости существуют два уравнения движения, которые описывают его движение в радиальном и тангенциальном направлении. Ими являются уравнения (17) и (19).

Отметим, что общие и частные расходы количества массы в тангенциальном и радиальном направлениях для плоского установившегося потока равны между собой,

$$\Delta M_r = \Delta M_{tg} \text{ и}$$
$$M_r = M_{tg}. \tag{20}$$

Это следует из условия движения потока.

Иметь просто количественные зависимости для плоского установившегося потока жидкости ещё недостаточно, чтобы иметь полное представление об его движении. Мы должны ещё найти для него линию и плоскость тока. Линия и плоскость тока для точки $A$ потока показаны на рис. 3. Как видите, линия тока этой точки показана определённой кривой линией, а плоскость образует по этой кривой поверхность в объёме потока. Совокупность поверхностей тока, когда эти поверхности плотно прижаты друг к другу, образует полный объём плоского установившегося потока. Поверхность тока означает, что жидкость потока точки $A$, двигаясь одновременно в радиальном и тангенциальном направлении, в конечном итоге совершает движение от наружной поверхности потока к его внутренней границе только по плоскости тока. Если представить себе точку $A$ в виде твёрдой точки и заставить её двигаться от внешней границы потока к внутренней, то она оставила бы после себя траекторию в виде линии тока. Наглядно такие линии тока мы можем наблюдать, например, при истечении жидкости в воронку ванны. На поверхности воды ванны в области плоского установившегося потока можно видеть характерные для линий тока возмущения. Они будут соответствовать нарисованной линии.

Из условия характера движения жидкости, которое определяется перпендикулярностью направления тангенциальной скорости $W_{tg}$ жидкости к радиусу потока в любой части его объёма, мы запишем линию тока с помощью математической зависимости. Она будет иметь такой вид:

$$r = ae^{\kappa\varphi}, \tag{21}$$

где φ – угол полярной системы координат; $a$ – некоторая величина, относящаяся к конкретной спирали;

$\kappa$ = ctg α – в нашем случае угол α равен 45°, поэтому ctg α = 1. При $\alpha = \frac{\pi}{2}$ кривая превращается в окружность.

Уравнением (21) записана кривая линия, которая носит наименование логарифмической спирали. Следовательно, линия тока плоского установившегося потока имеет форму логарифмической спирали.

Здесь зависимость линии тока дана без вывода потому, что её можно сравнительно легко получить, воспользовавшись положениями дифференциального исчисления. Поскольку вы ещё не знаете высшей математики, то вам бесполезно давать эти выводы.

Теперь перейдем к получению уравнений сил для плоского установившегося вида движения. Вывод уравнений сил для плоского установившегося вида движения мало отличается от вывода уравнения сил для установившегося вида движения, который мы приводили выше. Различие здесь будет заключаться в том, что в плоском установившемся потоке жидкость движется в двух направлениях, а в установившемся потоке – только в одном направлении. Поэтому для плоского установившегося движения нам придётся записать уравнения равновесия для радиальной и тангенциальной плоскости сечения потока. Поскольку мы уже знаем полную площадь сечения потока для одного и другого направления, то и запишем это условие для этих плоскостей, а не для точки $A$ (рис. 3), как мы это делали при выводе уравнений движения. В этом случае мы тоже соответственно компенсируем отброшенную часть потока принятыми радиальными $P_{\text{пр}\ r}$ и тангенциальными $P_{\text{пр}\ tg}$ силами давления. Тогда получим:

*для радиальной плоскости сечения:*

$$F_r P_{\text{пр}\ r} = F_r P_{\text{ст}\ r} + F_r P_{\text{дин}\ r}; \tag{22}$$

*для тангенциальной плоскости сечения:*

$$F_{tg} P_{\text{пр}\ tg} = F_{tg} P_{\text{ст}\ tg} + F_{tg} P_{\text{дин}\ tg}. \tag{23}$$

Далее с помощью второго закона механики безынертной массы мы заменяем динамические силы давления в уравнении (22) и (23) через характеристики потока. Тогда получим:

*для радиальной плоскости сечения:*

$$F_r P_{\text{пр}\ r} = F_r P_{\text{ст}\ r} + M_r W_r; \tag{24}$$

*для тангенциальной плоскости сечения:*

$$F_{tg} P_{\text{пр}\ tg} = F_{tg} P_{\text{ст}\ tg} + M_{tg} W_{tg}. \tag{25}$$

Затем с помощью уравнения движения заменяем расход массы в единицу времени через характеристики потока в уравнениях (24) и (25). Тогда получим:

*для радиальной плоскости сечения:*

$$F_r P_{\text{пр}\ r} = F_r P_{\text{ст}\ r} + F_r \rho W_r^2; \tag{26}$$

*для тангенциальной плоскости сечения:*

$$F_{tg} P_{\text{пр}\ tg} = F_{tg} P_{\text{ст}\ tg} + F_{tg} \rho W_{tg}^2. \tag{27}$$

Если мы теперь отнесём все эти силы к единице площади соответствующих плоскостей сечения, то есть мы разделим и левые, и правые части уравнений на соответствующие площади сечения потока, то получим:

*для радиальной плоскости сечения:*

$$P_{\text{пр}\,r} = P_{\text{ст}\,r} + \rho W_r^2\,; \tag{28}$$

*для тангенциальной плоскости сечения:*

$$P_{\text{пр}\,tg} = P_{\text{ст}\,tg} + \rho W_{tg}^2\,. \tag{29}$$

Начиная с уравнений (22) и (23) и кончая уравнениями (28) и (29), все они являются парными разновидностями полных уравнений сил плоского установившегося вида движения для соответствующих направлений движения жидкости в потоке.

Отметим, что при выводе уравнений движения мы установили, что тангенциальные скорости движения $W_{tg}$ являются постоянной величиной для всего потока в целом, а величины радиальных скоростей $W_r$ движения зависят от величины радиуса $r$ потока и изменяются с изменением радиуса. По этой причине величина радиальных динамических сил давления тоже будет зависеть от величины радиуса потока. Следовательно, каждая радиальная площадь сечения потока имеет определённое значение динамических сил давления отличное от любой другой радиальной плоскости сечения потока, находящейся в объёме потока, поскольку радиусы этих плоскостей будут различными. Затем, когда мы получим уравнения энергии для плоского установившегося вида движения, мы узнаем, что статические силы давления тоже изменяют величину своего значения в зависимости от величины радиуса потока, пропорционально динамическим силам давления.

При постоянстве тангенциальных сил давления в уравнении (29) статические тангенциальные силы давления имеют различную величину в тангенциальной площади сечения потока, которая тоже зависит от величины радиуса потока. Это связано с тем условием, что радиальные площади сечения потока пересекают тангенциальную площадь сечения в одной точке, а вернее, по одной прямой. В этом вы тоже можете убедиться, посмотрев на рис. 3. Поэтому каждая новая радиальная поверхность сечения потока будет пересекать одну и ту же тангенциальную площадь сечения потока в новой точке, поскольку радиусы новых радиальных площадей сечения будут различными. В точках пересечения радиальных и тангенциальных площадей сечения потока радиальные и тангенциальные статические силы давления имеют одинаковое значение. Например, для точки $A$ пересечения этих плоскостей (см. рис. 3) мы можем записать, что

$$P_{\text{ст}\,r(A)} = P_{\text{ст}\,tg(A)}\,. \tag{30}$$

Это условие вытекает из скалярности статических сил давления.

Эти положения вы просто должны представлять себе пока в первом приближении, так как мы на них ещё раз остановимся после вывода уравнений энергий.

Поскольку статические силы давления в плоском установившемся движении находятся в равновесном состоянии так же, как в установившемся потоке, то мы можем получить уравнения сил только для динамических сил давления таким же способом, как и для установившегося потока жидкости, то есть приняв за принятые силы давления отброшенной части потока только динамические силы давления. Тогда получим:

*для радиальной плоскости сечения:*

$$F_r P_{\text{пр}\,r} = F_r P_{\text{дин}\,r}; \tag{31}$$

*для тангенциальной плоскости сечения:*

$$F_{tg} P_{\text{пр}\,tg} = F_{tg} P_{\text{дин}\,tg}. \tag{32}$$

Затем делаем соответствующие преобразования по второму закону механики безынертной массы. Получим:

*для радиальной площади сечения:*

$$F_r P_{\text{пр}\,r} = M_r W_r; \tag{33}$$

*для тангенциальной площади сечения:*

$$F_{tg}P_{\text{пр}\,tg} = M_{tg}W_{tg}. \qquad (34)$$

Далее мы заменяем расход массы по уравнениям движения. Получим:

*для радиальной площади сечения:*

$$F_rP_{\text{пр}\,r} = F_r\rho W_r^2; \qquad (35)$$

*для тангенциальной площади сечения:*

$$F_{tg}P_{\text{пр}\,tg} = F_{tg}\rho W_{tg}^2. \qquad (36)$$

После всех преобразований мы можем отнести эти уравнения к единице площади, тогда получим:

*для радиальной площади сечения:*

$$P_{\text{пр}\,r} = \rho W_r^2; \qquad (37)$$

*для тангенциальной площади сечения:*

$$P_{\text{пр}\,tg} = \rho W_{tg}^2. \qquad (38)$$

Мы получили все необходимые зависимости для плоского установившегося вида движения жидкости и газа.

***в) Расходный вид движения***

Некоторое наглядное представление о расходном виде движения вы имеете. Например, вы можете его наблюдать, когда накачиваете велосипедную или футбольную камеру, при работе газа в цилиндрах двигателей внутреннего сгорания и т.д. Для выявления общих зависимостей расходного вида движения возьмём пример с камерой и дополнительно усложним его. Предположим, что жидкость в камеру качается несколькими насосами и что камера наша худая и имеет несколько отверстий, через которые вытекает какое-то количество жидкости. Покажем всё это на рис. 4

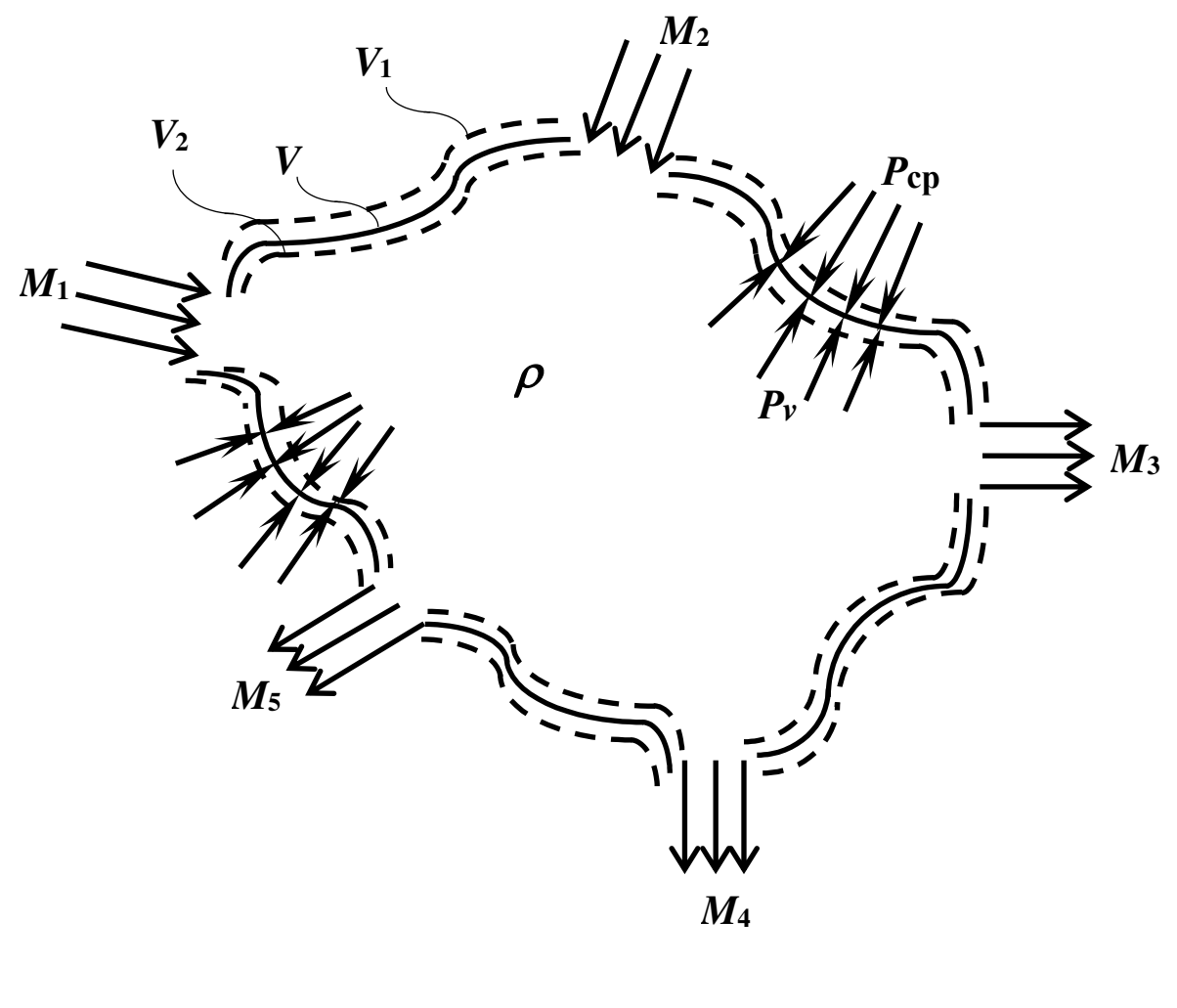

рис. 4

Границы этого объёма, которые, как и в остальных видах движения, определяются границами среды, не изменяют характера движения жидкости в расходном потоке жидкости. Вы понимаете, что через отверстия в объёме в каждый момент времени поступает и убывает различное количество массы. В результате чего в полном объёме *V* потока (в самой камере) количество жидкости тоже либо уменьшается, либо увеличивается. Изменение количества жидкости в объёме потока сказывается в изменении его границ. Поэтому с течением времени объём потока либо увеличивается, либо уменьшается. Таким способом выполняется условие третьего закона механики безынертной массы, который утверждает для расходного вида движения изменение во времени расхода количества массы в единицу времени.

В результате непрерывно изменяющихся величин поступающих в объём и вытекающих из него расходов жидкости количество массы жидкости в объёме *V* потока тоже будет непрерывно изменяться. В результате чего границы объёма будут изменяться, то есть объём потока будет непрерывно либо увеличиваться, либо уменьшаться. Всё это естественно. Поэтому поток, занимая в определённое время объём *V*, через некоторое время может занять объём $V_1$, а ещё через некоторое время займет, например, объём $V_2$ и так далее. Из рисунка 4 вы видите, что объёмы потока *V*, $V_1$ и $V_2$ разные по величине. Вот для этого движения потока нам нужно получить уравнение движения, чтобы его можно было оценить количественно.

Естественно, что ни какую систему координат для этого потока невозможно применить. Тогда поступим следующим образом: будем считать, что в какой-то момент времени нам удалось измерить объём потока и зафиксировать его границы. На рис. 4 таким объёмом потока будет объём *V*, изображенный сплошной линией. Тогда, рассматривая этот объём условно неподвижным, мы можем записать для него условия движения.

Фиксированный объём *V* потока будет содержать определённое количество массы жидкости *m*, плотность которой равна $\rho$. Этот объём массы жидкости накопился в потоке за время, которое отсчитывалось с начала поступления расходов массы в единицу времени $M_1$, $M_2$, $M_3$, $M_4$ и $M_5$ в объём потока до момента фиксации его объёма. Будем считать, что в этот период прошло время в количестве *t* единиц времени. Теперь мы можем это условие оформить математически. Получим такую зависимость:

$$m = V\rho = M_1t + M_2t - M_3t - M_4t - M_5t. \quad (39)$$

Отметим, что в правой части уравнения мы ставим положительные и отрицательные значения, хотя записали его без применения системы координат. Как вы знаете, положительные и отрицательные значения величин берутся в зависимости от расположения этих величин относительно системы координат. В нашем случае все величины уравнения (39) являются скалярными, то есть не имеющими направления относительно системы координат. Ибо любой расход, объём или плотность у нас всегда имеют материальное содержание. Поэтому мы приняли расходы массы, которые вытекают из объёма потока, отрицательными, а расходы массы жидкости, которые поступают в объём, положительными. Принимая таким способом положительные и отрицательные значения расходов массы жидкости, мы тем самым уравнением (39) записали условие сохранения массы жидкости в объёме потока. Для расходного потока жидкости уравнение (39) является уравнением движения.

Условие принятия положительных и отрицательных значений величин, которое мы здесь дали, не является частным условием. Оно считается общим для механики безынертной массы и должно применяться во всех её разделах и положениях, когда встаёт необходимость в выявлении положительных и отрицательных величин.

Далее, как мы полагаем, объём потока имеет множество отверстий, через которые жидкость притекает и вытекает из него. Тогда в правой части уравнения (39) нет смысла записывать каждый расход массы в отдельности. Мы просто можем записать разность сумм положительных и отрицательных расходов. Сумма обозначается знаком Σ (сигма). Тогда уравнение (39) примет вид:

$$m = V\rho = t\sum_{n=1}^{n} M_i - t\sum_{n=1}^{n} M_j\,. \quad (40)$$

Уравнение (40) будет являться общим уравнением движения расходного вида движения жидкости и газа.

В этом уравнении мы не расшифровываем его значения через характеристики потока лишь потому, что они определяются конкретными условиями движения. Например, объём потока *V* может быть цилиндрическим, шаровым или иметь какую-либо иную форму. Для всех этих конкретных объёмов существуют свои зависимости. Каждый расход массы в единицу времени $M$ мы можем выразить через уравнение движения установившегося вида движения, но с условием, что скорость движения в этом потоке непрерывно изменяется с течением времени. Изменение скорости движения жидкости $W$ в любом отверстии и его площадь сечения тоже будут зависеть от конкретных условий движения. Поэтому уравнение (40) лучше оставить в той форме, в которой оно у нас записано.

Мы получили уравнение (40) с помощью известных вам арифметических действий. Будем надеяться, что вы поняли сущность самого расходного движения и его количественную запись. Но с помощью дифференциального исчисления количественная зависимость для этого движения записывается в более простой и выразительной форме. Поскольку вы ещё не знаете этого исчисления, то можете не понять новую форму уравнения. Но мы всё равно дадим вывод уравнения движения в дифференциальной форме.

Масса жидкости потока у нас изменяется во времени, тогда мы её можем записать как зависящую от времени в таком виде: $m_t = f(t)$. Объём жидкости тоже изменяется с течением времени. Тогда он запишется в таком виде: $V_t = f(t)$. После чего мы можем записать равенство:

$$m_t = V_t\rho. \tag{41}$$

Затем продифференцируем это равенство по времени. Оно примет вид:

$$\frac{dm_t}{dt} = \rho\frac{dV_t}{dt}. \tag{42}$$

В уравнении (42) отношение $\frac{dm_t}{dt}$ есть ни что иное, как расход массы в единицу времени, то есть $\frac{dm_t}{dt} = M_t$. Тогда уравнение (42) примет вид:

$$M_t = \rho\frac{dV_t}{dt}. \tag{43}$$

Из уравнения (40) мы знаем, что расход массы $M_t$ в уравнении (43) является суммарной разностью расходов, то есть

$$M_t = \sum_{n=1}^{n} M_i - \sum_{n=1}^{n} M_j .$$

Подставим это уравнение в уравнение (43). Тогда получим:

$$\sum_{n=1}^{n} M_i - \sum_{n=1}^{n} M_j = \rho\frac{dV_t}{dt}. \tag{44}$$

Уравнение (44) тоже будет общей формой уравнения движения расходного вида движения.

*Примечание:* при исследовании, например, газа в расходном виде движения плотность $\rho$ является переменной величиной, так как газ сжимаем. Если принять для зависимости (41) переменными и объём потока $V_t$, и плотность $\rho_t$, то после дифференцирования и замены результирующего расхода массы $M_t$ на разность сумм расходов массы, как для уравнения (44), получим:

$$\rho\frac{dV_t}{dt} + V\frac{d\rho_t}{dt} = \sum_{n=1}^{n} M_i - \sum_{n=1}^{n} M_j . \tag{45}$$

Уравнение (45) является ещё более общей формой уравнения движения расходного вида движения жидкости и газа. Это уравнение дополнительно учитывает сжимаемость газа, но форму и сущность расходного вида движения оно не изменяет.

Для расходного вида движения различие связанное с сжимаемостью и несжимаемостью газов и жидкостей учитывается уравнением движения, а в выше рассмотренных видах движения газы ведут себя одинаково с жидкостями, то есть они не изменяют свою плотность.

Теперь попробуем получить уравнение сил расходного вида движения. Как мы выше установили, границы объёма потока непрерывно изменяются. Это значит, что силы давления в объёме потока (обозначим их как $P_V$) превышают силы давления, действующие на жидкость потока со стороны границ объёма (обозначим эти силы как $P_{ср}$). Различие по величине этих сил заставляет перемещаться границы потока (см. рис. 4). Разница в силах давления на границах потока будет компенсироваться динамическими силами давления $P_{дин}$, которые, как вы знаете, определяются вторым законом механики безынертной массы. Запишем это условие равновесия для расходного вида движения:

$$F_t P_{дин\, t} = F_t (P_{t\,V} - P_{t\,ср}) = \sum_{n=1}^{n} M_{ti} W_{ti} - \sum_{n=1}^{n} M_{tj} W_{tj}\ . \qquad (46)$$

Уравнение (46) будет уравнением сил расходного вида движения. Записано оно в общем выражении, как и уравнение движения этого вида движения жидкости. Потому что конкретность характеристик этого уравнения зависит от конкретных условий движения. Как эти конкретные условия движения реализуются, мы объяснили выше для уравнения движения этого вида движения жидкостей и газов. Мы получили необходимые зависимости для расходного вида движения.

### *г) Акустический вид движения*

Нам осталось получить уравнения движения и сил для четвёртого вида движения жидкостей и газов, который мы назвали акустическим. Акустический, или волновой, вид движения жидкостей и газов – наиболее трудный вид движения для вашего понимания. Ибо наглядно он выражен менее всего, когда мы его наблюдаем практически в естественных условиях. Для нас акустический вид движения является, прежде всего, звуком, хотя, например, сущность механического движения, как для звуковых, так и для штормовых волн одинакова.

Хотя мы не видим движения звуковых волн, но мы хорошо знаем, что звуковые волны создаются вибрирующей пластинкой, которую применяют и для радио, и для телефона. Колеблющаяся пластинка в микрофоне воспринимает звуковые колебания, подобная пластинка в радио воспроизводит их. Во всех этих случаях пластинка совершает возвратно-поступательное движение. Величина, или ход, прямолинейного перемещения пластинки настолько мала, что мы не в состоянии зрительно воспринять это движение.

Приступим к получению уравнений движения для акустического вида движения жидкостей и газов. Удобнее всего это сделать, рассматривая звуковое движение жидкостей и газов. Ибо в этом случае мы можем наблюдать начало и конец образования звуковых волн по движению вибрирующей пластинки. Поэтому начнём вывод уравнений движения с выбора вибрирующей пластинки и её движения. Покажем её на рис. 5.

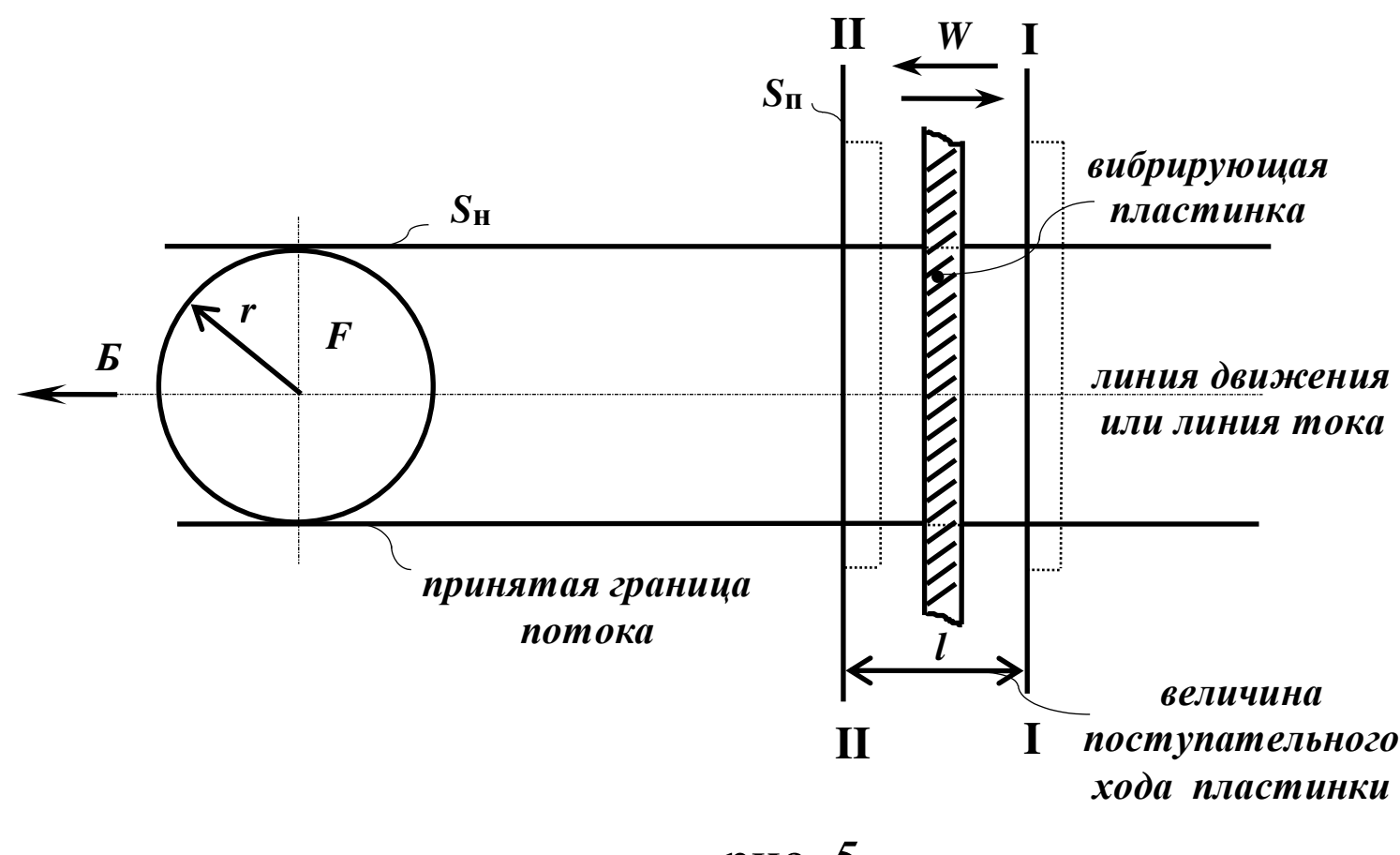

рис. 5

На рисунке 5 мы показали, что площадь вибрирующей пластины – бесконечно большая. Пластина непрерывно перемещается по стрелке *Б* из положения I в положение II и наоборот.

Величина хода пластины при её возвратно-поступательном движении равна величине $l$. Скорость движения пластины в обоих направлениях равна $W$. Далее, чтобы не рассматривать весь акустический поток, границы которого мы не всегда можем определить (хотя бы по той причине, что мы приняли площадь пластины бесконечно большой величиной), для своих исследований мы выделяем часть потока, которую создаёт круговая поверхность пластины площадью $F$, радиус окружности которой равен $r$. Этот поток ограничен цилиндрической поверхностью $S_н$, расположенной параллельно направлению скорости движения пластины. Цилиндрическая поверхность $S_н$ будет расположена симметрично линии тока. Радиус цилиндра будет равен $r$, а площадь его сечения будет равна величине $F$ (см. рис. 5). Поверхность $S_н$ будет являться одновременно границей потока, которая ограничивает исследуемую среду.

Вторую плоскость $S_п$ расположим перпендикулярно направлению скорости движения пластины $W$ и совместим её со вторым крайним положением вибрирующей пластинки (см. рис. 5).

Согласно третьему закону механики безынертной массы, акустическое движение зависит от параметров пространства и времени. При выводе уравнения движения мы опять не будем пользоваться привычной для вас системой координат. Поэтому все необходимые характеристики будем брать из условия движения акустического потока. На рисунке 6 покажем дополнительные упрощающие условия, которые способствуют лучшему пониманию этого вида движения.

Они будут заключаться в том, что выделенный участок площади вибрирующей пластины мы представим в виде поршня, торцевая поверхность которого совпадает с поверхностью движения пластины. Поршень этот будет двигаться в цилиндре с радиусом $r$, материальные границы которого будут располагаться в пределах длины хода $l$ вибрирующей пластины (см. рис.6). Дальше перейдем к поэтапному рассмотрению движения этого поршня и к записи уравнений движения жидкости при этом ходе поршня.

При возвратно-поступательном ходе поршня его движение будет состоять из двух этапов: первый этап – это поступательное движение из первого крайнего положения во второе крайнее положение с длиной пути равной $l$. Второй этап будет поступательным движением поршня из второго крайнего положения в первое крайнее положение, тоже с длиной пути равной $l$ (см. рис. 6.*а* и 6.*б*).

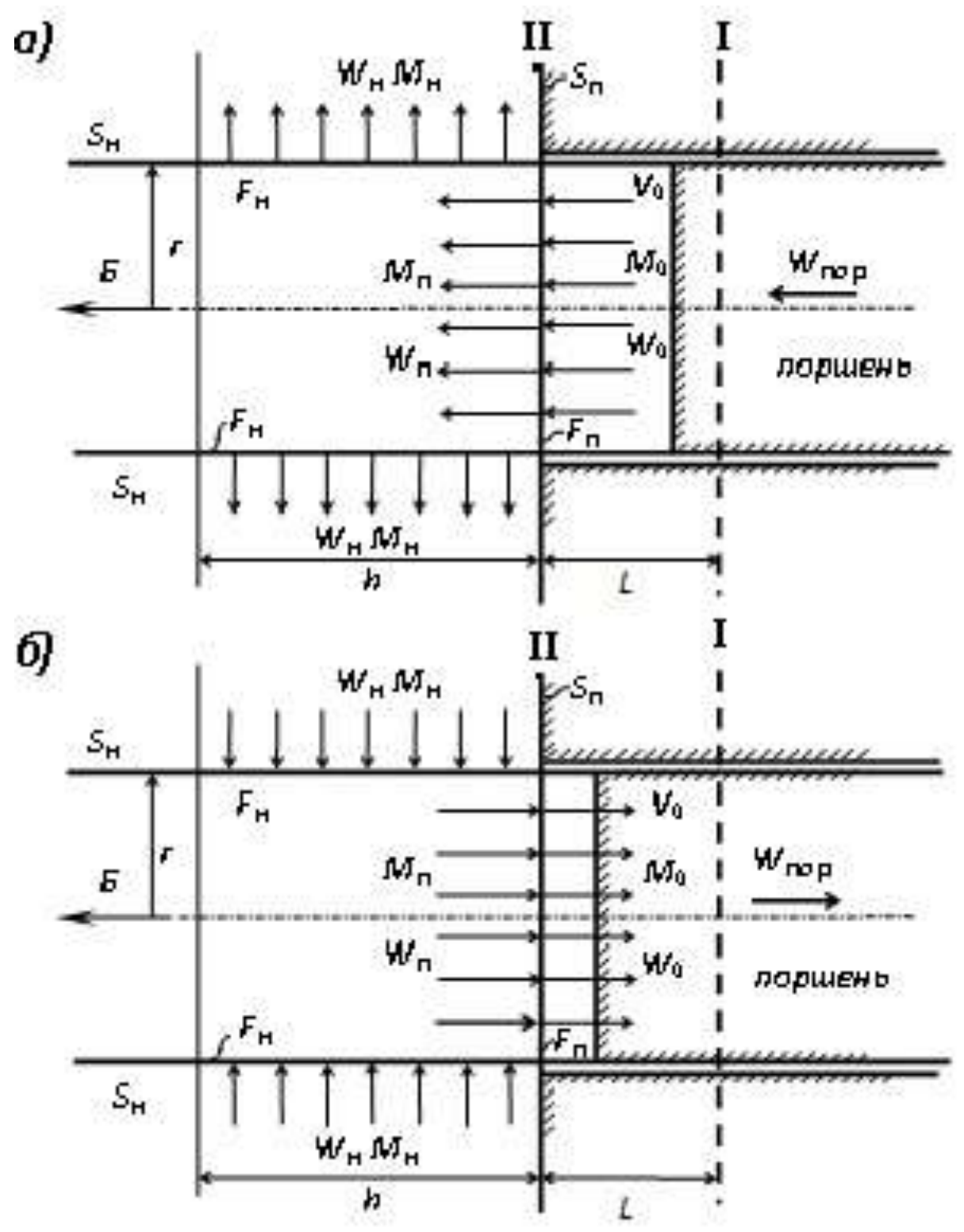

Рис. 6

**Рассмотрим первый этап движения поршня**.

Полагаем, что поршень находится в первом крайнем положении и из этого положения начинает своё движение со скоростью $W_{пор}$ во второе крайнее положение. В это время жидкость находится перед поршнем и занимает объём $V_0$, который располагается на длине $l$ между двумя крайними положениями поршня. Фактически такого объёма не существует перед поверхностью вибрирующей пластины. Здесь мы его придумали лишь для удобства вывода уравнений движения. Поэтому этот объём $V_0$ будем называть условным. Вы в дальнейшем поймете, для чего он нужен

При своём движении из первого положения во второе, поршень начинает выталкивать жидкость из условного объёма. Выталкиваемая жидкость будет поступать в среду, или объём потока, через площадь сечения потока $F_п$. У нас поршень движется со скоростью $W_{пор}$. Эта скорость тоже условная. Из площади сечения $F_п$ потока она поступает в поток как скорость $W_п$. По той причине, что жидкость несжимаема, скорость поршня $W_{пор}$ будет равна условной скорости движения жидкости $W_0$. А условная скорость движения $W_0$ одновременно со скоростью потока $W_п$ берёт начало из одной площади сечения $F_п$. По этой причине все три скорости потока будут равны между собой: $W_{пор} = W_0 = W_п$. Соответственно и расходы масс в единицу времени тоже будут равны между собой: $M_0 = M_п$.

Тогда мы можем записать уравнение движения для площади сечения $F_п$, которая находится в плоскости $S_п$. Это уравнение будет аналогично уравнению движения установившегося вида движения, но с той разницей, что скорость движения потока $W_п$ может быть переменной во времени, поскольку эта скорость равна скорости движения поршня, а эта скорость на этапе его движения из первого положения во второе может быть разнообразной. Коль скорости могут быть переменными, то расход массы жидкости тоже может быть переменным. Запишем это уравнение:

$$M_п(t) = \rho F_п W_п(t). \qquad (47)$$

Мы получили первое уравнение движения, которое является уравнением движения для потока в плоскости $S_п$ на первом этапе движения поршня.

Далее нам необходимо будет получить уравнение движения для нормальной поверхности $S_н$.

Отметим как аксиому один из важных принципов движения жидкостей и газов:

**жидкости и газы в любом потоке могут двигаться либо в прямолинейном направлении, либо одновременно в двух взаимоперпендикулярных направлениях, и только взаимоперпендикулярных.**[7]

Этот принцип следует из практических наблюдений за фактическим движением жидкостей и газов.

Это значит, что на первом этапе движения поршня одновременно с поступлением через плоскость $S_п$ массового расхода $M_п$ потока образуется нормальный массовый расход $M_н$ через нормальную плоскость $S_н$. Ведь расход основного потока $M_п$ поступает в среду потока, когда жидкость там находится в состоянии покоя. При своём движении вперёд этот расход $M_п$ начинает вытеснять жидкость среды в нормальном для себя направлении. То есть этот поступающий расход потока будет вытеснять из среды в нормальном направлении одинаковый расход с собственным: $M_п(t) = M_н(t)$. Ведь жидкость несжимаема.

Полная площадь сечения потока $F_н$ в нормальном направлении образуется не сразу, а постепенно, в течение всего времени первого этапа движения поршня, так как далее возмущение жидкости распространяется в прямом направлении в объёме невозмущённого потока не со скоростью движения поршня $W_{пор}$, а с иной скоростью, которая зависит от физических свойств жидкости. Эту скорость принято называть скоростью звука. Обычно она обозначается буквой $C$. Когда поршень за время $t$ достигает своего второго крайнего положения, пройдя расстояние $l$, протяжённость возмущения в потоке достигнет некоторой величины $h$ (см. рис. 6.*а*), которая будет равна произведению скорости звука $C$ на время движения поршня $t$, то есть

$$h = C \cdot t. \qquad (48)$$

При этой величине $h$ площадь сечения потока в нормальном направлении достигнет своей максимальной величины на первом этапе движения поршня. Эта поверхность сечения имеет форму цилиндра, радиус которого равен $r$, а высота равна высоте $h$. Тогда площадь сечения в

---

[7] Этот принцип относится к каждому из четырёх видов движения, включая установившееся движение по окружности, см. [1]. Движением реальной среды могут являться комбинации этих видов, но при этом данный принцип сохраняется для каждого вида (ред.).

нормальном направлении будет равна произведению длины окружности ($2\pi r$), умноженной на высоту $h$:

$$F_{\text{н}}(t) = 2\pi r h(t) = 2\pi r C t. \tag{49}$$

В уравнении (49) величины площади сечения и высота $h$ зависят от времени. Поэтому поступающий расход массы $M_{\text{п}}$ будет вытеснять в нормальном направлении себе подобный объём жидкости через непрерывно увеличивающуюся с течением времени нормальную площадь сечения $F_{\text{н}}$. Когда площадь сечения достигнет своего максимального значения, нормальная скорость движения $W_{\text{н}}$ достигнет своего минимального значения. Это значит, что нормальная скорость движения жидкости зависит от времени. Тогда уравнение движения в нормальном направлении будет иметь вид:

$$M_{\text{н}}(t) = \rho F_{\text{н}}(t) W_{\text{н}}(t)\ . \tag{50}$$

Подставим в уравнение (50) значение площади потока по уравнению (49). Получим:

$$M_{\text{н}}(t) = \rho \cdot 2\pi r C t \cdot W_{\text{н}}(t). \tag{51}$$

В уравнении движения (51) мы можем заменить время $t$ через длину $l$ хода поршня делённую на скорость движения поршня, как:

$$t = \frac{l}{W_{\text{пор}(t)}}\ .$$

Подставим это значение в уравнение (51), получим:

$$M_{\text{н}}(t) = \rho \cdot 2\pi r C \cdot l \frac{W_{\text{н}(t)}}{W_{\text{пор}(t)}}\ . \tag{52}$$

Поскольку нормальный расход выходит из нормальной плоскости потока, то мы будем вынуждены этот расход записать со знаком минус. Тогда мы будем иметь:

$$M_{\text{н}}(t) = -\rho \cdot 2\pi r C l \frac{W_{\text{н}(t)}}{W_{\text{пор}(t)}}\ . \tag{53}$$

Уравнение (53) будет уравнением движения жидкости в нормальном направлении на первом этапе движения поршня.

Мы получили уравнения движения (47) и (53) для первого этапа хода поршня. Теперь рассмотрим второй этап движения поршня, когда он будет двигаться из положения II в положение I (см. рис. 6.*б*).

Принцип движения и распространения возмущения в объёме потока жидкости остаётся таким же, как и на первом этапе движения поршня. Различие здесь заключается в том, что, двигаясь из положения II в положение I, поршень будет двигать жидкость из объёма $V_0$. Это значит, что скорости прямолинейного и нормального движения жидкости в потоке сменят направление на противоположное, то есть их направление будет противоположно направлению на первом этапе движения поршня. Соответственно изменится причина возмущения жидкости в объёме потока. На первом этапе движения поршня этой причиной были прямолинейные скорости, которые воздействовали на невозмущённый поток жидкости. На втором этапе этой причиной будут нормальные скорости $W_{\text{н}}$ потока. Ибо они компенсируют утечки жидкости в условный объём $V_0$. Отсюда следует, что уравнения движения для второго этапа движения поршня будут отличаться от уравнений движения для первого этапа только знаками. Поэтому перепишем уравнения движения первого этапа движения поршня с противоположными знаками и получим уравнения движения на втором этапе движения поршня. Тем самым мы избежим повторного объяснения движения жидкости в потоке.

$$M_{\text{п}}(t) = -\rho F_{\text{п}} W_{\text{п}}(t). \tag{54}$$

Уравнение движения для нормального потока при втором этапе движения поршня будет иметь вид:

$$M_{\text{н}}(t)=\rho\cdot 2\pi rCl\,\frac{W_{\text{н}}(t)}{W_{\text{пор}}(t)}\,. \qquad (55)$$

Как вы, наверное, догадались, волна состоит из двух полуволн. Первый этап движения поршня дал нам картину движения жидкости в первой полуволне, а второй этап дал нам картину движения во второй полуволне. В сумме они составят волну движения. Мы так и привыкли воспринимать для себя понятие волны, как состоящую из двух противоположных половинок. В действительности же эти противоположные половинки волны могут существовать каждая в отдельности не хуже, чем в соединении. Например, если мы совершим поршнем только первый этап движения или только второй, то образовавшаяся при этом полуволна будет распространяться с такой же скоростью и с таким же характером движения, как если бы она состояла в целой волне. Наш орган слуха способен различать характерные особенности самостоятельно существующих различных полуволн. Полуволну для первого этапа движения поршня мы можем воспроизвести, разорвав воздушный шарик. Тогда мы слышим резкий и сильный хлопок. Подобную половину волны мы слышим при полёте самолёта со сверхзвуковой скоростью или при выстреле орудия. Полуволну для второго этапа движения поршня мы можем воспроизвести, разбив обыкновенную электролампу. Ибо из объёма лампочки выкачан воздух и по отношению к атмосферному давлению она испытывает недостаток давления. Звук разбитой лампочки мы воспринимаем как глухой звук. Морские волны и им подобные, которые образуются от действия ветра, должны состоять тоже из полуволн первого этапа движения поршня, так как воды морей воспринимают воздействие ветра как первый этап движения поршня. Только распространение возмущения происходит не со скоростью звука, а в обычном порядке, который присущ механике безынертной массы. В этом заключается различие в движении морских волн и звуковых волн. В соответствии с этим различием претерпят изменения уравнения движения для обоих этапов движения поршня, то есть мы должны будем заменить в этих уравнениях скорость звука на скорость распространения возмущения при соответствующих динамических силах давления. Далее на рис. 7 мы изобразим распределение скоростей движения.

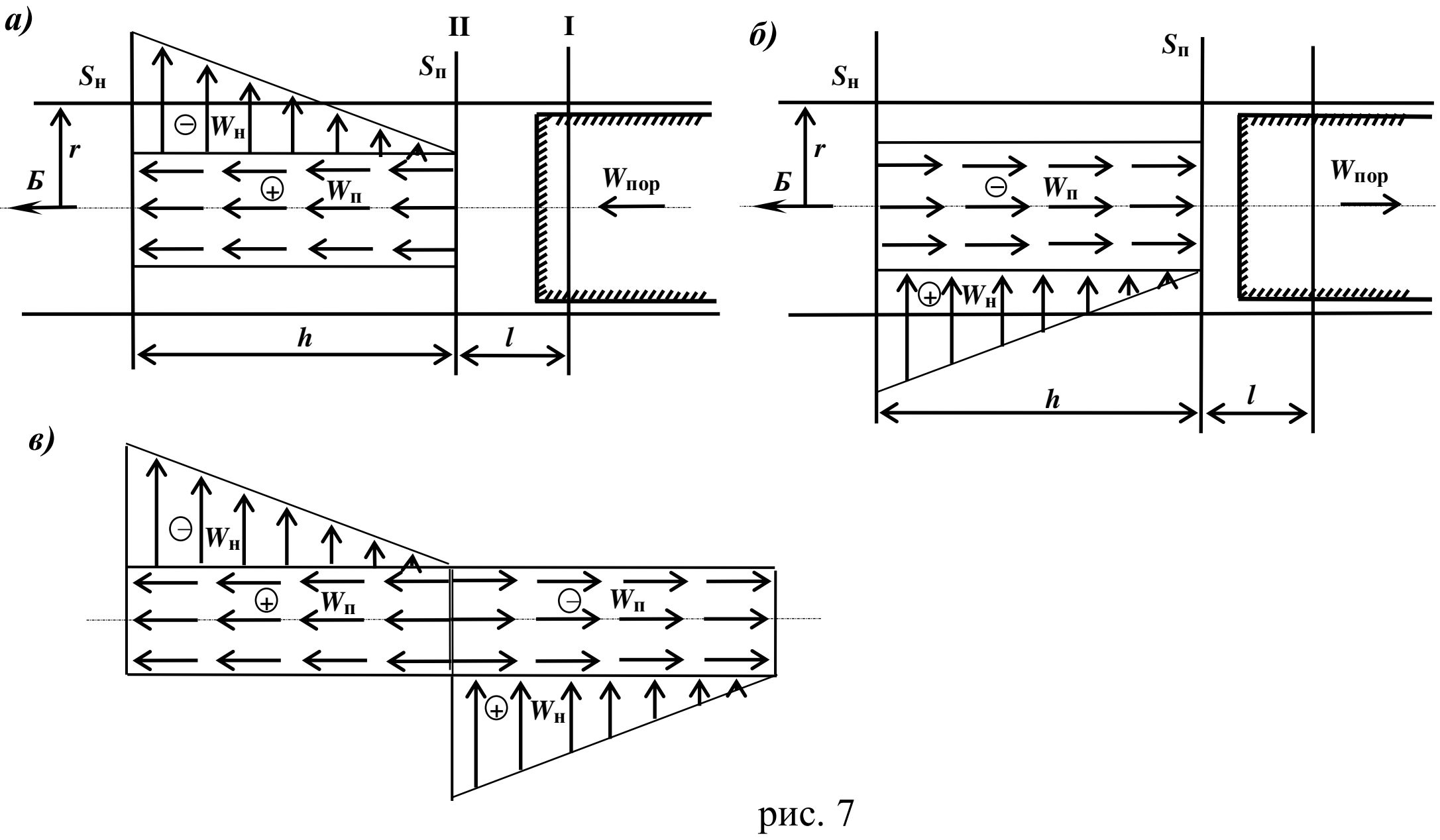

рис. 7

Здесь мы полагаем, что скорости поршня $W_{\text{пор}}$ на обоих этапах равны между собой и постоянны по величине. Тогда характер распределения нормальных скоростей $W_{\text{н}}$ будет именно таким, каким он показан на рис. 7. На рисунке 7.*а* дано распределение прямолинейных $W_{\text{п}}$ и нормальных $W_{\text{н}}$ скоростей на первом этапе движения поршня. На рисунке 7.*б* дано распределение прямолинейных и нормальных скоростей на втором этапе движения поршня. На рисунке 7.*в* дано распределение линейных и нормальных скоростей для двух этапов движения поршня, или для одной волны.

Если поршень будет непрерывно совершать возвратно-поступательное движение, то и волны будут образовываться непрерывно и последовательно удаляться, одна за другой, в пространство. Распространение звуковых волн мы воспринимаем как звучание различной продолжительности.

На рис. 7 мы не показали распределение ещё одной скорости – это скорости звука. Она – своеобразная скорость. Её своеобразие попробуем разъяснить на примере. Для чего нам придётся взять бильярдные шары и расставить их в ряд, как показано на рис. 8.

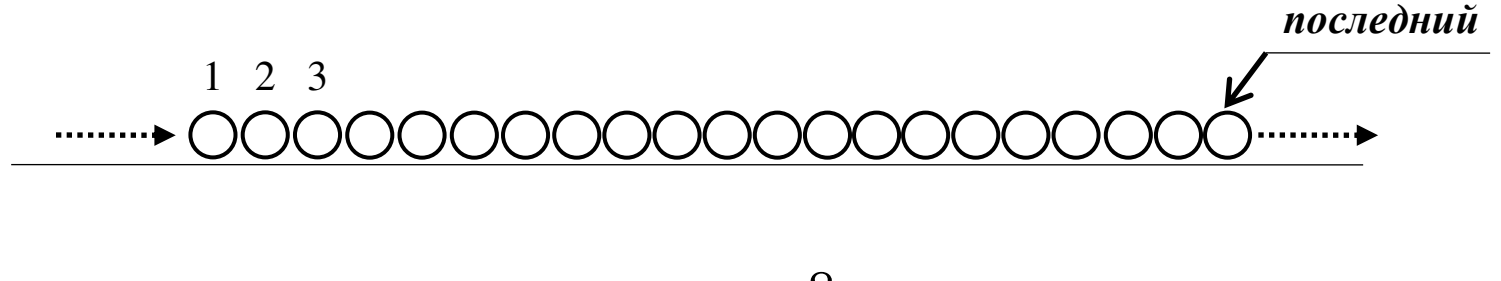

рис. 8

Если мы теперь слегка толкнём первый шар, то он ударится о второй шар и остановится. Второй шар, в свою очередь, начнет двигаться и ударится о третий шар и затем тоже остановится. В такой последовательности импульс первого шара за сравнительно небольшое время передастся в конечном итоге последнему. Что мы здесь заметим? Что каждый из шаров движется сравнительно медленно, а после передачи импульса даже останавливается, но скорость передачи импульса получается сравнительно высокой. Она во много раз превышает скорость любого из этих шаров. Так вот, в переводе на наш язык, скорость движения шаров будет нормальной и прямолинейной скоростью движения потока, а скорость передачи импульсов будет скоростью возмущения или скоростью звука. Поэтому она зависит от физических свойств жидкости.[8] В акустическом движении она определяет величину возмущённого объёма жидкости, которая определена нами величиной $h$. Это означает, что в этом участке объёма жидкость находится в движении, а в остальном объёме потока она находится в состоянии покоя.

Будем считать, что мы разобрались с распределением скоростей движения в акустическом потоке. На этом можно было бы закончить с уравнениями движения, но мы при своих упрощениях не учли ещё одного очень важного обстоятельства. Мы не учли в длине полуволн $h$ величину $l$ хода поршня или пластины. Мы ведь приняли условный объём $V_0$ потока с длиной $l$ и длину $h$ полуволны отсчитывали от неподвижной плоскости $S_п$. Теперь мы должны исправить это обстоятельство. Для чего нам придётся начало акустического потока брать не от плоскости $S_п$, а от поверхности поршня или пластины. Для чего обратимся к рис. 9, где показаны два этапа движения поршня.

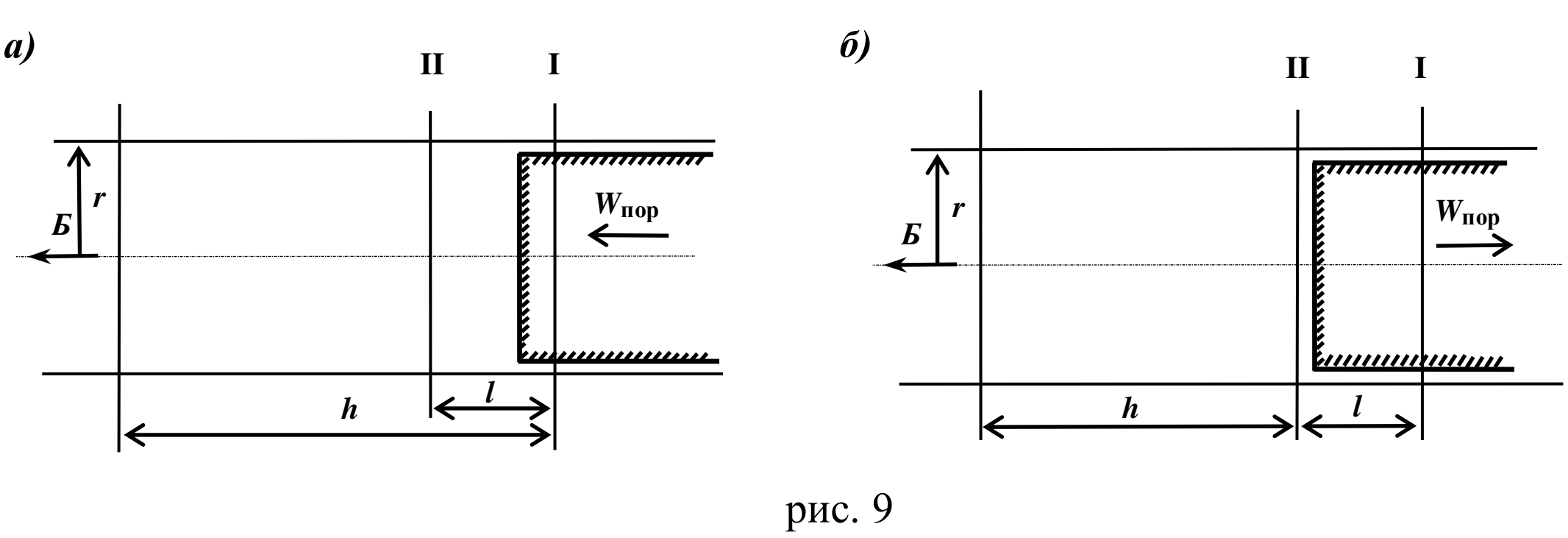

рис. 9

Начнём с первого этапа движения поршня. В реальных условиях, когда поршень находится в первом положении, начало отсчёта и для хода поршня $l$, и для длины $h$ распространения скорости возмущения мы должны будем взять от плоскости поршня, что мы и сделали на рис. 9.*а*. Как только поршень начинает движение из положения один в положение два, так сразу образуется скорость возмущения $C$. В акустическом виде движения её величина превосходит величину скорости движения поршня. Поэтому, опережая его, она будет распространяться самостоятельно.

---

[8] Т.к. ни о каком импульсе для безынертных структурных единиц среды не может быть и речи согласно первому закону механики безынертной массы, то сравнение скорости передачи импульса биллиардными шарами и скорости распространения возмущения является аналогией, позволяющей автору объяснить лишь различие между скоростью движения массы в поступательном направлении в акустическом потоке и скоростью образования этого потока, т.е. скоростью образования зоны возмущения среды, которая никак не зависит от характеристик источника возмущений, будь то пластина, поршень и что угодно. (ред.)

В конечном итоге, когда поршень пройдет расстояние $l$ и достигнет второго крайнего положения, за это же время скорость возмущения пройдет расстояние $h$. Поскольку мы взяли начало от плоскости поршня, который находился в положении I, то фактическая длина образовавшейся полуволны будет тоже отсчитываться от плоскости поршня. По этой причине фактическая длина полуволны первого этапа будет меньше величины распространения возмущения $h$ на величину хода поршня $l$, то есть

$$h_{\text{ф}}^{\text{I}} = h - l. \qquad (56)$$

В уравнении движения для нормального движения потока (53) площадь сечения зависит от величины возмущения потока $h$, то есть от фактической длины полуволны. Поэтому в уравнении (53) мы должны будем сделать соответствующую поправку. Тогда оно примет вид:

$$M_{\text{н}}^{\text{I}}(t) = -\rho \cdot 2\pi r\left(C\frac{l}{W_{\text{пор}}(t)} - l\right)W_{\text{н}}^{\text{I}}(t) \qquad (57)$$

Уравнение (67) будет действительным уравнением движения для нормального движения жидкости на первом этапе движения поршня.

Теперь рассмотрим движение поршня на втором этапе движения (см. рис. 9, *б*). Здесь мы берём начало отсчёта, когда с началом движения поршня в первое положение образовавшаяся скорость возмущения $C$ начнет распространяться в сторону противоположную движению поршня. Когда поршень достигнет первого положения, то за это время длина зоны возмущения достигнет величины $h$. Поэтому фактическая длина полуволны второго этапа движения поршня будет больше на величину хода поршня $l$, то есть

$$h_{\text{ф}}^{\text{II}} = h + l. \qquad (58)$$

Поэтому в уравнение движения (55) для нормального движения потока при втором этапе движения поршня мы должны будем внести соответствующую поправку. Тогда оно примет вид:

$$M_{\text{н}}^{\text{II}}(t) = \rho \cdot 2\pi r\left(C\frac{l}{W_{\text{пор}}(t)} + l\right)W_{\text{н}}^{\text{II}}(t)\ . \qquad (59)$$

Уравнение (59) будет действительным уравнением движения для нормального потока на втором этапе движения поршня.

На этом можно закончить разбор движения акустического потока и перейти к получению для него уравнений сил. Зная уравнения движения потока, составить уравнения сил для него не представляет большого труда. Для этого лишь надо записать условие равновесия на площадях сечения потока. Для акустического потока эти площади лежат в нормальной плоскости $S_{\text{н}}$ и поступательной плоскости $S_{\text{п}}$.

Начнём опять с первого этапа движения поршня (см. рис. 6.*а*). Сначала запишем условие равновесия для площади сечения потока $F_{\text{п}}$ в прямом направлении. Получим:

$$F_{\text{п}}P_{\text{пр.п}}(t) = F_{\text{п}}P_{\text{ст.п}}(t) + F_{\text{п}}P_{\text{дин.п}}(t). \qquad (60)$$

В уравнении (60) заменим динамические силы давления по второму закону механики безынертной массы. Получим:

$$F_{\text{п}}P_{\text{пр.п}}(t) = F_{\text{п}}P_{\text{ст.п}}(t) + M_{\text{п}}W_{\text{п}}(t). \qquad (61)$$

Можно также уравнение (61) отнести к единице площади сечения потока, разделив его члены на площадь сечения потока $F_{\text{п}}$, и преобразовать расход массы по уравнению движения. Тогда получим:

$$P_{\text{пр.п}}(t) = P_{\text{ст.п}}(t) + \rho W_{\text{.п}}^{2}(t) \qquad (62)$$

Уравнения (60), (61) и (62) являются разновидностями уравнения сил при движении потока в прямом направлении на первом этапе движения поршня. Они собой представляют равенство принятых сил давления $P_{\text{пр.п}}$ сумме статических $P_{\text{п.ст}}$ и динамических $P_{\text{п.дин}}$ сил давления потока. Все члены уравнений имеют знак времени $(t)$, который означает, что силы давления потока могут быть переменными во времени. Переменность сил давления потока будет зависеть от изменения во времени сил давления на поршне движущемся на первом этапе своего движения.

Затем для первого этапа движения поршня (см. рис. 6.*а*) записывается условие равновесия для нормальной площади сечения потока. Оно будет иметь вид:

$$F_{\text{н}}(t)P_{\text{пр.н}}(t) = F_{\text{н}}(t)P_{\text{ст.н}}(t) + F_{\text{н}}P_{\text{н.дин}}(t). \qquad (63)$$

В уравнении (63) заменим динамические силы давления по второму закону механики безынертной массы. Получим:

$$F_{\text{н}}(t)P_{\text{пр.н}}(t) = F_{\text{н}}(t)P_{\text{ст.н}}(t) + M_{\text{н}}(t)W_{\text{н}}(t). \qquad (64)$$

Оставим уравнение (64) в таком виде. Уравнения (63) и (64) являются уравнениями сил для нормального потока на первом этапе движения поршня, или, проще, они являются уравнениями сил в нормальном направлении для первой полуволны. В таком виде мы их оставим лишь потому, что нормальные силы давления изменяются во времени не только в зависимости от характера движения поршня, но ещё зависят от изменения площади сечения потока в нормальном направлении, которая изменяется во времени независимо от характера движения поршня. По этой причине в более развернутой форме уравнения сил лучше применять для конкретных условий движения. Таким образом, мы получили уравнения сил для первого этапа движения поршня, или для первой полуволны.

Для второго этапа движения поршня, или для второй полуволны (см. рис. 6.*б*), уравнения сил будут аналогичными. Поэтому мы их не будем специально выводить, а лишь укажем, что форма записи поступательных и нормальных сил давления для второго этапа движения поршня одинакова с уравнениями сил (60) – (64) для первого этапа движения поршня. Так что мы можем в одинаковой мере пользоваться уравнениями сил (60) – (64) как для первой полуволны, так и для второй.

Мы получили всё необходимые зависимости для характеристики акустического, или волнового, движения жидкостей и газов.

## Глава 3. РАБОТА И ЭНЕРГИЯ

Переходим к зависимостям работы и энергии.

Работа означает новое качественное понятие, которое независимо от переменности движения жидкостей и газов определяет прошедшее движение в общей сумме. Поэтому мы выше определили работу как относящуюся к прошлому движению жидкостей и газов. Это новое понятие не даётся с помощью каких-либо выводов, а берётся непосредственно из практических наблюдений. Как мы знаем, действующими силами в жидкостях и газах являются силы давления *P*, или просто давление. С помощью этих сил происходит проталкивание или выталкивание жидкостей и газов через определённые площади сечения потока. Вытолкнутый объём жидкости или газа является вещественной величиной, свидетельствующий как факт о произошедшем ранее движении жидкостей и газов. Поэтому для работы можно будет дать следующее определение:

***Работой L сил давления P на неподвижной поверхности или плоскости потока называется произведение объёма жидкости V на силы давления P, или***

$$\boldsymbol{L = VP} \qquad (65)$$

Отсюда следует, что работа *L* не зависит ни от плотности жидкостей и газов, ни от времени, ни от других каких-либо характеристик потока. Следовательно, работа является удобной величиной для практического применения. Практически этой величиной пользуются не тогда, когда жидкость разольётся, вытолкнутая из потока силами давления, или когда газ улетучится, тоже вытолкнутый из потока силами давления. В этом случае трудно собрать эти выброшенные из потока объёмы. Гораздо проще замерять силы давления и проходящие объёмы жидкостей и газов, установив для этой цели необходимые датчики и приборы в этом потоке. А затем по записям или по непосредственному наблюдению за приборами определяют количество совершенной работы. Или определяют саму её величину, которую человек способен извлекать из движущихся потоков жидкостей и газов. Так практически выражается основное понятие работы.

Работу, совершаемую в настоящий момент, мы назвали мощностью. Мощность определяется количеством работы в единицу времени. Если обозначить работу буквой $L$, а время буквой $t$, то мощность $N$ можно будет выразить следующей зависимостью:

$$N = \frac{L}{t} = \frac{VP}{t}. \tag{66}$$

Объём мы можем представить как площадь $F$, умноженную на длину $K$, то есть $V = F \cdot K$, а затем подставить в уравнение (66). Тогда получим:

$$N = \frac{FKP}{t}. \tag{67}$$

Длина $K$ делённая на время $t$, есть ни что иное, как скорость $W$. Тогда уравнение (67) будет иметь вид:

$$N = FWP. \tag{68}$$

Уравнение (68) является наиболее удобной формой для выражения мощности, или работы в единицу времени. Ибо для конкретного потока площадь $F$ будет площадью сечения потока, а скорость $W$ будет скоростью потока. В таком виде уравнение (68) становится практически необходимым.

Энергия, как мы выше её назвали, это будущая или располагаемая работа. Как вы сами понимаете, знать её величину просто необходимо. В противном случае, не зная величины энергии того или иного жидкостного или газового потока, мы просто не сможем его использовать. Не зная энергии, мы бы не смогли понять многих явлений природы. Впервые уравнение энергии было получено и сформулировано Даниилом Бернулли. Поэтому оно известно всем как уравнение энергии Бернулли. Запишем его для жидкостного потока:

$$VP_{\text{пол}} = VP_{\text{ст}} + \frac{1}{2} VP_{\text{дин}}, \tag{69}$$

где $V$ обозначает полный объём потока, $P_{\text{пол}}$ – полную величину сил давления потока, $P_{\text{ст}}$ обозначает статические силы давления, а $P_{\text{дин}}$ – динамические силы давления потока.

В уравнении (69) произведение объёма $V$ на полные силы давления $P_{\text{пол}}$, как видите, обозначает работу, но коль эта работа ещё не совершилась, то мы назовем её энергией. Поскольку в этом произведении мы имеем дело с полным объёмом потока и с полными силами давления $P_{\text{пол}}$, то назовем его полной энергией потока жидкости.

Второе произведение уравнения (69) – это объём потока жидкости $V$, умноженный на статические силы давления потока $P_{\text{ст}}$, является тоже будущей работой. Название *статические* обозначает, что эти силы давления являются стационарными, или неподвижными, силами потока. В механике твёрдого тела энергию неподвижных, или стационарных, сил принято называть потенциальной. По аналогии и мы назовем эту часть энергии потока потенциальной энергией.

Третье произведение уравнения (69) представляет собой половину объёма потока $V$, умноженного на динамические силы давления $P_{\text{дин}}$. В механике принято называть энергию движения тела кинетической. Мы эту часть энергии потока тоже назовем кинетической энергией потока.

Тогда смысл уравнения (69) будет выражаться в том, что полная энергия любого потока жидкости ($VP_{пол}$) равна сумме его потенциальной ($VP_{ст}$) и кинетической ($\frac{1}{2}VP_{дин}$) энергий. Затем мы можем преобразовать уравнение (69), заменив в нем динамические силы давления по второму закону механики безынертной массы как $P_{дин} = \frac{MW}{F}$. Тогда получим:

$$VP_{пол} = VP_{ст} + \frac{1}{2}V\frac{MW}{F}. \qquad (70)$$

В уравнении (70) мы можем заменить расход массы в единицу времени по уравнению движения установившегося вида движения как $M = F\rho W$. Тогда получим:

$$VP_{пол} = VP_{ст} + \frac{1}{2}V\rho W^2. \qquad (71)$$

Но более удобно пользоваться уравнением (71), если отнести его к единице объёма. Для чего разделим и правую, и левую части на объём потока $V$, получим:

$$P_{пол} = P_{ст} + \frac{\rho W^2}{2}. \qquad (72)$$

Все эти уравнения являются разновидностями уравнения энергии любого потока жидкости. Поэтому каждое из этих уравнений в равной степени пригодно для практических целей.

Для газового потока уравнение энергии будет иным, так как газ сжимаем. На сжатие газа тоже затрачивается работа, которая затем выделяется при его расширении. Поэтому мы должны будем учесть это обстоятельство в уравнении энергии. В термодинамике работа, затрачиваемая на сжатие газа, учитывается таким уравнением:

$$L = \int_{V_1}^{V_2} PdV. \qquad (73)$$

Будущая или располагаемая работа, записанная уравнением (73), будет уже энергией. Это интегральная форма записи работы и энергии. Вы скажите, что в этом вы ничего не понимаете. Пока вам не нужно разбираться в интегралах. В будущем вы в них разберётесь и поймете, что это довольно простая вещь. Здесь интегральная форма уравнения (73) есть, ни что иное, как тоже произведение объёма потока на силы давления, то есть $VP$. Подобную интегральную форму можно записать для любого уравнения работы или энергии жидкости, которые мы получили выше. Поэтому вы, наверное, догадались, что мы могли бы записать уравнение (73) просто как произведение объёма потока на его силы давления и дополнить им, например, уравнение (71), и тогда бы мы получили уравнение такого вида:

$$VP_{пол} = VP_{ст} + \frac{1}{2}V\rho W^2 + VP_{тер}. \qquad (74)$$

В уравнении (74) мы обозначили силы давления последнего члена уравнения как термодинамические силы давления $P_{тер}$. Или мы это же уравнение (74) можем записать ещё таким способом:

$$VP_{пол} = VP_{ст} + \frac{1}{2}V\rho W^2 + \int_{V_1}^{V_2} PdV. \qquad (75)$$

Уравнения (74) и (75) будут равнозначны. Они представляют собой полное уравнение энергии газового потока. Тогда полная энергия потока газа будет состоять из потенциальной энергии, кинетической и энергии сжатия, или термодинамической. Термодинамическая энергия зависит как от механического сжатия газа, так и от температуры или теплового влияния. Ибо вы знаете, что при нагревании газы расширяются, а при охлаждении сжимаются. В зависимости от этого газ может либо испытывать дополнительное давление, либо терять часть давления. С точки зрения энергетического баланса поток газа либо получает дополнительную энергию при нагреве, либо теряет её при охлаждении. Вот эту потерю или прирост энергии и отражает интегральный член уравнения (75). Зависимость между давлением, температурой и объёмом газа записывается иными зависимостями, которых нет в механике жидкости и газа. Этими зависимостями занимается иная наука, которая называется термодинамика. Поэтому при решении задач нам придётся брать её зависимости и подставлять их в термодинамическую составляющую полной энергии газового потока. Таким образом, мы получили искомую энергию газового потока.

Будем считать, что вы разобрались с уравнениями работ и энергий в общем плане. Для большей ясности рассмотрим их относительно уже известных нам четырёх видов движения жидкостей и газов.

Начнём с установившегося вида движения жидкостей и газов.

С точки зрения энергии этот вид движения жидкостей и газов предназначен или приспособлен для передачи или переноса энергии на расстояние. В идеальном случае, когда нет потерь в потоке, связанных, например, с трением, обтекаемостью или потерей тепла в потоке, полная энергия потока должна полностью сохраняться, на какое бы расстояние мы её не транспортировали, то есть она не зависит от длины потока. Мы же в своей работе рассматриваем идеальный случай движения жидкостей и газов. Основной особенностью установившегося потока жидкости является то обстоятельство, что сечение этого потока может быть разным на его длине. Покажем часть установившегося потока на рис. 10.

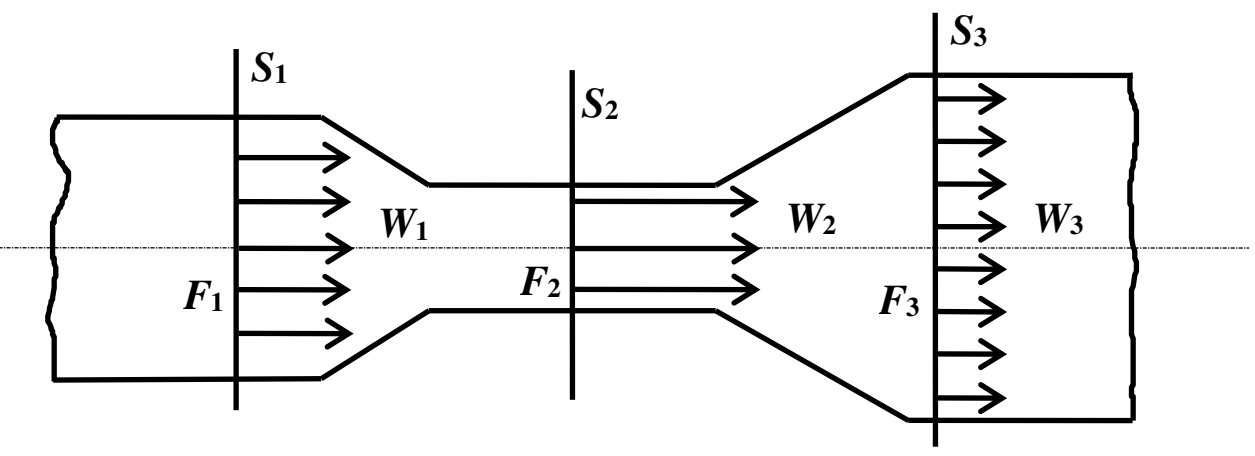

рис. 10

Как мы знаем, в любом сечении этого потока расход массы в единицу времени есть величина постоянная. Поэтому постоянство расхода массы в потоке с различными площадями сечения поддерживается различными скоростями движения в этих площадях сечения потока. На рис. 10 вы видите, что площадь сечения $F_1$ потока больше площади сечения $F_2$, площадь сечения $F_3$ больше $F_1$ и $F_2$, то есть

$$F_2 < F_1 < F_3.$$

Согласно этим площадям сечения потока устанавливаются скорости движения потока. Это вы можете проверить по уравнению движения установившегося вида движения. Тогда максимальная скорость движения жидкости будет в сечении $F_2$, а минимальная – в сечении $F_3$, так как площадь сечения потока $F_2$ имеет минимальную величину, а $F_3$ – максимальную величину.

$$W_3 < W_1 < W_2.$$

Для установившегося потока жидкости величина полной энергии потока тоже есть величина постоянная, то есть

$$E_{пол} = VP_{пол} = \text{const}.$$

Полная энергия потока, как мы знаем, состоит из суммы потенциальной и кинетической энергий. Для наглядности ещё раз выпишем уравнение (71):

$$E_{\text{пол}} = VP_{\text{пол}} = VP_{\text{ст}} + \frac{1}{2}V\rho W^2 = \text{const}.$$

.

Как мы выше установили, скорость движения потока непрерывно изменяется от сечения к сечению. Это значит, что кинетическая энергия в уравнении (71) тоже изменяется от сечения к сечению потока. Так, в сечении $F_2$ она будет максимальной, а в сечении $F_3$ – минимальной, то есть:

$$\frac{1}{2}\rho W_3^2 \langle \frac{1}{2}\rho W_1^2 \langle \frac{1}{2}\rho W_2^2.$$

При изменении кинетической энергии от сечения к сечению потока полная энергия потока ($VP_{\text{пол}}$) будет оставаться постоянной величиной. Чтобы полная энергия потока оставалась постоянной при изменении кинетической энергии, должна от сечения к сечению соответственно изменяться и потенциальная энергия, ведь эти энергии в сумме составляют полную энергию потока. Тогда потенциальная энергия от сечения к сечению потока будет распределяться в такой последовательности:

$$VP_{\text{ст}2} < VP_{\text{ст}1} < VP_{\text{ст}3}\,.$$

Для максимальной площади сечения потока максимальной будет величина потенциальной энергии и минимальной – величина кинетической энергии. Для минимальной площади сечения потока минимальной будет величина потенциальной энергии и максимальной – величина кинетической. Зависимость энергии для каждого сечения потока будет выглядеть таким образом:

$$VP_{\text{пол}} = VP_{\text{ст}1} + \frac{1}{2}V\rho W_1^2 = VP_{\text{ст}2} + \frac{1}{2}V\rho W_2^2 = VP_{\text{ст}3} + \frac{1}{2}V\rho W_3^2 = \text{const}.$$

Таким способом мы выразили неизменность величины полной энергии установившегося потока жидкости при изменении от сечения к сечению потока её составляющих, то есть потенциальной и кинетической энергий.

Теперь рассмотрим предельные случаи изменения площади сечения установившегося потока жидкости.

1. Полагаем, что на какой-то длине установившегося потока площадь сечения потока $F$ имеет бесконечно большую величину. Эту ситуацию легко можно представить себе практически. Полагаем, что на какой-то длине трубопровода установлен резервуар большого объёма. Тогда, согласно уравнению движения установившегося вида движения: $M=\rho FW=\text{const}$, скорость потока в этом сечении упадет до нуля. Выражаясь математически, она будет стремиться к нулю.

В уравнении энергий (71) установившегося движения жидкости ($VP_{\text{пол}}=VP_{\text{ст}}+\frac{1}{2}V\rho W^2$) кинетическая энергия тоже будет равна нулю, то есть $\frac{1}{2}V\rho W^2 = 0$, поскольку скорость в этом сечении потока равна нулю ($W = 0$), а умножение на ноль даёт в результате ноль. Тогда уравнение (71) примет вид:

$$VP_{\text{пол}} = VP_{\text{ст}} + 0. \qquad (76)$$

Отсюда следует, что в бесконечно большой площади сечения потока полная энергия потока будет равна потенциальной энергии потока. Таким образом, мы выяснили, что при увеличении площади сечения установившегося потока жидкости кинетическая энергия уменьшается, а потенциальная возрастает, и в предельном случае, когда площадь становится бесконечно большой, кинетическая энергия становится равной нулю, а потенциальная – полной энергии потока.

2. Возьмём второй предельный случай и определим, до какого предела можно уменьшать площадь сечения установившегося потока жидкости.

Ответ на этот вопрос напрашивается сам собой. Уменьшать площадь сечения потока можно лишь до того предела, когда потенциальная энергия потока в уравнении (71) будет равна кинетической энергии потока, то есть:

$$VP_{\text{ст}} = \frac{1}{2} V\rho W^2.$$

Это объясняется тем, что жидкость проталкивается в потоке потенциальной энергией потока. В этом случае и потенциальная, и кинетическая энергия потока равны по величине половине полной энергии потока:

$$\frac{1}{2} VP_{\text{пол}} = VP_{\text{ст}} = \frac{1}{2} V\rho W^2. \qquad (77)$$

Практически получается, что потенциальная энергия всегда в меньшей или большей степени превращается в кинетическую энергию потока, то есть

$$VP_{\text{ст}} \geq \frac{1}{2} V\rho W^2.$$

По уравнению (77) определяют возможные максимальные скорости потока, а затем по уравнению движения определяют минимальную площадь сечения установившегося потока.

Если мы захотим уменьшить минимальную площадь сечения потока, то в этом случае у нас должен будет уменьшиться расход массы в единицу времени $M$. Если мы захотим оставить расход массы соответствующим нашему первоначальному потоку, то мы будем вынуждены поднять давление в потоке. То есть мы будем вынуждены увеличить полную энергию потока, и тогда равенство (77) снова восстановится. Но в этом случае мы будем иметь иной поток жидкости, отличный от первоначального по уровню энергии. Это значит, что мы в любом случае не сможем нарушить равенство (77). По этой причине минимальные площади сечения потока при сохранении равенства (77) принято называть критическими $F_{\text{кр}}$, а скорости потока, соответствующие движению жидкости в этом сечении, тоже критическими $W_{\text{кр}}$.

Уравнения движения, сил и энергий для установившегося потока будут одинаковыми как для жидкости, так и для газов. Различие заключается в том, что в критическом сечении установившегося потока газа критические скорости $W_{\text{кр}}$ будут равны скорости звука $C$ для этого газа, то есть $W_{\text{кр}}=C$. Отметим, что это очень важные энергетические условия широко применяются в современной технике.

**Далее перейдем к плоскому установившемуся потоку жидкости**.

Объём потока плоского установившегося вида движения имеет форму цилиндра, в котором жидкость движется по логарифмической спирали либо от внешней границы потока к центру потока, либо, наоборот, от центра потока к его внешней границе. Далее мы представили движение по логарифмической спирали в виде двух взаимно перпендикулярных движений: радиального и тангенциального. Затем мы установили, что тангенциальные скорости потока $W_{tg}$ одинаковы для всего сечения потока и равны определённой величине. Радиальные скорости потока $W_r$ имеют различные величины. На внешней границе потока они имеют минимальную величину, а на внутренней поверхности потока, расположенной в непосредственной близости от оси потока, они имеют максимальную величину.

Теперь снова представьте себе ванну, где открыта воронка для стока воды. Вы увидите, что вода в ванне спокойна, то есть почти без движения, а у воронки ванны образуется цилиндрический поток плоского установившегося вида движения. Этот поток образуется в результате разницы энергий в самой ванне и в воронке отверстия. Обозначим энергию воды в ванне через $E_{\text{ван}}$, а энергию потока в воронке через $E_{\text{вар}}$. Поток образуется из-за того, что $E_{\text{ван}} > E_{\text{вор}}$. Тогда в потоке будет расходоваться энергия, равная разности между энергией воды ванны и воронки, то есть

$$E_{\text{пот}} = E_{\text{ван}} - E_{\text{вор}}. \qquad (78)$$

Разность энергий уравнения (78) даст нам полную энергию для образования потока, которая, согласно уравнению (71), будет равна:

$$E_{\text{пот}} = VP_{\text{пол}} = VP_{\text{ст}} + \frac{1}{2} V\rho W^2. \qquad (79)$$

Уравнение (79) в таком виде будет выражать общую, не конкретизированную ещё энергию плоского установившегося потока. Затем мы поступим с уравнением (79) следующим образом. Здесь задача будет заключаться в том, что на длине линии тока потока тангенциальные и радиальные скорости потока будут разными (см. рис. 11), а статическое давление должно быть одинаковым в точках *A* даже при разных скоростях. Поэтому нам необходимо будет распределить полную энергию потока $VP_{\text{пол}}$ по её составляющим частям, то есть по кинетической и потенциальной энергии.

Мы знаем, что полная энергия потока сохраняется без изменения по всему объёму потока, то есть полная энергия потока есть величина постоянная для каждого конкретного потока. Следовательно, сумма потенциальной и кинетической энергий тоже есть величина постоянная.

Мы знаем, что тангенциальные скорости движения потока $W_{tg}$ имеют постоянную величину для сечения потока. До какого-то радиального сечения потока, например $F_{r1}$ с радиусом $r_1$, тангенциальные скорости движения $W_{tg}$ будут превосходить по величине радиальные скорости $W_r$ движения потока. Тогда основная доля кинетической энергии будет приходиться на тангенциальное движение, то есть величина потенциальной энергии потока будет зависеть от величины тангенциальной кинетической энергии потока. Поэтому полную энергию потока распишем для тангенциального движения. Получим:

$$E_{\text{пот}} = VP_{\text{пол}} = VP_{\text{ст}} + \frac{1}{2}V\rho W_{tg}^2. \qquad (80)$$

Границы применимости уравнения (80) будут распространяться в объёме потока от внешних границ потока до некоторой площади радиального сечения $F_{r1}$ (см. рис. 11). Эта радиальная площадь сечения потока $F_{r1}$ характерна тем, что её величина будет равна по величине тангенциальной площади сечения потока $F_{tg}$. По этой причине и согласно уравнению движения в точках площади радиального сечения $F_{r1}$, радиус которой равен $r_1$, радиальные и тангенциальные скорости будут равны между собой по величине, то есть $W_r = W_{tg}$. По этой причине мы назвали радиальную площадь сечения потока $F_{r1}$ с радиусом $r_1$ площадью равных скоростей.

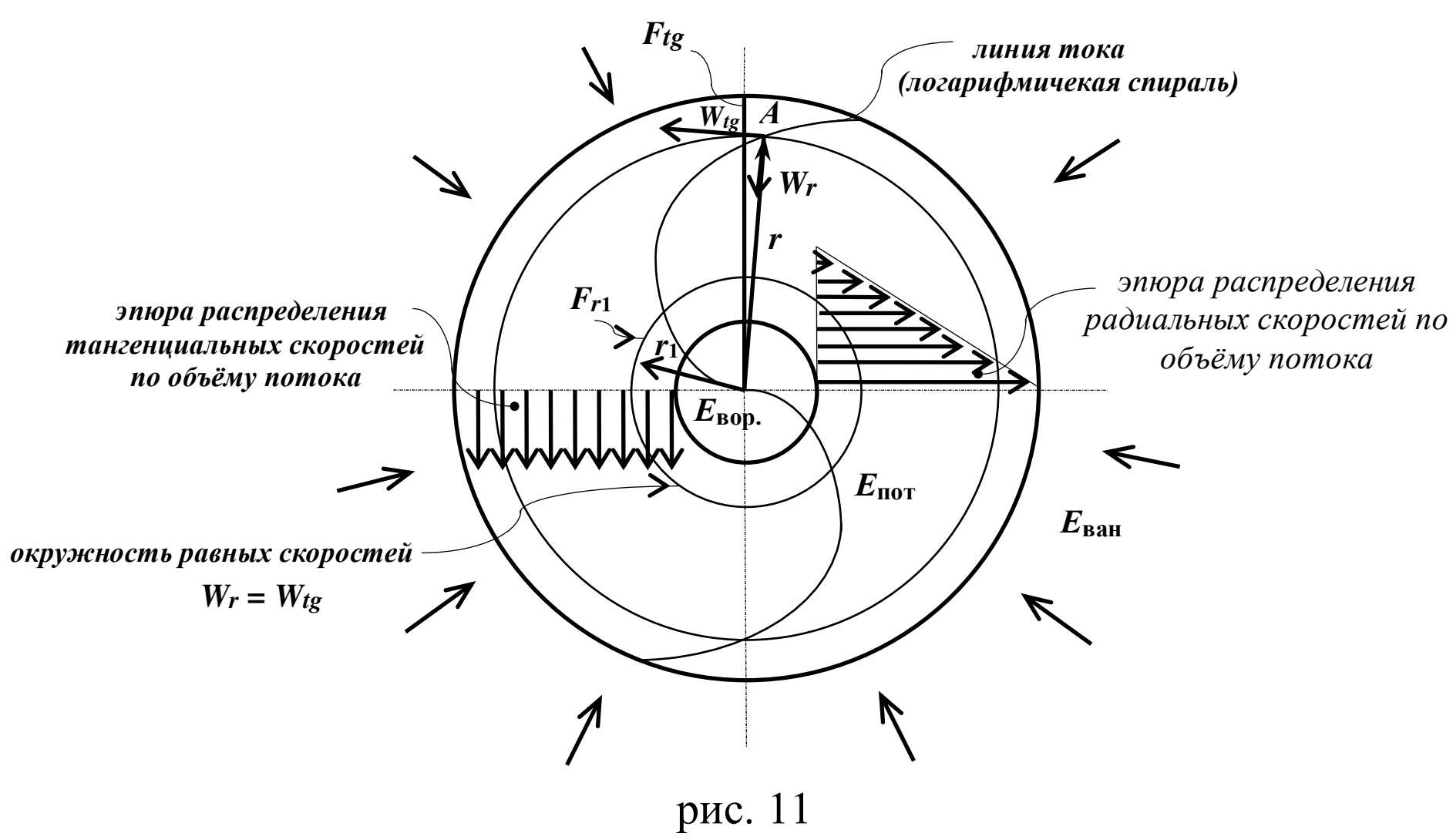

рис. 11

Отметим, что на окружности этой площади располагаются замечательные точки, как принято называть их в математике, относительно которых производят построение логарифмической спирали. По уравнению (80) мы в последующем сможем определить истинную величину потенциальной энергии в части объёма потока, расположенного от внешних границ до площади равных скоростей.

Если нам необходимо будет в этой части объёма потока определить радиальную кинетическую энергию, то её можно получить, воспользовавшись радиальным уравнением движения (17), по которому мы определяем необходимую нам скорость в любом радиальном сечении потока. После чего мы сможем определить радиальную кинетическую энергию:

$$E_{r\ \text{кин}} = VP_{r\ \text{дин}} = \frac{1}{2} V\rho W_r^2 . \qquad (81)$$

Таким образом, мы получили зависимости для полной энергии потока в одной части объёма. Это уравнения (80) и (81).

В следующей части потока, которая расположена от окружности равных скоростей до внутренней границы потока у оси потока, радиальные скорости потока $W_r$ будут больше по величине тангенциальных скоростей движения $W_{tg}$. При этом радиальные скорости увеличиваются по величине при переходе от окружности равных скоростей к внешней границе потока. Это происходит потому, что радиальные площади сечения непрерывно уменьшаются в этом направлении. Этим мы просто подчеркнули, что величина потенциальной энергии в этой части плоского установившегося потока будет зависеть от величины радиальной кинетической энергии. Тогда уравнение энергии примет вид:

$$E_{\text{пот}} = VP_{\text{пол}} = VP_{\text{ст}} + \frac{1}{2} V\rho W_{r_i}^2 . \qquad (82)$$

Долю тангенциальной кинетической энергии можно будет определить по следующему уравнению:

$$E_{tg\ \text{кин}} = VP_{tg\ \text{дин}} = \frac{1}{2} V\rho W_{tg}^2 . \qquad (83)$$

Таким образом, мы получили все необходимые зависимости для распределения энергии в плоском установившемся потоке жидкости.

Просто в установившемся потоке жидкости мы установили, что минимум площади сечения потока $F$ ограничивается количественным распределением между потенциальной и кинетической энергией. Это условие мы записали такой зависимостью:

$$VP_{\text{ст}} \geq \frac{1}{2} V\rho W^2 .$$

Эта зависимость говорит о том, что потенциальная энергия потока всегда либо больше, либо равна кинетической энергии потока. Это условие будет действительным и для плоского установившегося потока. В этом потоке радиальные площади сечения $F_r$, которые имеют цилиндрическую форму, непрерывно уменьшаются от периферийной границы потока к внутренней. Уменьшение величины этой радиальной площади сечения потока может быть сколь угодно большим, так как её пределом служит ось потока. По этой причине радиальные скорости потока могут быть сколь угодно большими. Это условие выполняется теоретически, практически оно ограничивается распределением потенциальной и радиальной кинетической энергии потока, то есть:

$$VP_{\text{ст}} \geq \frac{1}{2} V\rho W_{r\ \text{гран}}^2 . \qquad (84)$$

По зависимости (84) мы можем определить максимально возможные в потоке радиальные скорости движения, и из радиального уравнения движения (17) мы сможем найти радиальную площадь сечения $F_r$, на которой располагаются эти максимально возможные скорости движения. Вот эта радиальная поверхность определит нам внутреннюю границу плоского установившегося потока.

В установившемся потоке мы назвали подобную площадь сечения потока критической. В плоском установившемся потоке она носит название внутренней границы плоского установившегося потока. Там мы назвали скорости движения в этой площади тоже критическими, а для плоского установившегося движения мы назвали радиальные скорости движения $W_{r\ \text{гран}}$. Для газового потока критические скорости равны скорости звука в этом газе, соответственно и радиальные граничные скорости тоже будут равны скорости звука в соответствующем газе.

Отметим ещё одну особенность. Полную энергию плоского установившегося потока мы выше определили как разность энергий жидкости, находящейся в ванне, и энергией открытой воронки ванны. Таким способом мы определили, что поток жидкости организуется при наличии двух источников с различной величиной энергии. Полная энергия потока $E_{\text{пол}}$ соответствует лишь разнице величин этих двух источников энергии. Предельное распределение между кинетической и потенциальной энергией в потоке идёт, например, по условию (84) в пределах разницы двух источников энергий, между которыми образуется соответствующий поток. Это условие движения жидкостей и газов в установившемся и плоском установившемся потоках – очень важное!

**Далее рассмотрим акустический, или волновой, вид движения жидкостей и газов.**

Предварительно вспомним, что наглядной формой образования акустических волн служит вибрирующая пластинка. При возвратно-поступательном её движении образуются волны первого и второго этапов движения этой пластины, которые распространяются в окружающем пространстве в среде воздуха или жидкости. Дальше посмотрим, каким образом образуется здесь энергия и работа. Для чего на рис. 12, как мы раньше показывали на рис. 6, изобразим первый этап движения поршня как части вибрирующей пластины.

Перед поршнем мы показали условный объём $V_0$ потока, который располагается на длине $l$ хода поршня. Мы рассматриваем только первый этап движения поршня, из первого положения во второе. При движении поршня в нулевом объёме образуется поток жидкости. Переместившись из первого положения во второе, поршень совершит работу, вытеснив из условного объёма $V_0$ жидкость. Как вы уже знаете, работа $L$ равна объёму жидкости, умноженному на давление. Объём вытесненной жидкости нам известен: он будет равняться $V_0$, то есть условному объёму. Давление будет равняться динамическому давлению.

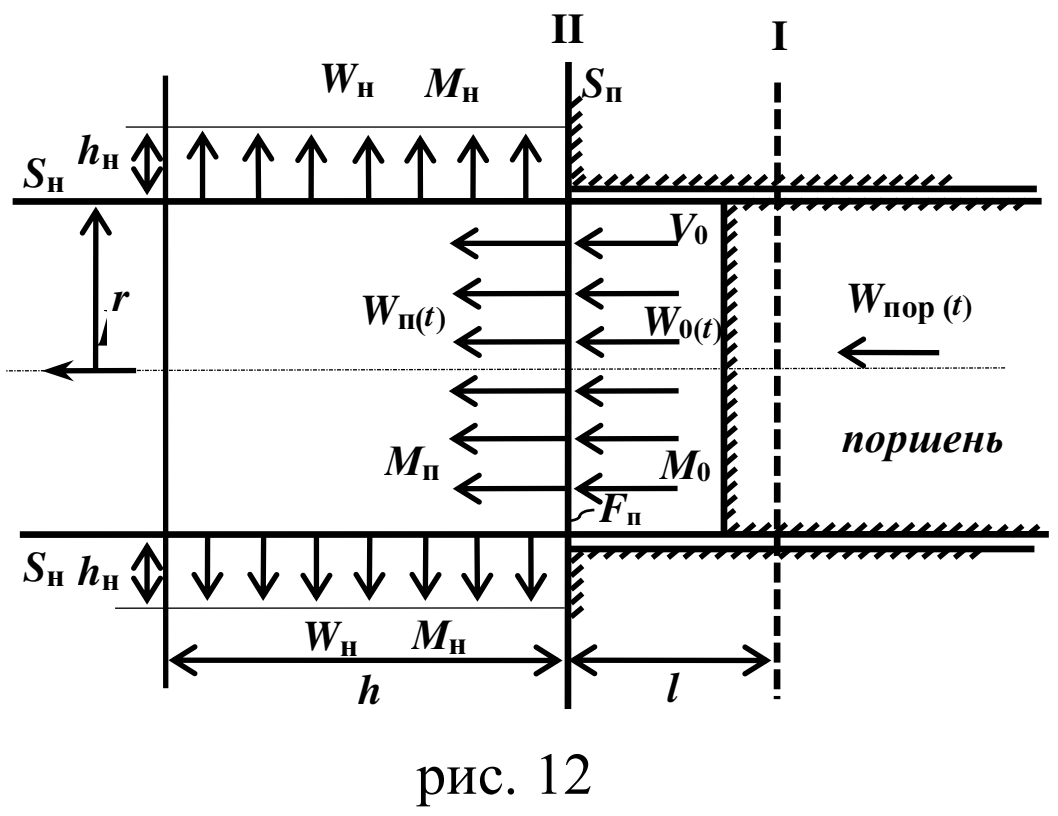

рис. 12

Тогда работу, совершенную поршнем на первом этапе его движения, можно будет записать в таком виде:

$$L_{\text{пор}} = V_0 P_{\text{п.дин.}} \tag{85}$$

Динамические силы давления $P_{\text{п.дин}}$ в прямом направлении, согласно уравнению (61), будут равны:

$$P_{\text{п.дин}} = \frac{M_{\text{п}} W_{\text{п}}(t)}{F_{\text{п}}}$$

Подставим их в уравнение (85), получим:

$$L_{\text{пор}} = V_0 \frac{M_{\text{п}} W_{\text{п}}(t)}{F_{\text{п}}}. \tag{86}$$

По уравнению движения заменим расход массы $M_{\text{п}}$ через характеристики потока, получим:

$$L_{\text{пор}} = V_0 \rho W_{\text{п}}^2(t). \tag{87}$$

В наших уравнениях к поступательной скорости $W_{\text{п}}$ приписан знак $t$. Он обозначает, что скорость потока может изменяться во времени.

В уравнении (87) поступательную скорость $W_п(t)$ надо заменить скоростью движения поршня. Ибо жидкость или газ будет двигаться со скоростью поршня $W_{пор}(t)$. Тогда уравнение (87) примет вид:

$$L_{пор} = V_0 \rho W_{пор}^2(t)\ . \qquad (88)$$

Уравнение (88) выражает работу, совершенную поршнем по выталкиванию жидкости в нулевом объёме $V_0$.

Работа поршня будет переходить в энергию прямолинейного потока жидкости. Энергия такого потока записывается уравнением (71), где полная энергия потока ($VP_{пол}$) равна сумме потенциальной ($VP_{ст}$) и кинетической ($1/2\ V\rho W^2$) энергий. Теперь надо применить это уравнение к нашим условиям движения потока. Полная энергия потока в нашем случае будет равна работе поршня, записанной уравнением (88), то есть

$$VP_{пол} = V_0 \rho W_{пор}^2(t)\ . \qquad (89)$$

Статическое давление так и остаётся статическим давлением, но оно уже будет статическим давлением прямолинейного потока $P_{п.ст}$. Объём потока $V_п$ будет равен площади сечения потока $F_п$, умноженной на длину потока $h$. Как мы выше установили, действительная длина потока при первом этапе движения поршня будет меньше на величину хода поршня $l$. Тогда объём потока будет равен:

$$V_п = F_п\ (h - l).$$

Потенциальную энергию потока мы сможем записать в таком виде:

$$E_{пол.\ п} = P_{ст.п} V_п = P_{ст.п} F_п\ (h - l). \qquad (90)$$

.

Для кинетической энергии объём потока будет таким же, как и для потенциальной энергии. Поступательная скорость будет равна скорости движения поршня, то есть $W_п(t) = W_{пор}(t)$. Так что для кинетической энергии потока можно записать любую из этих скоростей. В уравнении движения мы записали её как поступательную, поэтому и для кинетической энергии потока запишем её как поступательную. Тогда кинетическая энергия потока будет равна:

$$E_{кин.\ п} = \frac{1}{2} F_п (h - l) \rho W_п^2(t)\ . \qquad (91)$$

Теперь мы можем полностью записать уравнение энергии прямолинейного потока жидкости для первого этапа хода поршня. Оно будет иметь вид:

$$VP_{пол.п} = V_0 \rho W_{пор}^2(t) = P_{ст.п} F_п (h - l) + \frac{1}{2} F_п (h - l)\ \rho W_п^2(t)\ . \qquad (92)$$

Уравнение (92) будет уравнением энергий прямолинейного движения потока на первом этапе движения поршня. Это уравнение записано для полной волны первого этапа (по привычной классификации), то есть оно записано для полуволны первого этапа хода поршня.

До начала хода поршня в прямолинейном объёме потока жидкость будет находиться в состоянии покоя. В состоянии покоя она тоже будет обладать вполне определённым количеством энергии. Эту энергию можно себе представить, например, как энергию установившегося потока с бесконечно большой площадью сечения. В этом случае, как мы уже знаем, полная энергия будет равна потенциальной энергии потока, так как его кинетическая энергия равна нулю. Принимая объём невозмущённой среды равным объёму потока, то есть

$$V_{ср} = F_п (h - l),$$

а статическое давление невозмущённой среды будет соответствующим давлением среды $P_{ср}$, в этом случае мы получим энергию невозмущённой среды потока, которую можем записать:

$$E_{ср} = V_{ср}P_{ср} = P_{ср}F_{п}(h - l). \quad (93)$$

Поскольку объём потока невозмущённой среды обладает определённым количеством энергии, то первый этап движения поршня дополняет это количество энергии своей работой по вытеснению объёма жидкости. Поэтому мы будем обязаны прибавить к полной энергии среды полную энергию потока, то есть

$$E_{ср} + E_{пот}. \quad (94)$$

В развернутом виде полная энергия среды записана уравнением (93), а полная энергия потока – уравнением (92). Мы можем эти два уравнения подставить в уравнение (94):

$$V_0\rho W_{пор}^2(t) + P_{ср}F_{п}(h - l) = P_{ст.п}F_{п}(h - l) + \frac{1}{2}F_{п}(h - l)\,\rho W_{п}^2(t) + P_{ср}F_{п}(h - l). \quad (95)$$

Преобразуем правую часть уравнения (95). Для чего вынесем за знак скобок общий член $F_{п}(h - l)$. Получим:

$$V_0\rho W_{пор}^2(t) + P_{ср}F_{п}(h - l) = F_{п}(h - l)\cdot[P_{ст.п} + P_{ср} + \frac{1}{2}\rho W_{п}^2(t)]. \quad (96)$$

Уравнение (96) выражает действительно полную энергию потока жидкости в прямолинейном направлении. Непосредственно энергия потока, записанная уравнением (92), выглядит по отношению к полной энергии невозмущённой среды, записанной уравнением (93), дополнительным энергетическим импульсом. Это значит, что невозмущённая среда в результате первого этапа хода поршня в прямолинейном потоке получает прирост энергии, количество которой равно работе, совершенной поршнем при вытеснении жидкости на длине своего хода $l$. Следовательно, объём прямолинейного потока содержит в себе больше энергии, чем окружающая среда, на величину, равную работе поршня.

Как мы выше установили, поток жидкости образуется между двумя источниками с различной величиной энергии. В нашем случае источником с повышенным уровнем энергии является объём прямолинейного потока жидкости, а источником с пониженным уровнем энергии является окружающая среда. По этой причине между ними образуется поток жидкости, который мы назвали нормальным потоком. Энергию такого потока мы определяем непосредственно на нормальной площади сечения потока $F_{н}$. Полная энергия этого потока нам известна. Она равна полной энергии невозмущённой среды, то есть

$$E_{ср} = V_{н}P_{ср}. \quad (97)$$

Потенциальная энергия нормального потока будет:

$$E_{н.пот} = V_{н}P_{н.ст}. \quad (98)$$

Кинетической энергией нормального потока будет:

$$E_{н.кин} = \frac{1}{2}V_{н}P_{дин}. \quad (99)$$

Теперь постараемся расшифровать все составляющие части энергии нормального потока, после чего запишем их одним уравнением энергии. Начнём с нормального объёма потока $V_{н}$. Площадь сечения нормального потока $F_{н}$ представлена нами в виде цилиндра, поэтому её величина будет равна:

$$F_{н} = 2\pi r\left(C\frac{l}{W_{пор}(t)} - l\right).$$

Нормальная площадь сечения потока $F_н$ одновременно является началом движения нормальных скоростей потока. Следовательно, за время $t$, за которое поршень проходит расстояние $l$, нормальные скорости $W_н$ проделают путь равный $h_н$. Тогда время движения нормальных скоростей будет равно:

$$t = \frac{l}{W_{пор}(t)} .$$

Отсюда величина $h_н$ будет равна:

$$h_н = W_н(t) \frac{l}{W_{пор}(t)} .$$

Теперь мы можем определить объём нормального потока. Он будет равен:

$$V_н = F_н h_н = 2\pi r(C\frac{l}{W_{пор}(t)} - l)W_н(t) \frac{l}{W_{пор}(t)} . \qquad (100)$$

Уравнение (100) будет выражать объём нормального потока жидкости за полное время движения поршня на первом этапе.

Отметим, что величины для записей всех форм зависимостей и уравнений мы берём в усреднённом виде. Ибо все эти зависимости служат для количественной расшифровки движения жидкостей и газов. Поэтому их надо отнести к разряду базовых уравнений, относительно которых в последующем эти же уравнения расписываются для конкретных условий движения жидкостей и газов с учётом необходимой точности расчётов.

Уравнением (100) мы определили объём нормального потока жидкости на первом этапе движения поршня и выяснили, что он зависит от нормальных скоростей $W_н$ движения потока, а не от скорости возмущения $C$, как объём прямолинейного потока. Этим мы просто подчеркнули характер нормального движения. Но вы должны помнить, что площадь сечения нормального потока полностью зависит от скорости возмущения среды $C$.

Динамическую энергию для нормального потока легко можно определить, взяв динамические силы давления по уравнению (64), а нормальные скорости – по уравнению движения (59). Если всё это расписать в развернутом виде, то мы получим довольно длинную зависимость. Поэтому запишем уравнение энергии нормального потока, используя зависимости (97), (98) и (99). Мы получим:

$$V_н P_{ср} = V_н P_{н.ст} + \frac{1}{2} V_н P_{дин}. \qquad (101)$$

Уравнение (101) является уравнением энергии нормального потока жидкости на первом этапе движения поршня. Полная энергия потока здесь определяется полной энергией находящейся в состоянии покоя среды. Объём нормального потока определяется по уравнению (100). Кинетическую энергию мы можем получить, используя уравнения (59), (64) и (99). Зная полную энергию потока и кинетическую энергию, мы легко можем определить по уравнению (101) потенциальную энергию потока.

Мы получили все необходимые зависимости для энергии на первом этапе движения поршня.

На втором этапе движения поршня, из второго положения в первое, сущность изменится лишь с учётом обратного движения хода поршня. Все объяснения и дополнения останутся одинаковыми как для первого, так и для второго этапов его движения. Поэтому мы отметим лишь различие и запишем зависимости с учётом этого различия. Различие, связанное с длиной волны, или длиной возмущённого прямолинейного движения, которая будет отличаться на две длины хода поршня, мы учли выше, в уравнениях движения и сил.

Энергия потока, образовавшаяся при движении поршня в обратном направлении, будет иметь отрицательную величину. Поэтому она будет не увеличивать полную энергию жидкости объёма прямолинейного потока, находящегося в состоянии покоя, а уменьшать. Поэтому мы должны будем вычесть полную энергию потока из полной энергии среды. Покажем это на зависимостях. Зависимость для полной энергии будет иметь вид:

$$VP_{\text{пол.п}} = V_0\rho W_{\text{пор}}^2(t) = P_{\text{ст.п}}F_{\text{п}}(h + l) + \frac{1}{2}F_{\text{п}}(h + l)\rho\, W_{\text{п}}^2(t)\,. \tag{102}$$

Зависимость полной энергии для неподвижной среды будет иметь вид:

$$E_{\text{пол.п}} = V_{\text{п}}P_{\text{ст.п}} = P_{\text{ст.п}}F_{\text{п}}(h + l). \tag{103}$$

Теперь запишем зависимость абсолютно полной энергии для прямолинейного потока на втором этапе хода поршня. Она будет иметь вид:

$$V_0\rho W_{\text{пор}}^2(t) - P_{\text{ст.п}}F_{\text{п}}(h + l) = F_{\text{п}}(h + l)\cdot[P_{\text{ст.п}} - P_{\text{ср}} + \frac{1}{2}\rho W_{\text{п}}^2(t)\,]. \tag{104}$$

По этой причине объём прямолинейного потока станет источником с меньшим количеством энергии, а неподвижная среда – источником с большим количеством энергии. Тогда нормальный поток жидкости образуется как движение из неподвижной среды в прямолинейный объём потока через нормальную площадь сечения. Объём нормального потока будет определяться теми же условиями, что и объём нормального потока первого этапа движения поршня. Поэтому без дополнительных объяснений запишем этот объём. Он будет равен:

$$V_{\text{н}} = F_{\text{н}}h_{\text{н}} = 2\pi r(C\frac{l}{W_{\text{пор}}(t)} + l)W_{\text{н}}(t)\;\frac{l}{W_{\text{пор}}(t)}\;. \tag{105}$$

Основные зависимости для полной энергии нормального потока получаются из тех же условий, что и для нормального потока первого этапа движения поршня. Запишем эту зависимость:

$$V_{\text{н}}P_{\text{ср}} = V_{\text{н}}P_{\text{н.ст}} + \frac{1}{2}V_{\text{н}}P_{\text{дин}}. \tag{106}$$

Уравнение (106) является уравнением энергии нормального потока для второго этапа движения поршня.

В этом уравнении полная энергия потока тоже равна полной энергии среды, жидкость которой находится в состоянии покоя. Затем определяется кинетическая энергия по уравнениям движения, сил и кинетической энергии. Будем считать, что мы получили все необходимые зависимости для нормального потока жидкости на втором этапе движения поршня.

Теперь определим крайние возможные пределы для уравнений энергии акустического вида движения. В какой-то степени они нам должны быть известны, вернее, нам уже должен быть понятен метод определёния пределов распределения энергии.

Опять начнём с первого этапа движения поршня. В прямолинейном потоке жидкости работа хода поршня преобразуется в энергию. Из уравнения энергии преобразования работы поршня следует, что, какая бы ни была скорость движения поршня, мы всё равно не нарушим предельное распределение между потенциальной и кинетической энергией в этом потоке, которое определяется неравенством:

$$V_{\text{п}}P_{\text{ст.п}} \geq \frac{1}{2}V_{\text{п}}\rho W_{\text{п}}^2(t)\,,$$

так как сам поток выражает разницу в приросте энергий между двумя источниками энергии: работой поршня и полной энергией покоящейся среды. Это условие сохраняется именно для разницы энергий источников с различной её величиной. Именно движением поршня создаётся эта разница. По этой причине движение поршня, какова бы ни была бы скорость его движения, не нарушит неравенства распределения кинетической и потенциальной энергий. Это значит, что с ростом скорости поршня происходит прирост энергии в объёме потока относительно покоящейся среды.

Здесь мы можем, как вы понимаете, изменить площадь сечения потока, так как движение поршня происходит в бесконечно большом пространстве среды. Среда же показывает, что прямолинейный поток ограничивается нормальной поверхностью $S_н$, через которую образуется нормальный поток жидкости. В нормальном потоке, полная энергия которого определяется полной энергией неподвижной среды, распределение между потенциальной и кинетической энергиями зависит от прироста энергии в прямолинейном потоке. Для этого потока запишем условие распределения потенциальной и кинетической энергии:

$$V_н P_{ст.н} \geq \frac{1}{2} V_н P_{дин}. \qquad (107)$$

В прямолинейном потоке полная энергия потока растет с ростом скорости поршня. В нормальном же потоке полная энергия имеет определённую постоянную величину, которая равна полной энергии покоящейся среды. Нормальная скорость $W_н$ и кинетическая энергия нормального потока находятся в прямой зависимости от скорости поршня $W_{пор}$. С ростом скорости поршня растет скорость нормального потока и его кинетическая энергия $E_{кин}$. В этом легко убедиться, так как разность, или прирост, энергий между движущимся поршнем и неподвижной средой нормального потока одинакова и равна по величине. По этой причине скорость $W_н$ нормального потока тоже будет равна скорости движения поршня $W_{пор}$, то есть

$$W_н(t) = W_{пор}(t).$$

Поэтому мы можем проследить изменение нормальной скорости по изменению движения поршня, и в зависимости от этого мы можем проследить изменение энергии в прямом и нормальном направлении движения потока жидкости на первом этапе хода поршня.

В прямолинейном потоке при любой скорости движения поршня всегда происходит прирост энергии относительно покоящейся среды невозмущённого потока. Вот этот прирост энергии покоящейся среды с физической точки зрения может происходить либо за счёт температурных изменений жидкости, либо за счёт сжатия жидкости в объёме прямолинейного потока, которое тоже характеризуется изменением, вернее, повышением температуры жидкости в пределах этого же объёма. Все эти физические характеристики состояния жидкостей и газов определяются термодинамическими зависимостями. По этой причине в уравнение энергии прямолинейного потока следовало бы ввести третий член, который учитывает термодинамические изменения среды, как мы это сделали в уравнении (75), а сама зависимость для этого члена записана у нас уравнением (73). Вам это пока просто надо запомнить, а при практическом применении этих зависимостей вы самостоятельно сможете дополнить уравнения энергий этой термодинамической частью, которая учитывает термодинамические изменения энергии по физическому состоянию жидкостей и газов. Практическое увеличение скорости движения поршня будет ограничиваться только этими состояниями жидкостей и газов.

Для нормального потока, когда скорости движения поршня лежат в пределах неравенства (107), в объёме жидкости нормального потока происходит простое перераспределение между потенциальной и кинетической энергией в пределах полной энергии потока, или полной энергии невозмущённой среды.

При дальнейшем увеличении скорости движения поршня, когда происходит нарушение неравенства (107), доля кинетической энергии начинает преобладать над долей потенциальной энергии нормального потока жидкости, то есть величина кинетической энергии становится больше величины потенциальной энергии. В этом случае доля потенциальной энергии тоже начинает расти за счёт физических изменений состояния жидкостей и газов. Что в сумме приводит к росту полной энергии нормального потока, то есть тоже происходит прирост энергии потока. Этот прирост энергии учитывается уже термодинамическими зависимостями, так как он идёт за счёт увеличения температур жидкостей и газов в объёме нормального потока. По этой причине мы должны будем дополнить уравнение энергий (106) нормального потока жидкости термодинамическими составляющими. Прирост полной термодинамической энергии обозначим как

$$E_{ст.\,тер} = V_н P_{тер}. \qquad (108)$$

Прирост термодинамической энергии в правой части уравнения (108) обозначим уравнением (73), то есть

$$E_{\text{тер}} = \int_{V_{\text{н}}}^{V_2} P_{\text{тер}}\, dV.$$

Дополним этими составляющими уравнение (106). Тогда получим:

$$V_{\text{н}}P_{\text{ср}} + V_{\text{н}}P_{\text{тер}} = V_{\text{н}}P_{\text{н.ст}} + \frac{1}{2}V_{\text{н}}P_{\text{дин}} + \int_{V_{\text{н}}}^{V_2} P_{\text{тер}}\, dV. \qquad (109)$$

Уравнение (109) учитывает прирост термодинамической энергии в объёме нормального потока жидкости на первом этапе движения поршня, когда его увеличивающаяся скорость нарушает неравенство (107).

В то же время неравенство (107), в конечном итоге, не нарушается, а видоизменяется, что мы видим из уравнения (109). Оно лишь принимает иную форму, то есть

$$V_{\text{н}}P_{\text{н.ст}} + \int_{V_{\text{н}}}^{V_2} P_{\text{тер}}\, dV \geq \frac{1}{2}V_{\text{н}}P_{\text{дин}}, \qquad (110)$$

поскольку мы наблюдаем одновременно прирост полной энергии нормального потока.

Дополним, что термодинамическая энергия относится к разряду потенциальной энергии, которая реализуется через кинетическую энергию. На этом, будем считать, мы полностью рассмотрели предельные возможности распределения и прироста энергии для прямолинейного и нормального потоков жидкости и газов на первом этапе движения поршня или вибрирующей пластинки.

Отметим, что сам первый этап движения поршня и присущие ему потоки жидкостей и газов могут существовать сколь угодно долгое время. Например, картину первого этапа движения поршня со сколь угодно долгим временем движения мы можем наблюдать при движении самолёта в воздухе или при движении корабля в воде. В этом случае в лобовой части как самолёта, так и корабля организуются и прямолинейные, и нормальные потоки, присущие первому этапу движения поршня. В этом случае предельные состояния потоков будут выглядеть следующим образом: при полёте самолёта со скоростями в пределах неравенства (107) в прямолинейном потоке происходит незначительный разогрев газов в соответствии с термодинамическими законами, так как воздух объёма этого потока получает дополнительный прирост энергии. В нормальном потоке идёт простое распределение полной энергии покоящегося воздуха или полной энергии нормального потока между кинетической и потенциальной.

Когда скорость самолёта достигает скорости звука, или критической скорости, в прямолинейном потоке происходит более интенсивный разогрев воздуха, соответствующий скорости движения. В нормальном потоке скорости воздуха достигают тоже скорости звука, а неравенство (107) находится на предельной границе. При дальнейшем увеличении скорости самолёта, которая будет в той или иной степени больше скорости звука, для нормального потока вступает в силу неравенство (110).

При движении корабля в воде со скоростями в пределах неравенства (107), в нормальном потоке не учитывается термодинамическая часть энергии. При движении корабля в воде за пределами неравенства (107) в нормальном потоке должна увеличиваться термодинамическая составляющая энергии потока. Граничным условием неравенства (107) соответствует скорость движения корабля порядка 40 км/час или 10 м/сек, поскольку атмосферное давление у поверхности воды равно приблизительно 1 кг/см$^2$. Вот эта скорость будет критической скоростью для воды в нашем случае. Эта критическая скорость воды не характеризуется особо видимым эффектом. По этой причине она выпадает из поля наблюдения исследователей. Просто после этой скорости начинается более интенсивный разогрев нормального потока жидкости при дальнейшем увеличении скорости корабля, то есть в этом случае мы должны будем пользоваться неравенством (110).

При скорости корабля свыше 150 км/час или 40 м/сек, температура в потоке жидкости достигает такой величины, что начинается кипение жидкости в объёме нормального потока. Это кипение жидкости при подобных условиях называют кавитацией. Ибо давление в паровых пузырьках делается настолько большим, что оно нарушает целостность металла обшивки корабля. Это будет вторая замечательная точка, относящаяся к физическому состоянию жидкостей. В этом случае мы тоже будем пользоваться неравенством (110).

При увеличении скорости корабля свыше 150 км/час, или 40 м/сек, вода в прямом и нормальном потоках будет переходить сразу в парообразное состояние. С увеличением скорости движения корабля температура пара в потоке будет расти до определённого предела, на этом уровне она будет держаться в определённом диапазоне повышения скорости корабля. Следующая замечательная точка наступит тогда, когда скорость корабля достигнет скорости звука для парового потока. Приблизительно эта скорость будет порядка 800 км/час. На этом счёт замечательным точкам физического состояния заканчивается. Во всех этих случаях мы должны руководствоваться неравенством (110). Так выглядят замечательные точки для жидкостей и газов при распределении энергии в прямолинейном и нормальном потоках.

Нам осталось рассмотреть замечательные точки для второго этапа движения поршня - из второго положения в первое. При данном движении поршня происходит уменьшение энергии неподвижной среды. Поэтому для прямого и нормального потоков этого этапа движения поршня мы должны будем руководствоваться неравенством (107). Оно говорит о том, что при достижении поршнем критической скорости наступают конечные граничные условия неравенства (107), то есть в этом случае скорости жидкостей и газов потока достигают своих максимальных величин и дальнейшее увеличение этих скоростей невозможно. По этой причине дальнейшее увеличение скорости движения поршня не изменит результатов движения жидкостей и газов в потоках, то есть они останутся неизменными. Следовательно, и динамические силы взаимодействия между поршнем и жидкостью должны оставаться неизменными при дальнейшем росте скоростей поршня.

На этом закончим разбор замечательных точек распределения энергии при акустическом, или волновом, движении жидкостей и газов.

**Нам осталось рассмотреть расходное движение жидкостей и газов**.

Вспомним, что расходное движение наглядно представляет собой некоторый объём, куда втекают и вытекают многие потоки. В общем, это вроде нашего сердца, которое стараются понять, и на самом понятном месте всё становится снова непонятным. Потому что всё сводят к клапанам, которые научились заменять пластмассой. Радости ведь в этом мало. Было бы лучше, если бы клапаны работали всегда нормально. У нас задача потруднее, коль мы не можем свести всё к обыкновенным пластмассовым или железным клапанам. У нас так и остался некоторый объём, куда втекают и вытекают многие потоки без всяких клапанов, да и границы объёма могут изменяться или оставаться неподвижными в зависимости от силовых условий границы, вернее, в зависимости от его энергетических возможностей. Ибо сила есть проявление энергии.

По отношению к втекающим и вытекающим потокам объём расходного потока представляет собой результат их движения в течение определённого времени. Поэтому он относится к разряду совершённого. Это значит, что основной объём потока мы должны воспринимать как работу (см. рис. 4). Работа, как вы знаете, равна произведению объёма потока на давление. Объём потока непрерывно изменяется во времени – $V_{\text{пот}}(t)$. Давление потока тоже зависит от времени, тогда обозначим его как $P_{\text{пот}}(t)$. Теперь мы можем записать работу расходного потока:

$$L_{\text{пот}} = V_{\text{пот}}(t)\, P_{\text{пот}}(t)\,. \tag{111}$$

Работа расходного потока есть результат переноса энергии потоками в течение какого-то времени в него втекающими и вытекающими из него. Потоки, втекающие в объём жидкости, должны будут иметь энергию, превышающую работу расходного потока. Одновременно все составляющие её части энергий будут зависеть от времени. Поэтому обозначим параметры этого уравнения значком времени $t$, а то, что поток втекающий, обозначим значком $_{\text{вт}}$. Тогда для одного из втекающих потоков мы можем записать:

$$V_{\text{вт}}(t)\, P_{\text{вт}}(t) = V_{\text{вт}}(t)\, P_{\text{вт.ст}}(t) + V_{\text{вт}}(t)\, P_{\text{вт.дин}}(t)\,. \tag{112}$$

Но этот поток втекает в объём расходного потока в течение какого-то времени. По этой причине мы должны будем поставить в зависимость от времени. Для этого поступим следующим образом. Представим объём потока как площадь потока $F_{\text{вт}}$, умноженную на длину $l$, то есть

$$V_{\text{вт}}(t) = F_{\text{вт}} l_{\text{вт}}(t)\,. \tag{113}$$

Длину потока $l_{\text{вт}}$ мы, в свою очередь, можем представить как скорость потока $W_{\text{вт}}$, умноженную на время $t$, то есть

$$l_{\text{вт}} = W_{\text{вт}} t. \tag{114}$$

Заменим в уравнении (112) объём потока через зависимости (113) и (114). Получим:

$$F_{\text{вт}} W_{\text{вт}} \cdot t \cdot P_{\text{вт}}(t) = F_{\text{вт}} W_{\text{вт}}(t) \cdot t[P_{\text{вт.ст}}(t) + P_{\text{вт.дин}}(t)]. \tag{115}$$

Уравнением (115) мы записали энергию втекающего потока в зависимости от времени.

Точно таким же способом мы можем записать энергию вытекающего потока в зависимости от времени, но только со значком $_{\text{ист}}$, что обозначает истекающий поток. Тогда уравнение будет иметь вид:

$$F_{\text{ист}} W_{\text{ист}}(t) \cdot t \cdot P_{\text{ист}}(t) = F_{\text{ист}} W_{\text{ист}}(t) \cdot t[P_{\text{ист.ст}}(t) + P_{\text{ист.дин}}(t)]. \tag{116}$$

Уравнение (116) обозначает энергию истекающих расходов.

Коль втекающих и вытекающих потоков из объёма расходного потока может быть очень много, то мы запишем работу расходного потока как сумму разностей втекающих и вытекающих потоков. Тогда мы получим такое уравнение:

$$L_{\text{пот}} = V_{\text{пот}}(t) P_{\text{пот}}(t) = \sum_{1}^{n} F_{i\ \text{вт}} W_{i\,\text{вт}}(t)[\,P_{i\,\text{вт.ст}}(t) + P_{i\,\text{вт.дин}}(t)] - \sum_{1}^{n} F_{j\ \text{ист}} W_{j\,\text{ист}}(t) t[P_{j\,\text{ист.ст}}(t) + P_{j\,\text{ист.дин}}(t)]. \tag{117}$$

Уравнение (117) является уравнением работы расходного вида движения жидкостей.

Уравнения (111) – (117) тоже являются базовыми уравнениями, которые должны расписываться для каждых конкретных условий движения жидкостей и газов. Ещё можно отметить, что эти уравнения при необходимости дополняются третьим членом – термодинамической частью энергии.

## Глава 4. СОСТОЯНИЕ ПОКОЯ ЖИДКОСТЕЙ И ГАЗОВ

В своей жизненной практике вам часто приходится встречаться с жидкостями и газами, находящимися в состоянии покоя. Примером могут служить озера, воздух в летний тихий день, сжатый воздух, находящийся в баллоне под давлением и т.д. Состояние покоя жидкостей и газов мы можем определить как состояние застывшего движения. По этой причине законы механики безынертной массы будут приемлемы и для состояния покоя жидкостей и газов.

Согласно первому закону механики безынертной массы, жидкости и газы в состоянии покоя сохраняют энергию силового поля. Силовые поля, как вы знаете, бывают векторными и скалярными. Рассмотрим сначала состояние покоя жидкостей и газов в скалярном силовом поле. Объём покоящейся жидкости или газа в этом случае мы можем принять за объём установившегося потока, площадь сечения которого очень велика. Поэтому мы можем принять скорости движения потока равными нулю. В этом случае динамическая часть энергии потока будет равна нулю, а полная энергия потока будет равна полной статической энергии того же потока. Тогда энергию этого потока можно записать в таком виде:

$$E_{\text{п}} = VP_{\text{п}} = VP_{\text{ст}}. \tag{118}$$

Уравнение (118) выражает полную энергию покоящейся в скалярном силовом поле жидкости или газа. Здесь, как мы видим, энергия поля $E_{\text{п}}$ равна произведению объёма покоящейся жидкости $V$ на давление $P_{\text{п}}$ покоящейся жидкости, или на статическое давление $P_{\text{ст}}$ застывшего установившегося потока. Если сделать этот объём проточным, то жидкость начнёт двигаться в соответствии с зависимостями установившегося потока, то есть, действительно, состояние покоя жидкостей и газов в скалярном силовом поле представляет собой застывший поток установившегося вида движения.

Если этот объём жидкости или газа нагревать или охлаждать, то соответственно его энергия будет либо увеличиваться, либо уменьшаться. По этой причине значительную часть полной энергии покоя жидкостей и газов составляет ещё термодинамическая энергия жидкостей и газов. Тогда мы должны будем дополнить энергию покоя жидкостей и газов, записанную уравнением (118), дополнительным членом, который выражает содержание термодинамической энергии в покоящихся жидкостях и газах как

$$E_{\text{терм}} = VP_{\text{терм}}. \quad (119)$$

Эту часть термодинамической энергии, выраженную уравнением (119), мы должны будем подставить в уравнение (118).

*Примечание:* при выводе уравнений состояния покоя жидкостей и газов в скалярном силовом поле мы везде говорим, что рассматриваем застывшее движение установившегося потока. Это не верно. Мы здесь рассматриваем застывшее движение расходного потока жидкостей и газов. Просто мы здесь оговорились. Поэтому вам необходимо будет учитывать это обстоятельство[9]

Будем считать, что мы получили необходимое понятие о состоянии покоя жидкостей и газов в скалярном силовом поле. Теперь рассмотрим состояние покоя жидкостей и газов в векторном силовом поле. Таким силовым полем, обладает, например, наша Земля. Поэтому, как вы знаете, с увеличением глубины погружения в любом бассейне нашей планеты давление тоже увеличивается пропорционально величине этого погружения. Поэтому будем считать, что мы в этом случае имеем дело тоже с застывшей формой установившегося потока, направление движения которого совпадает с направлением земного радиуса, то есть движение направлено от точки погружения к центру Земли по прямой радиуса Земли. Тогда уравнением движения застывшего установившегося движения будет уравнение такого вида:

$$M = \rho FhW. \quad (120)$$

Мы записали уравнение (120) без вывода, который вы самостоятельно можете сделать без всякого труда. Оно обозначает, что застывший расход массы в единицу времени $M$ равен произведению плотности жидкости $\rho$ умноженной на площадь сечения потока $F$, умноженной на глубину погружения $h$ и умноженной на застывшую скорость движения потока $W$. Здесь под площадью сечения $F$ потока понимается поверхность, расположенная либо на одинаковой глубине погружения $h$, либо все точки которой имеют одинаковое давление.

Застывшие динамические силы давления берутся по второму закону механики безынертной массы, который можно записать следующей зависимостью:

$$P = \frac{M}{F}W. \quad (121)$$

Уравнением (121) записано, что застывшие динамические силы давления $P$ равны застывшему расходу массы $M$, делённому на площадь сечения $F$ потока и умноженному на застывшие скорости движения $W$ потока. В уравнении (121) по уравнению (120) заменим расход массы через характеристики потока. Тогда получим:

$$P = \rho hW^2. \quad (122)$$

---

[9] По *мнению редактора*, т.к. скалярные силовые поля образуют сферу действия сил давления в виде точек среды, а сфера является объёмом по определению, то состояние покоя среды в скалярном силовом поле действительно надо рассматривать как застывшую форму расходного вида движения. Должны разбираться специалисты. Из примечания автора видно, что ненормальные условия создания этого труда оказывали влияние и на стиль изложения, и на содержание, и что автор это знал, но не имел возможности перепечатать.

В уравнениях движения и сил нам известны все характеристики, которые мы всегда можем замерить, кроме одной из них – это застывшая скорость движения $W$ потока.

Чтобы определить её величину, мы поступим следующим образом. Выделим в каком-либо бассейне жидкости определённый столб жидкости с высотой $h$ и площадью сечения $F$, направление высоты которого совпадает с направлением застывшего движения. Согласно законам механики твёрдого тела, этот объём выделенной жидкости будет иметь вес, или силу веса $R$, которая будет равна массе $m$ выделенного объёма, умноженной на ускорение земного тяготения $g$. Величина этого ускорения нам известна: она равна $g$ = 9,81 м/сек$^2$. Тогда вес выделенного объёма будет равен:

$$R = m \cdot g. \tag{123}$$

Вес $R$ выделенного объёма потока давит на основание потока $F$, расположенного на глубине $h$, не сосредоточенной силой, а распределённой. Поэтому распределённые силы веса $R_{\text{расп}}$ мы можем записать как вес $R$, делённый на площадь сечения $F$ потока. Мы получим:

$$P_{\text{расп}} = \frac{R}{F} = \frac{mg}{F}. \tag{124}$$

Распределённые силы веса должны будут равняться застывшим динамическим силам давления $P$ на площади сечения $F$ , расположенной на глубине $h$, то есть

$$P = P_{\text{расп}}. \tag{125}$$

Массу выделенного объёма потока в уравнении (124) можно представить как произведение площади сечения потока $F$ глубины $h$ погружения и плотности жидкости $\rho$, то есть

$$m = \rho F h.$$

Подставим её в уравнение (124). Получим:

$$P_{\text{расп}} = \rho h g. \tag{126}$$

Теперь в уравнение (125) подставим значение застывших динамических сил давления (122) и значение распределённых сил веса выделенного объёма потока (126). Получим:

$$\rho h W^2 = \rho h g. \tag{127}$$

Равенство (127) мы можем сократить на произведение $\rho h$. Тогда оно примет вид:

$$W^2 = g.$$

Отсюда следует, что квадрат застывших скоростей движения потока равен ускорению в поле земного тяготения $g$ = 9,81 м/сек$^2$. Отсюда можно получить величину застывшей скорости движения. Она будет равна:

$$W = \sqrt{9{,}81} = 3{,}132(\text{м/сек}).$$

Эта величина застывших скоростей движения потока есть величина постоянная для жидкостей и газов [среды] в поле земного тяготения, как и величина ускорения [тел] в поле земного тяготения.

Отметим, что в размерности ускорения $g$ поля земного тяготения, которое имеет размерность: метр делённый на секунду в квадрате, не извлекается квадратный корень для размерности скорости застывшего движения потока, которая имеет размерность: метр делённый на секунду. Вы скажете, что из секунды в квадрате можно извлечь корень квадратный, а из метра – нет. Противоречие. На самом деле здесь противоречия нет. Каким способом оно устраняется, смотри *Механика жидкости и газа, или механика безынертной массы*. Мы просто не даём здесь этого объяснения, чтобы не отнимать лишнего времени. Чтобы выделить застывшую скорость движения жидкостей и газов в поле земного тяготения, обозначим её буквой $w$, $w$ = 3,132 м/сек.

Для состояния покоя жидкостей и газов в векторном силовом поле тоже есть зависимости, которые определяют величину их энергии $E_{в}$. Энергия состояния покоя в этом случае записывается относительно площади сечения $F$ застывшего потока расположенной на глубине $h$. В поле земного тяготения эта энергия записывается такой зависимостью:

$$E_{в}=V\rho w^2 h. \qquad (128)$$

Уравнение (128) означает, что энергия состояния покоя $E_{в}$ для площади сечения застывшего потока, расположенной на глубине $h$, равна произведению объёма $V$ потока, который расположен выше площади сечения застывшего потока, плотности жидкости или газа $\rho$, квадрату скорости поля земного тяготения $w$ и глубины потока $h$.

***

Мы рассмотрели все необходимые законы и положения механики безынертной массы, которые нам будут нужны для нашего путешествия.

Отметим только, что виды движения жидкостей и газов, которых мы насчитываем всего четыре, в потоках жидкостей и газов могут существовать как в чистом виде, так и в смешанном виде, где в одном потоке одновременно могут существовать два и более вида движения жидкостей и газов. Например, дует ветер, и мы разговариваем. Здесь на установившийся вид движения накладывается акустический вид движения.

Реальные жидкости и газы обладают вязкостью и при своём движении испытывают сопротивление, которое уменьшает энергию потока. Нам этого знать пока не нужно, поэтому мы об этом лишь упоминаем.

Полная и потенциальная энергия, полное и статическое давление жидкостей и газов измеряется приборами и датчиками манометрического типа, где рабочим органом служит некоторый объём. Например, в манометрах рабочим органом является изогнутая металлическая трубка, которая связана со стрелкой манометра. Для измерения динамических сил давления пригодны датчики и приборы, рабочим органом которых является либо плоскость, либо поверхность и на которые действуют непосредственно динамические силы давления. Это замечание сделано потому, что в настоящее время принято считать, что динамические и статические силы давления одинаковы по своей природе. Поэтому величину тех и других сил давления определяли по уравнению энергии Бернулли. При этом величина статических сил давления поддаётся такому методу измерения, то есть замеренные и действительные величины совпадают. Величину динамических сил давления при таком способе замера уменьшили ровно в два раза по сравнению с действительной. Поэтому я обращаю внимание на этот факт[10].

<…>[11]

---

[10]1. Конструкция прибора для измерения динамических сил давления дана в работе «Механика жидкости и газа, или механика безынертной массы».

[11] Далее по тексту оригинала идёт расчёт крыльчатки центробежного насос. Расчёт перенесён редактором в «Приложение» к работе «Механика жидкости и газа, или механика безынертной массы»